%% file: main.tex
\documentclass[11pt]{article}

\usepackage{amsmath, amssymb, amsthm}
\usepackage{bbm}
\usepackage{graphicx}
\usepackage{fullpage}
\usepackage[round]{natbib}
\usepackage[colorlinks=true, citecolor=blue, linkcolor=blue, urlcolor=blue]{hyperref}
\usepackage{xcolor}
\usepackage{adjustbox}
\usepackage{booktabs}

\title{Does Rerandomization Help Beyond Covariate Adjustment? \\ 
A Review and Guide for Theory and Practice}
\author{
    Ant\^onio Carlos Herling Ribeiro Junior\thanks{Department of Statistics and Data Science, Carnegie Mellon University}
    \and
    Zach Branson$^*$
}
\date{}

\input{premble}

\begin{document}

\maketitle

\input{text/abstract.tex}

\section{Introduction}\label{sec:introduction}
\input{text/introduction.tex}

\section{History}\label{sec:history}
\input{text/history.tex}

\section{Setup and Notation}\label{sec:setup}

\input{text/notation}

\section{Rerandomization}\label{sec:rerandomization}
\input{text/rerandomization.tex}

\section{Theoretical Developments}\label{sec:theory}
\input{text/theory.tex}

\section{Simulations}\label{sec:simulations}
\input{text/simulations.tex}

\section{A Guide for Practitioners}\label{sec:practice}
\input{text/practice.tex}

\section{Conclusion}\label{sec:conclusion}
\input{text/conclusion.tex}

\section*{Acknowledgments}
The authors would like to thank Fan Li and Bryan Keller for helpful comments that improved the presentation of this work.

\bibliographystyle{plainnat}
\bibliography{main}

\clearpage
\begin{appendix}
\input{text/appendix}
\end{appendix}

\end{document}

%% file: premble.tex
\newcommand{\Q}{\boldsymbol{Q}}
\newcommand{\I}{\boldsymbol{I}}
\newcommand{\1}{\mathbbm{1}}
\newcommand{\Ical}{\mathcal{I}}

\newcommand{\R}{\mathbb{R}}
\newcommand{\Prob}{\mathbb{P}}
\newcommand{\E}{\mathbb{E}}

\newcommand{\var}{\text{Var}}
\newcommand{\cov}{\text{cov}}

\newcommand{\A}{\mathcal{A}}

\newcommand{\obs}{\text{obs}}
\newcommand{\adj}{\text{adj}}
\newcommand{\rr}{\text{rr}}
\newcommand{\eif}{\text{EIF}}

\newcommand{\converge}{\stackrel{\text{d}}{\to}}
\newcommand{\tauest}[1]{\widehat{\tau}_{\text{#1}}}

\newcommand{\tauestrr}{\widehat{\tau}_{\text{X}^{\text{rr}}}}

\newcommand{\mean}[1]{\overline{#1}}

%% file: text/abstract.tex
Rerandomization is a modern experimental design technique that repeatedly randomizes treatment assignments until covariates are deemed balanced between treatment groups. This enhances the precision and coherence of causal effect estimators, mitigates false discoveries from p-hacking, and increases statistical power. Recent work suggests that balancing covariates via rerandomization does not alter the asymptotic precision of covariate-adjusted estimators, thereby making it unclear whether rerandomization is worthwhile if adjusted estimators are used. However, these results have two key caveats. First, these results are asymptotic, leaving finite sample performance unknown. Second, these results focus on precision, while other potential benefits, such as increased coherence among flexible estimators, remain understudied. Hence, in this paper we provide three main contributions: (i) a comprehensive review of the rerandomization literature, covering historical foundations, theoretical developments, and recent methodological advancements, (ii) an extensive simulation study examining finite-sample performance, and (iii) a practical guide for practitioners. Our study compares precision, coherence, power, and coverage of various estimators under rerandomization versus complete randomization. We find rerandomization to be a complementary design strategy that enhances the precision, robustness, and reliability of causal effect estimators, especially for smaller sample sizes.

%% file: text/introduction.tex
Randomized experiments are considered the ``gold standard" for estimating causal effects in a wide range of disciplines, such as education \citep{mosteller2002evidence, raudenbush2020randomized}, economics \citep{athey2017econometrics, list2024experimental}, political science \citep{druckman2012cambridge, gerber2012field}, and medicine \citep{rosenberger2002randomization, carlin2007introduction}, because covariates are balanced on average and thus many causal effect estimators are unbiased, including the simple difference in mean outcomes. However, chance alone can result in covariate imbalance between treated and control units. Hence, conditional on this imbalance, estimators may become biased and lose precision, thus weakening the validity of statistical inference \citep{ding2021two}.

To address covariate imbalance, one could consider alternative randomized experimental designs, such as blocking or stratified randomization. These designs involve grouping units with similar covariate values into strata or blocks and then assigning treatment at random, thus ensuring more comparable treatment groups within each block and ultimately increasing precision of causal effect estimators \citep{fisher1935design, imai2008variance, miratrix2013adjusting, pashley2021insights, bai2022optimality, pashley2022block, tabord2022stratification}. However, these designs depend on discretized covariates, and it can be unclear how to implement blocking when covariates are continuous or high-dimensional. Instead of discretizing covariates, units could first be organized into matched pairs, and then treatment could be randomized within pairs \citep{imai2008variance, bai2022inference, bai2022optimality}. If units are observed sequentially, one option to address covariate imbalance is adaptive randomization designs \citep{hu2006theory, rosenberger2002randomization}. These designs change treatment allocation probability sequentially to maintain balanced treatment groups. Another alternative to address covariate imbalance, which is the one we focus on in this paper, is rerandomized designs. It is worth acknowledging that in the biostatistics and medical literature, particularly in the context of cluster-randomized trials, these designs are often referred to as constrained randomization \citep{li2016evaluation, li2017evaluation, zhou2022constrained}.

Rerandomization repeatedly randomizes treatment assignment until a prespecified covariate balance criterion is met. Although the concept of rerandomization has been commented about at least as soon as the 1950s \citep{grundy1950restricted, jones1958inadmissible, savage1962foundations}, and informally used by many practitioners for decades \citep{bruhn2009}, it was novelly formalized in \cite{morgan2012} as randomizing treatment assignment until the Mahalanobis distance between treatment and control covariate means is small. 
It has since been extended to a variety of experimental design settings, such as 2$^K$ factorial, stratified, sequential, survey, cluster, and split-plot designs \citep{brason2016improving, li2020factorial, zhou2018, wang2021, yang2021rejective, lu2023, shi2024rerandomization}.
Since rerandomization ensures comparable treatment groups not just in expectation but also in realization, a common result in the literature is that rerandomization increases precision and retains unbiasedness of causal effect estimators.

On the other hand, practitioners can also address covariate imbalance in the analysis stage using covariate-adjusted estimators \citep{fisher1935design, rosenbaum2002covariance}. In randomized experiments, it is well-known that covariate-adjusted estimators can increase precision while retaining unbiasedness \citep{lin2013, li2020rerandomization, ding2024, wang2024}. Thus, results have been developed that account for covariate imbalance either in the design, analysis, or both stages of an experiment. If all covariates used for rerandomization in the design stage are also used for adjustment in the analysis stage, 
then adjusted estimators' asymptotic marginal distributions are the same under complete randomization or rerandomization \citep{li2020rerandomization, wang2024}.
At first, this equality seems to suggest that rerandomizing in the design stage is not necessary as long as one adjusts for covariates in the analysis stage. However, this equality only holds asymptotically, only considers the marginal distributions of estimators, and depends on the amount of covariates available in the design and analysis stage.
Thus, several works have conjectured that rerandomization may still improve precision for adjusted estimators in finite samples \citep{yang2021rejective, zhu2024design, wang2024}. Furthermore, rerandomization may yield other potential benefits beyond increased precision.

For example, rerandomization has also been shown to increase the statistical power of hypothesis tests \citep{branson2024power}, thus lowering the sample size needed to achieve high-powered experiments. Furthermore, rerandomization makes different linearly adjusted causal estimators asymptotically more similar or ``coherent" with each other, in the sense that they tend to yield similar point estimates, thereby reducing sensitivity to model specification \citep{zhao2024b}. This increased coherence implies that rerandomization can help mitigate p-hacking, as the choice of estimator and covariates becomes less consequential for inference \citep{lu2025rerandomization}. That said, these other potential benefits from rerandomization have also focused on asymptotic results, as well as difference-in-means or linear regression estimators.


To clarify whether rerandomization is helpful for various settings and estimators, we make three contributions. First, we provide a comprehensive review of the rerandomization literature and its open problems, such that researchers can understand the directions and statistical tools needed to make further contributions in the literature. Second, we provide practical guidance on each step of the rerandomization process, such that practitioners can better understand how to leverage rerandomization for their own experiments. Third, we conduct several simulation studies to assess the performance and relationship among parametric and nonparametric estimators under rerandomization in finite samples. We find that rerandomization (i) improves the precision of adjusted estimators in finite samples and (ii) decreases sensitivity to model choices in the analysis stage (e.g., parametric versus nonparametric specifications, or the use of covariate selection procedures) by increasing the coherence among estimators.

We structure the remainder of this review as follows. In Section \ref{sec:history} we situate rerandomization within its historical context, and contrast it with classical randomized and deterministic experimental designs. In Sections \ref{sec:setup} and \ref{sec:rerandomization}, we establish the formal notation and detail the rerandomization algorithm itself. In Section \ref{sec:theory}, we review the key theoretical properties of rerandomized experiments. In Section \ref{sec:simulations}, we present our simulation studies, where we consider the precision and coherence of parametric and nonparametric estimators under rerandomization in finite samples. In Section \ref{sec:practice}, we provide a practical guide for how to implement rerandomization, including how to address computational difficulties, issues of missing data, and other real-data challenges. In Section \ref{sec:conclusion}, we conclude with a discussion of recommendations and open problems for rerandomization.

%% file: text/history.tex
The idea of improving the quality of randomized experiments by selectively choosing among randomizations has a long history and has sparked substantial debate over the decades \citep{gosset1938, yates1939, greenberg1951, kiefer1959, taves1974, harville1975, arnold1986, kempthorne1986, treasure1998}. While early designs, such as those advocated by \cite{fisher1926}, emphasized randomization as essential for ensuring unbiasedness and valid inference, there was also an implicit recognition that some randomizations could lead to more desirable covariate balance than others. This led some researchers to advocate for procedures that deterministically minimized an imbalance measure \citep{taves1974, treasure1998}; for instance, \cite{treasure1998} suggested that if randomization is the gold standard, a platinum standard would be based on the \textit{minimization} of the covariate differences between treatment and control groups. However, such deterministic approaches can lack robustness to imbalance in unmeasured variables and may invalidate the assumptions underlying standard statistical inference, which relies on a known, random-assignment mechanism \citep{yates1939, harshaw2024balancing}.

The foundational tension between randomized designs and optimally balanced designs continues to provoke debate, and recent contributions have revisited this issue from theoretical and empirical perspectives \citep{kasy2016, kallus2018, banerjee2020, kapelner2020, kallus2021, johansson2021, kapelner2022, wang2022, jerzak2023degrees, harshaw2024balancing}. While the purpose of this paper is not to settle this debate, we will focus on \textit{rerandomized} designs that sit in between random and deterministic designs. \cite{wang2022} make this intermediate position precise by studying rerandomization when the imbalance threshold decreases with the sample size, which gives a formal sense in which rerandomization interpolates between complete randomization and an optimally balanced design.

The work of \cite{morgan2012} formally defined rerandomization as a procedure in which units are repeatedly randomized until a prespecified criterion for covariate balance is met. Their framework brought rigor to the informal practice of discarding ``bad” randomizations, while preserving the validity of statistical inference under certain conditions. Thus, rerandomized experiments act as a deterministic design by discarding clearly imbalanced randomizations while also being able to conduct randomization-based inference as randomized designs. This sparked interest in extending rerandomization to many settings, including but not limited to, different experimental designs \citep{zhou2018, branson2021, wang2021, lu2023, zhang2023quasi, zhao2024c}, theoretical properties of rerandomization in general \citep{li2018, li2020rerandomization, wang2024, branson2024power, zhao2024b, wang2025asymptotic}, and other covariate balance criteria \citep{morgan2015, zhao2024b, schindl2024, zhang2024}.



In this paper, we review recent developments on rerandomization so that others can understand how to leverage its benefits and contribute further to this literature.

%% file: text/notation.tex
For ease of exposition, we focus on two-arm randomized experiments involving a treatment and control group. In Section \ref{sec:practice:multi-arm}, we discuss multi-arm experiments; however, it is still unclear how to conduct rerandomization over continuous treatments, such as dose-response. 

Consider an experiment with $n$ units, from which $n_1$ units are randomly assigned to treatment and the remaining $n_0$ to control. We observe a $n \times k$ covariate matrix $X = (X_{1}, \dots, X_{n})'$, from which $d$ covariates denoted by $X^{\rr} \subseteq X$ are considered for rerandomization in the design stage, and $p$ covariates denoted by $X^{\adj} \subseteq X$ are considered for covariate adjustment in the analysis stage. Let $Z = (Z_1, \dots, Z_n)'$ be the vector of treatment assignments, where $Z_i = 1$ indicates treatment and $Z_i = 0$ control for unit $i$. We use the potential outcomes framework of \cite{rubin1974} to define causal effects, with $Y(1) = (Y_1(1), \dots, Y_n(1))'$ and $Y(0) = (Y_1(0), \dots, Y_n(0))'$ being the potential outcomes under treatment and control. We consider a finite-population setting where $X, Y(1), Y(0)$ are fixed and the sole source of randomness is in the treatment assignment.

Once treatments are assigned the observed outcome for unit $i$ is defined as $Y_i = Y_i(1)Z_i + Y_i(0)(1-Z_i)$, and the vector of observed outcomes $Y = (Y_1, \dots, Y_n)'$. This notation assumes no interference, meaning that one unit's outcome does not depend on another unit's treatment. There are various experimental designs that allow for interference \citep{rosenbaum2007interference, aronow2017estimating, basse2018analyzing, basse2024randomization, choi2023new}; and as we revised this paper, some works have considered rerandomized experiments with interference \citep{yuan2026rerandomization, zhang2026rerandomization}. In the end, the observed data is $(X, Y, Z)$. This setup follows most of the literature assuming that there is no missingness in covariates nor outcomes. See Section \ref{sec:practice:missing} for a discussion and practical recommendations when there is missing data.

So far, we have described the setup for a two-arm, completely randomized experiment, where a fixed number of subjects are assigned to treatment and control. Thus, in what follows, we will compare rerandomization to a completely randomized design, which is the typical comparison in the literature. That said, rerandomization can be applied to other settings, such as 2$^K$ factorial designs, sequential designs, stratified experiments, survey experiments, clustered designs, and split-plot designs \citep{brason2016improving, li2020factorial, zhou2018, wang2021, yang2021rejective, lu2023, shi2024rerandomization}.


The overarching goal of rerandomization is to determine how we can randomize $Z$ such that causal effects can be well-estimated. By far the most common estimand in the literature is the average treatment effect (ATE), defined as:
\begin{equation}\label{eq:ate}
    \tau = \frac{1}{n}\sum_{i=1}^{n} Y_i(1) - Y_i(0) = \frac{1}{n}\sum_{i=1}^{n}\tau_i.
\end{equation}

As one exception, \cite{wang2024} consider a more general set of estimands: those defined by any function of average potential outcomes,  $f\left[\E(Y(1)), \E(Y(0))\right]$. Thus, \cite{wang2024} consider a super-population setting, where covariates and potential outcomes are also considered random \citep{ding2017bridging}. In this case, the ATE would be defined as $\tau = \E\left[Y(1) - Y(0)\right]$. Although this super-population ATE differs slightly from the finite-population ATE in Equation (\ref{eq:ate}), the results in \cite{wang2024} are analogous to other rerandomization results that focus on a finite-population framework, as we discuss in the next section. On the other hand, it is still an open problem on how to design and analyze rerandomized experiments when other estimands are of interest, such as quantile effects and conditional average treatment effects. Indeed, while we revised this paper, \cite{han2026rerandomization} considered the problem of quantile treatment effect estimation after rerandomization.

%% file: text/rerandomization.tex
In general, rerandomization involves collecting covariate data $X$, and then randomizing the treatment $Z$ until some covariate balance criterion (based on $X$ and $Z$) is fulfilled. Thus, rerandomization methods are uniquely characterized by (1) the balance criterion, and (2) how treatment is randomized to ensure the criterion is fulfilled. We discuss these two characteristics below.




\subsection{Covariate balance criteria}\label{sec:criteria}
\input{text/rerandomization/criteria.tex}

\subsection{Randomization of treatment allocation}
\input{text/rerandomization/allocation.tex}

%% file: text/rerandomization/criteria.tex
Prior to conducting the experiment, we need to define a measure $\varphi$ of the covariate balance between treated and control units, where
\begin{equation*}
    \varphi(X,Z) =
    \begin{cases}
        1, \;\;\text{if}\;Z\;\text{is an acceptable randomization,}\\
        0, \;\;\text{if}\;Z\;\text{is \textbf{not} an acceptable randomization.}\\
    \end{cases}
\end{equation*}

To ensure that estimators are unbiased after rerandomization, this measure must be symmetric in $Z$, meaning $\varphi(X,Z) = \varphi(X, 1-Z)$. This ensures that if a treatment assignment is acceptable under rerandomization, it will also be acceptable if we flip every treatment and control label for that assignment. When, in addition to symmetry, $n_1 = n_0$, then the difference-in-means estimator---defined in the next section---is unbiased in finite samples \citep{morgan2012}. When $n_1 \neq n_0$, finite-sample unbiasedness is no longer guaranteed, but the estimators are still asymptotically unbiased if the symmetry condition holds \citep{li2018}, which is the case for the Mahalanobis distance defined below. The set of randomizations that have $\varphi(X,Z) = 1$ will be deemed acceptable to run the experiment, formally $\A = \{Z : \varphi(X, Z) = 1\}$.

Although $\varphi$ can be defined in various ways, which we explore in Section \ref{sec:practice:criterion}, we focus on the most common measure chosen in the literature, which depends on the Mahalanobis distance, 
\begin{equation} \label{eq:notation:mahalanobis}
    M = \tauestrr^\prime\Sigma^{-1}\tauestrr,
\end{equation}
where, 
\begin{equation*}
    \tauestrr = \overline{X}^\rr_1 - \overline{X}^\rr_0 = \frac{1}{n_1}\left(X^{\rr^\prime} Z\right) - \frac{1}{n_0}\left(X^{\rr^\prime}(1_n - Z)\right),
\end{equation*}
$1_n \in \R^n$ is a vector whose coefficients are all equal to one, and $\Sigma = \cov\left(\tauestrr\right)$ is the $d \times d$ covariance matrix for the vector of covariates' difference-in-means under complete randomization. Note that the only random variable in this covariance is the treatment assignment $Z$. As shown by \cite{morgan2012}, under complete randomization, $\Sigma = \frac{n}{n_1 n_0} S^2_X$, where $S^2_X$ is the sample covariance matrix of the covariates.

For the Mahalanobis distance, we define the set of acceptable randomizations as those that have $M < a$, where $a > 0$ is a pre-specified threshold, i.e. $\A = \{Z: M < a\}$. The most common way to define the threshold $a$ is based on the proportion $p_a$ of randomizations to be considered acceptable, $p_a = \Prob(M < a)$. Although there is no definitive method to set $p_a$, some common choices are $p_a = 0.01$ and $p_a = 0.001$ \citep{li2018, branson2021, zhang2024, schindl2024}. We provide a more practical discussion on the choice of $p_a$ in Section \ref{sec:practice:threshold}. However, it is important to point out that a small $p_a$ leads to more balanced randomizations, but it reduces the number of accepted randomizations. The extreme case where $\A$ comprises a single treatment allocation is analogous to a deterministic design and thus makes randomization-based inference impossible.

%% file: text/rerandomization/allocation.tex
Despite its theoretical appeal and intuitive simplicity, rerandomization often faces practical limitations due to its computational demands for randomly drawing acceptable randomizations. The classical rerandomization scheme, as introduced by \cite{morgan2012}, relies on acceptance-rejection sampling: treatment allocations are generated at random and only accepted as long as a prespecified covariate balance criterion is satisfied.

The computational burden is amplified in randomization-based inference, where balanced treatment assignments must be repeatedly sampled to approximate the null distribution of the test statistic. Recognizing this challenge, recent work has focused on developing computationally efficient alternatives to rejection sampling. These include methods that, given a randomization, switches units from the treatment groups until balance is achieved (e.g., pair-switching algorithms \citep{zhu2023}) or avoid rejection sampling entirely (e.g., importance sampling \citep{branson2019randomization} or integer programming approaches \citep{lu2025fast}). 
We return to these computational strategies in depth in Section \ref{sec:practice}.

Recently, \cite{wang2025asymptotic} proposed to generate a number of randomizations, such that the number of randomizations generated does not increase too quickly with sample size, and accepting the most balanced randomization, thereby reducing the computational demands of rejective sampling; this scheme was discussed earlier by \cite{bruhn2009}. This new procedure has similar theoretical results to the usual rerandomization procedure, which we review now.

%% file: text/theory.tex
In this section, we review the theoretical properties of three causal effect estimators based on a rerandomization scheme using the Mahalanobis distance, defined in Equation (\ref{eq:notation:mahalanobis}),  based on covariates $X^\rr$.

The first estimator we consider is the difference-in-means estimator,
\begin{equation}\label{eq:theory:estimator-D}
    \tauest{D} = \frac{1}{n_1}\sum_{i=1}^{n}Y_iZ_i - \frac{1}{n_0}\sum_{i=1}^{n}Y_i(1-Z_i),
\end{equation}
which is by far the most common estimator in the rerandomization literature \citep{morgan2012, morgan2015, li2018, li2020rerandomization, branson2021, zhu2023, branson2024power, zhang2024, schindl2024, wang2025asymptotic, lu2025fast}.

The second estimator, considered in \cite{lin2013, li2020rerandomization}, is a linearly adjusted estimator
\begin{equation}\label{eq:theory:estimator-L}
    \tauest{L} = \frac{1}{n}\sum_{i=1}^{n}\left(\widehat{\mu}_1\left(X^\adj_i\right) - \widehat{\mu}_0\left(X^\adj_i\right)\right),
\end{equation}
where $X^\adj$ are the covariates used for adjustment in the linear outcome model $\widehat{\mu}_z\left(X_i\right) = \widehat{\E}\left(Y_i|X_i,Z_i=z\right) = \widehat{\beta}_{0z} + \widehat{\beta}_{1z}X_i$, for $z \in \{0,1\}$.\footnote{Some papers use the notation based on R code, see \cite{lin2013, zhao2024a, zhao2024b, zhao2024c, lu2025rerandomization}. In this case, $\tauest{L}$ is numerically the same as the coefficient of $Z$ in $\texttt{lm}(Y_i \sim 1+Z_i+\tilde{X}_i+\tilde{X}_iZ_i)$, with $\tilde{X}_i = (X_i - \mean{X})$.}

Lastly, we consider the doubly-robust estimator studied in \cite{wang2024},
\begin{equation}\label{eq:theory:estimator-DR}
    \tauest{DR} = \frac{1}{n}\sum_{i=1}^{n}\left(\widehat{\eta}_1\left(X_i^\adj\right) - \widehat{\eta}_0\left(X_i^\adj\right)\right)
\end{equation}
with
\begin{equation*}
    \widehat{\eta}_z\left(X_i\right) = \frac{\1(Z_i = z)}{\pi^z(1-\pi)^{1-z}}\left(Y_i - \widehat{\mu}_z\left(X_i\right)\right) + \widehat{\mu}_z\left(X_i\right),
\end{equation*}
where $\widehat{\mu}_z\left(X_i\right)$ is any (possibly nonparametric) estimator of $\E\left[Y_i(z)|X_i\right]$ for $z=0,1$, $\pi = n_1/n$ is the propensity score, and $\1$ is the indicator function\footnote{Equation~(\ref{eq:theory:estimator-L}) is the special case of Equation~(\ref{eq:theory:estimator-DR}) in which $\hat{\mu}_z$ is a linear regression fitted within treatment group $z$.}.

Treatment effect estimation through $\tauest{DR}$ can suffer from bias due to overfitting. This occurs when estimation of the nuisance parameter $\mu_z$ and computation of $\widehat{\eta}_z$, for $z \in \{0,1\}$, happen on the same dataset. One way to mitigate this bias is through cross-fitting. This technique separates data into independent folds, in which all but one fold are used to estimate $\mu_z$ and the remaining fold is used to compute $\widehat{\eta}_z$, thereby decoupling the dependency between estimation of $\mu_z$ and computation of $\widehat{\eta}_z$ \citep{kennedy2024semiparametric}. Moreover, since this is a randomized experiment, the true propensity score $\pi$ is known. However, in observational studies, the outcome models $\mu_z$ and propensity score $\pi$ are nuisance parameters that must be estimated. More details on cross-fitting and how to apply it for $\tauest{DR}$ are discussed in Appendix \ref{app:estimators}.

Now we explore the main theoretical developments for these three estimators under rerandomization. In particular, we discuss two aspects that make rerandomization a promising approach. First, we discuss the asymptotic distributions of the above estimators, which justifies confidence intervals that can be used for inference and are narrower than analogous confidence intervals under complete randomization, hence increasing statistical power of hypothesis testing. Second, we discuss the extent that rerandomization increases the coherence of different estimators, in the sense that they yield similar point estimates, thereby reducing sensitivity to model specification. We end this section with a discussion of an ongoing debate and open problems about the equality between rerandomization and covariate adjustment. The simulation study in Section \ref{sec:simulations} is meant to address some of those open problems.

\subsection{Asymptotic Distributions}\label{sec:theory:distributions}

\input{text/theory/distributions.tex}

\subsection{Coherence}\label{sec:theory:coherence}
\input{text/theory/coherence.tex}


\subsection{Debates, Caveats, and Conjectures}\label{sec:theory:debates}

\input{text/theory/caveats.tex}

%% file: text/theory/distributions.tex
To start, under completely randomized experiments, we have that  
\begin{equation*}
    \sqrt{n}(\tauest{*} - \tau) \converge \sqrt{V_{\tau\tau, *}}\epsilon,
\end{equation*}
where $\epsilon \sim N(0,1)$, and $* = \text{D, L, DR}$, refer to the difference-in-means, linear regression, and doubly robust estimators in Equations (\ref{eq:theory:estimator-D}), (\ref{eq:theory:estimator-L}), and (\ref{eq:theory:estimator-DR}), respectively. 
The asymptotic normality of $\tauest{D}$ was proven in \cite{neyman1990application}, and later extended for $\tauest{L}$ and $\tauest{DR}$ \citep{lin2013, li2020rerandomization, wang2024}.  Here, $V_{\tau\tau, \text{D}} = \frac{n}{n_1}S^2_{Y(1)} + \frac{n}{n_0}S^2_{Y(0)} - S^2_{\tau}$ such that $S^2_{Y(z)}$
are the finite population variances of the outcomes in treatment and control and $S^2_{\tau}$ is the finite-population variance of the individual treatment effects. The variances $S^2_{Y(z)}$ can be unbiasedly estimated with sample analogs, whereas the treatment effect heterogeneity term $S^2_{\tau}$ can typically not be estimated, although it can be bounded \citep{ding2018decomposing}. The variances of $\tauest{L}$ and $\tauest{DR}$ are similar to $V_{\tau\tau, \text{D}}$ but after accounting for adjustment on $X^\adj$. See Appendix \ref{app:estimators} for the exact definition of $V_{\tau\tau, *}$ for $* =$ L, DR and estimators for the three variances.

\cite{morgan2012} established two main properties for $\tauest{D}$ under rerandomization and when the number of treated and control units is the same, i.e. $n_1 = n_0$. First, the estimator is unbiased under finite samples. Second, under additional assumptions of additive treatment effect and normality of covariates and outcome means, then the finite-sample variance of the estimator is reduced by $1-(1-v_{d,a})R_{\text{D}, X^\rr}^2$. Under finite-population inference, this normality can be justified by the finite-population central limit theorem \citep{li2017}. Here,  $v_{d,a} = \Prob(\chi^2_{d+2} \leq a)/\Prob(\chi^2_{d} \leq a) \leq 1$, which gets closer to 1 as $d$ and $a$ increase. Intuitively, this means that if we attempt to balance many covariates at once, or if we are not strict in our acceptance criterion, rerandomization will not provide much precision improvements compared to complete randomization. On the other hand, $R^2_{\text{D}, X^\rr}$ denotes the proportion of the potential outcomes' variance that can be explained by a linear projection on $\tauestrr$, formally defined as:
\begin{equation*}
    R^2_{\text{D}, X^\rr} = \frac{\frac{n}{n_1}S^2_{Y(1)|X^\rr} + \frac{n}{n_0}S^2_{Y(0)|X^\rr} - S^2_{\tau|X^\rr} }{ \frac{n}{n_1}S^2_{Y(1)} + \frac{n}{n_0}S^2_{Y(0)} - S^2_{\tau} },
\end{equation*}
where $S^2_{Y(z)|X^\rr}, z = 0,1$ and $S^2_{\tau|X^\rr}$ are the finite population variance of the linear projections of the potential outcomes and individual treatment effects on the covariates used for rerandomization. Their mathematical definitions are presented in Appendix \ref{app:statistics}. Under the additive treatment effect assumed in \cite{morgan2012}, $R^2_{D, X^\rr}$ is the squared multiple correlation between $Y(0)$ and $X^\rr$; the expression above is from \cite{li2018} and does not require additivity. Notice that $R^2_{D, X^\rr}$ has the same interpretation as $R^2$ in linear regression models, however $\tauest{D}$ does not fit a linear model and there is no assumption that the covariates and potential outcomes follow a linear model. Moreover, when $d$ decreases, $R^2_{D, X^\rr}$ also decreases, meaning that the precision benefits of rerandomization diminish. This is the opposite consequence observed for $v_{d,a}$, therefore there is a trade-off on the amount of covariates one should choose so that it maximizes $R^2_{D, X^\rr}$ while minimizing $v_{d,a}$. We discuss how to choose the set of covariates in more depth in Section \ref{sec:practice}.


Assuming that the finite-population variance and covariance of the potential outcomes, individual treatment effects, and covariates are well-defined and have finite limiting values, \cite{li2018, li2020rerandomization, wang2024} extend these results by deriving the asymptotic distribution of $\tauest{D}, \tauest{L}$, and $\tauest{DR}$, respectively, under rerandomization. We leave the mathematical definition of these assumptions in Appendix \ref{app:statistics}. These works established that the asymptotic distributions of these estimators after rerandomization follow a convolution of a normal distribution and a component of a truncated multivariate normal distribution:
\begin{equation}\label{eq:theory:rerand-distribution}
    \sqrt{n}(\tauest{*} - \tau)|\A \converge \sqrt{V_{\tau\tau, *}}\left(\sqrt{1-R_{*, X^\rr}^2}\epsilon + \sqrt{R_{*, X^\rr }^2}L_{d, a}\right),
\end{equation}
where $\A$ indicates that the distribution is conditioned on balanced randomizations, $L_{d, a} \sim Q_1 | \Q'\Q \leq a$, with $\Q = (Q_1, \dots, Q_d)' \sim N(0, \I_d)$, is a truncated normal distribution centered at $0$ with $\var(L_{d,a}) = v_{d,a}$. 
Intuitively, $R^2_{\text{L}, X^\rr}$ and $R^2_{\text{DR}, X^\rr}$ are analogous to $R^2_{\text{D}, X^\rr}$, but they apply to residuals of the potential outcomes after adjusting for $X^\adj$ through a linear regression model or a more flexible regression model, respectively.
Thus, $R^2_{*, X^\rr}$ also depends on the covariates $X^\adj$ used for adjustment in the linear regression or doubly robust estimator. See Appendix \ref{app:estimators} for the mathematical definition of $R^2_{*, X^\rr}, * =$ L, DR, and their respective estimators. Note that the results for $\tauest{D}$ and $\tauest{L}$ are under a finite-population framework \citep{li2018, li2020rerandomization}, whereas the results for $\tauest{DR}$ are from \cite{wang2024}, who consider a super-population framework. We did not extend the results of \cite{wang2024} to the finite-population framework, which, to our knowledge, is still an open problem.

Since both $\epsilon$ and $L_{d,a}$ are symmetric and unimodal at zero, their linear combination is also symmetric and centered at zero. Hence the estimator $\tauest{*}$ is asymptotically unbiased and consistent for the ATE. Furthermore, this distribution is more ``peaked" than a normal distribution, resulting in a tighter variance and more precise estimators. In fact, the ratio of the variance of the estimators under rerandomization and under complete randomization is given by
\begin{equation*}
    \frac{\var(\tauest{*}|\A)}{\var(\tauest{*})} = 1-(1-v_{d,a})R^2_{*, X^\rr},
\end{equation*}
for $*=$ D, L, DR.  \cite{li2018} presents more details on the geometry, shape, and peakedness of the resulting distribution. This increased precision in the estimators' distributions improves statistical power, meaning that tests in rerandomized experiments typically require fewer subjects than completely randomized experiments to achieve the same power \citep{branson2024power}. In other words, the confidence intervals for estimators under rerandomization are narrower. For instance, a $(1-\alpha)$-level confidence interval is given by
\begin{equation*}
    \tauest{*} \pm \widehat{v}_{1-\alpha/2}\left(\widehat{R}^2_{*, X^\rr}\right)\widehat{V}_{\tau\tau, *}^{1/2}n^{-1/2},
\end{equation*}
where $* =$ D, L, DR and $v_{1-\alpha/2}\left(R^2_{*, X^\rr}\right)$ is the $(1-\alpha/2)$-quantile of the asymptotic distribution in Equation (\ref{eq:theory:rerand-distribution}). This quantile can be estimated through rejection sampling, that is, by generating random draws from $\Q'\Q$ and retaining only those that satisfy $\Q'\Q \leq a$, then approximating the desired quantile of the convolution using these accepted values. In Section \ref{sec:practice:inference:ci} we detail the rejection sampling algorithm to estimate the quantile of the asymptotic distribution of $\tauest{*}$ and construct confidence intervals.






\subsection{Dependence of the asymptotic distributions in terms of \texorpdfstring{$X^\rr$}{Xrr} and \texorpdfstring{$X^\adj$}{Xadj}}\label{sec:theory:scenarios}

Depending on the covariates used for the analysis and design stages, there exists clear and intuitive relationships between the marginal distributions of estimators under complete randomization and rerandomization. \cite{li2020rerandomization} cover the relationship between $\tauest{D}$ and $\tauest{L}$ while \cite{wang2024} cover the relationship between $\tauest{D}$ and $\tauest{DR}$. We summarize them now.

\textit{First scenario:} if the same covariates are used in the design and analysis stages, $X^\rr = X^\adj$, rerandomization can be viewed as linear covariate adjustment in the design stage,
\begin{equation*}
    \sqrt{n}(\tauest{D} - \tau)|\A \dot\sim \sqrt{n}(\tauest{L} - \tau)
\end{equation*}
when $a \to 0$. We use the tilde with a dot to denote two sequences of random vectors converging weakly to the same distribution. In other words, for sufficiently small threshold $a$, the difference-in-means estimator after rerandomization is asymptotically equivalent to the linear regression estimator after complete randomization.

\textit{Second scenario:} if the design stage covariates (and possibly other covariates) are adjusted for in the analysis stage, $X^\rr \subseteq X^\adj$, then $R^2_{*, X^{\rr}} = 0$, for $* = $ L, DR, and the asymptotic distribution of adjusted estimators under rerandomization and complete randomization are the same,
\begin{equation*}
    \sqrt{n}(\tauest{*} - \tau)|\A \dot\sim \sqrt{n}(\tauest{*} - \tau),
\end{equation*}
for $*= $ L, DR. This scenario can happen if only a subset of covariates are used when randomizing, or if additional baseline covariates are gathered after the experiment.

\textit{Third scenario:} if at least as many covariates are used in the design stage as in the analysis stage, $X^\rr \supseteq X^\adj$, rerandomization provides precision benefits based on the variance of $\tauest{*}$ that can be linearly explained by $X^\rr$ and is not explained by $X^\adj$, hence $R^2_{*, X^\rr} \geq 0$ for $* =$ L, DR in Equation (\ref{eq:theory:rerand-distribution}) \citep{li2020rerandomization, wang2024}. On top of that, \cite{li2020rerandomization} show that we can more explicitly rewrite
\begin{equation*}
    R_{\text{L}, X^\rr}^2 = \frac{R^2_{D, X^\rr} - R^2_{D, X^\adj}}{1-R^2_{D, X^\adj}}.
\end{equation*}
However, there is no clear form to rewrite $R^2_{\text{DR}, X^\rr}$ since the projection on $X^\adj$ is non-linear \citep{wang2024}.

This scenario can happen if, for instance, a variable selection procedure is performed in the analysis stage such that only a subset of $X^\rr$ are selected for adjustment, or if units' privacy is a concern after the experiment, such that only some covariates are available.

\textit{Fourth scenario:} in more general scenarios, in which neither $X^\rr$ nor $X^\adj$ are a subset of the other, there are no simple formulae or interpretation to relate the asymptotic distributions of $\tauest{D}$ and $\tauest{L}$, $\tauest{DR}$ \citep{li2020rerandomization, wang2024}. The $R^2$ term in the asymptotic distribution would depend on the covariance structure of $X^\rr$ and $X^\adj$, as well as the amount of overlap between $X^\rr$ and $X^\adj$.


All these scenarios and their results are summarized in Table~\ref{tab:asymptotic-summary}. We assess to what extent these results hold in finite samples in our simulation study (Section~\ref{sec:simulations}). 

\begin{table}[htbp]
\centering
\caption{Summary of asymptotic relationships between estimators under complete randomization and rerandomization. We categorize these relationships according to how the sets of covariates used for rerandomization ($X^\rr$) and for adjustment ($X^\adj$) overlap. For each scenario, we summarize (i) which covariate set is larger or whether they coincide, (ii) the resulting asymptotic distribution of the difference-in-means ($\tauest{D}$), linear adjusted ($\tauest{L}$), and doubly robust ($\tauest{DR}$) estimators, and (iii) the theoretical implications for estimator precision  from rerandomization.
}
\begin{tabular}{@{}cl@{}}
\toprule
 \textbf{Scenario} & \textbf{Key Asymptotic Result} \\
\midrule
    \textbf{1.} \adjustbox{valign=c}{\includegraphics[width=2.2cm]{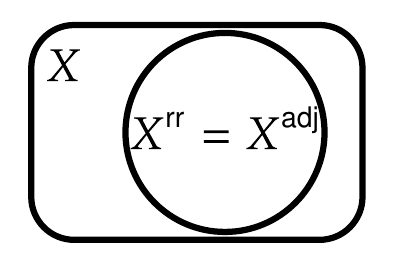}} & 
    \begin{tabular}[c]{@{}l@{}}Rerandomization is equivalent to linear adjustment. \\ $\sqrt{n}(\tauest{D} - \tau)|\A \dot\sim \sqrt{n}(\tauest{L} - \tau)$.
    \end{tabular}  \\ 
\addlinespace 
    \textbf{2.}\adjustbox{valign=c}{\includegraphics[width=2.2cm]{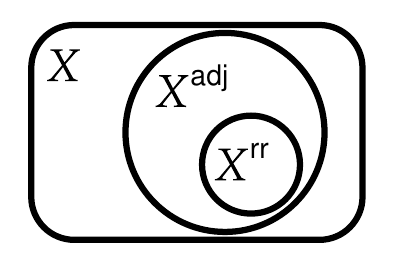}} & 
    \begin{tabular}[c]{@{}l@{}}Rerandomization provides no additional\\ asymptotic gain for adjusted estimators.\\ $\sqrt{n}(\tauest{*} - \tau)|\A \dot\sim \sqrt{n}(\tauest{*} - \tau), *=$ L, DR.
    \end{tabular} \\
\addlinespace
    \textbf{3.}\adjustbox{valign=c}{\includegraphics[width=2.2cm]{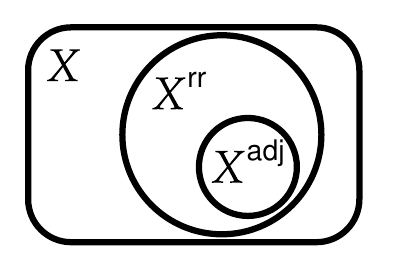}} & 
    \begin{tabular}[c]{@{}l@{}}Rerandomization provides additional \\ asymptotic gain for adjusted estimators.\\
    $R^2_{*, X^\rr} \geq 0$, for $*=$ L, DR.
    \end{tabular} \\
\addlinespace 
    \textbf{4.} \adjustbox{valign=c}{\includegraphics[width=2.2cm]{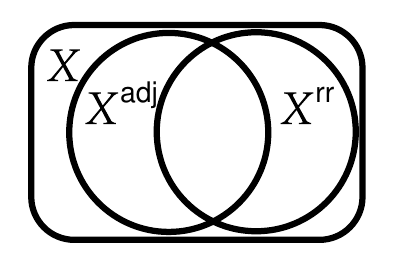}} & 
    \begin{tabular}[c]{@{}l@{}}No simple formulae or interpretation to relate \\ the asymptotic distributions of $\tauest{D}$ and $\tauest{L}$, $\tauest{DR}$.\end{tabular}\\
\bottomrule
\end{tabular}
\label{tab:asymptotic-summary}
\end{table}

Although these results are about the marginal distributions of each estimator, the results suggest, for instance, that the difference-in-means estimator is more similar to the linear estimator when rerandomization is applied in the design stage. This similarity is formalized by the concept of ``coherence", defined in \cite{zhao2024b}. We now dive into coherence.

%% file: text/theory/coherence.tex
Coherence was first mentioned and defined in \cite{zhao2024b}. It measures how coherent, or similar, two estimators $\tauest{*}$ and $\tauest{**}$ are. Formally, the coherence between two estimators under complete randomization and rerandomization are defined as
\begin{equation*}
    \E\left\{(\tauest{*} - \tauest{**})^2\right\} \;\;\;\text{and}\;\;\;\; \E\left\{(\tauest{*} - \tauest{**})^2|\A\right\}.
\end{equation*}

The smaller the mean squared difference is, the more coherent the estimators are.

\cite{zhao2024b} derives that asymptotically, with $X^\rr = X^\adj$, and under rerandomization based on the Mahalanobis distance, the estimators $\tauest{D}, \tauest{L}$ are more coherent under rerandomization, more specifically
\begin{equation*}
    \frac{\E\left\{(\tauest{D} - \tauest{L})^2|\A\right\}}{\E\left\{(\tauest{D} - \tauest{L})^2\right\}} = v_{d,a}.
\end{equation*}

Contrary to the precision results, this result does not depend on any $R^2_{*, X^\rr}$, $* = $ D, L.

Coherence is directly related to p-hacking, which refers to practices researchers may use to obtain more significant p-values \citep{simmons2011false, brodeur2020methods}. One example of p-hacking involves estimating linear regression models across various subsets of available covariates and reporting only those specifications that yield the desired p-values for the average treatment effect. In principle, if multiple regression specifications were coherent with one another—that is, if they produced similar point estimates regardless of which covariates were included—then the choice of model specification would be inconsequential for inference. Under such coherence, different covariate adjustments would converge to the same estimate, rendering results more robust to modeling choices.

Building on this, \cite{lu2025rerandomization} extend the idea of coherence between estimators to study p-hacking within rerandomization. They show that strategically selecting covariates for regression adjustment after rerandomization does not significantly distort p-values due to the higher coherence between the resulting estimators.

Formally, define $p_L = \underset{X^{\text{hack}} \subseteq X^{\adj}}{\min} p_{L,X^{\text{hack}}}$ to be the smallest two-sided p-value obtained by searching across every subset of covariates $X^{\text{hack}} \subseteq X^\adj$ used for adjustment on $\tauest{L}$. Then
under the null hypothesis of zero average treatment effect,
\begin{equation*}
    \lim_{n\to \infty}\Prob(p_L \leq \alpha) \geq \alpha,
\end{equation*}
where $\alpha$ is the significance level. On the other hand, under rerandomization based on all available covariates and the same null hypothesis,
\begin{equation*}
    \lim_{n\to \infty}\Prob(p_L \leq \alpha|\A) \leq \alpha,
\end{equation*}
as $a \to 0$. Hence, asymptotically, rerandomization controls the type I error rate, thus mitigating p-hacking. The intuition for this is that as $a$ gets closer to $0$, there is no remaining imbalance for the linear estimator to adjust for, and hence prevents the practitioner from choosing the linear adjustment that yields the desired p-values.

In our simulation study, we aim to expand the coherence discussion to doubly-robust estimators and the finite sample setting.

%% file: text/theory/caveats.tex
To wrap up this section, we discuss key takeaways from the above results which have led to debates about whether rerandomization is helpful if adjusted estimators are used. However, this debate comes with important caveats, which have sparked conjectures about the benefits of rerandomization in practice.

\cite{li2018, li2020rerandomization} highlight the asymptotic equivalence between rerandomization and linear regression adjustment: both aim to exploit covariate information, one through design and the other through analysis. This raised a debate around a fundamental question: \textit{Does rerandomization help beyond covariate adjustment in the analysis stage?}

To understand this debate, consider the implications of the asymptotic results in Scenarios 1 and 2 (discussed in Section \ref{sec:theory:scenarios}). When all covariates from the design stage are adjusted for in the analysis stage, rerandomization provides no additional asymptotic precision gains for adjusted estimators. This suggests that rerandomization may be unnecessary if one simply uses adjusted estimators for inference.

However, this argument relies on asymptotic results. Thus, researchers have conjectured that rerandomization might still provide finite-sample benefits, even when asymptotic theory suggests equivalence \citep{yang2021rejective, zhu2024design, wang2024}, and hence rerandomization would act as a ``safeguard" even if analysis-stage adjustment is performed \citep{li2020rerandomization, zhu2024design, wang2024}.

\begin{figure}[ht]
\centering
\includegraphics[width=0.70\textwidth]{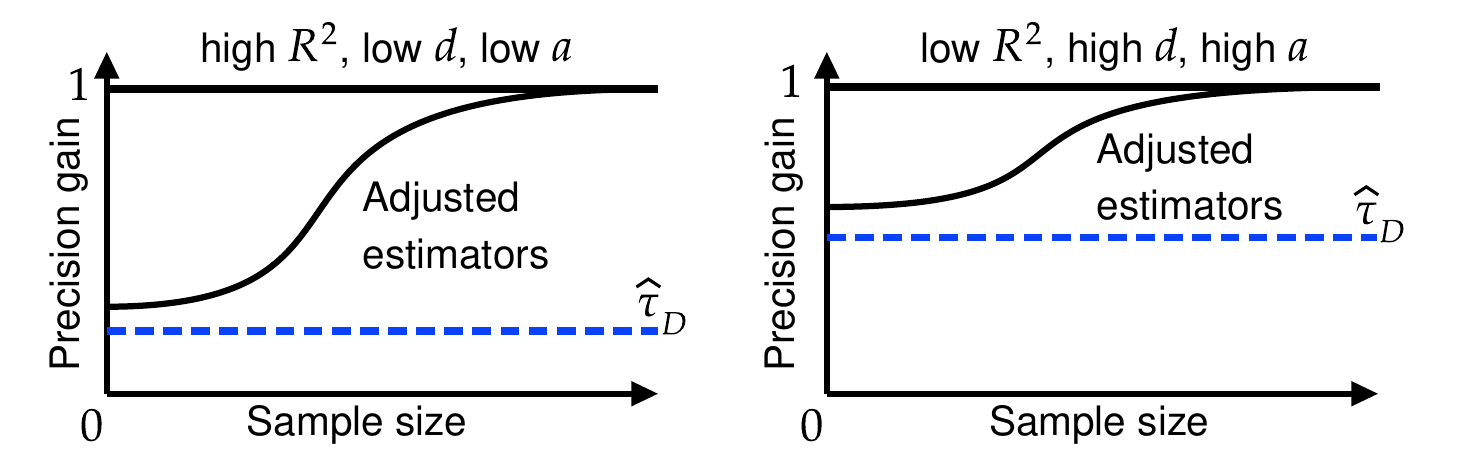}
\caption{Bounds on the conjectured variance ratios of adjusted estimators under rerandomization across sample sizes. Each plot illustrates the anticipated variance
ratio, defined as the ratio of an estimator's variance under rerandomization versus complete randomization, $\var(\tauest{*}|\A)/\var(\tauest{*})$, as a function of sample size. The upper bound (solid bold black line at 1) represents the asymptotic variance ratio for covariate-adjusted estimators when all rerandomized covariates are used in adjustment, indicating no precision gain. The lower bound (blue dashed line) corresponds to the asymptotic variance ratio of the unadjusted difference-in-means estimator, which depends on the number of covariates ($d$), linear predictive strength ($R^2_{D, X^\rr}$), and the rerandomization threshold ($a$). We conjecture that in finite samples and certain scenarios, adjusted estimators achieve variance ratios between these two bounds when the lower bound holds (see the text). The left plot shows when the potential benefits from rerandomization are greater for smaller sample sizes, in this case the covariates are highly predictive of the potential outcomes (high $R^2$) covariates are low-dimensional (low $d$), and high balance is enforced in the randomization (low $a$). The right plot shows when there are less potential benefits from rerandomization in finite sample sizes, in this case the covariates are not predictive of the potential outcomes (low $R^2$), covariates are high-dimensional (high $d$), and balance is barely enforced in the randomization (high $a$).}
\label{fig:precision}
\end{figure}

Building on this, we conjecture on the anticipated variance ratio of adjusted estimators under rerandomization. We define the variance ratio as the ratio of an estimator's variance under rerandomization to that under complete randomization $\var(\tauest{*}|\A)/\var(\tauest{*})$, such that a ratio below 1 indicates that rerandomization makes the estimator more precise. Figure \ref{fig:precision} illustrates the conjectured variance ratios as a function of sample size. The key insight is that the variance ratios we would anticipate in practice are bounded by two extremes. At the upper bound, the variance ratio equals 1, indicating no benefit from rerandomization;  this represents the asymptotic behavior of adjusted estimators when $X^\rr \subseteq X^\adj$. At the lower bound, represented by the dashed blue line, we have the asymptotic variance ratio for $\tauest{D}$, given by $1-(1-v_{d,a})R^2_{D, X^\rr}$. For $\tauest{L}$, this lower bound holds when one set of covariates contains the other (Scenarios 1–3), because in this case adjustment only removes variation that rerandomization also targets. The same holds for $\tauest{DR}$ in Scenarios 1 and 2, and in Scenario 3 when the conditional mean of the potential outcomes is linear in $X^\adj$ (and thus $\tauest{DR}$ is numerically equivalent to $\tauest{L}$). Otherwise, the lower bound may not hold: adjusting for covariates that are not in $X^\rr$ (Scenario 4), or for non-linear functions of $X^\adj$ ($\tauest{DR}$ in Scenario 3), can remove variation that is not linearly related to $X^\rr$, such that rerandomization can help the adjusted estimator more, in relative terms, than it helps $\tauest{D}$. This does not happen in our simulations in Section~\ref{sec:simulations}, where the covariates are strongly correlated.

When the lower bound holds, for adjusted estimators in finite samples we would expect the variance ratio to lie between these two bounds, gradually approaching its asymptotic value as the sample size increases. Importantly, the position of the lower bound—and thus the potential magnitude of finite-sample gains—depends on three factors: the number of covariates $d$, the coefficient $R^2_{D, X^\rr}$, and the threshold $a$. As $d$ increases, $R^2_{D, X^\rr}$ decreases, and $a$ increases, the blue line rises closer to 1, narrowing the gap between the two bounds. Consequently, there is ``less room" for rerandomization to help adjusted estimators in this case. Conversely, as the number of covariates decrease, the covariates have a strong linear predictive power of the potential outcomes, and we enforce more balance on the covariates in the design stage, we conjecture that the potential finite-sample gains from rerandomization are larger. When the lower bound does not hold, these three factors still govern how much rerandomization can help, but the variance ratio of adjusted estimators is no longer guaranteed to stay above the dashed line in Figure~\ref{fig:precision}.

Although most of the literature has focused on the precision gains (or lack thereof) for rerandomization, some recent works have shown that rerandomization can improve coherence of estimators and thus mitigate $p$-hacking \citep{zhao2024b, lu2025rerandomization}. However, these theoretical results are again asymptotic and focus on the specific case where $X^\rr = X^\adj$ and the comparison is between $\tauest{D}$ and $\tauest{L}$. It remains an open question whether similar coherence gains extend to other estimators, such as $\tauest{DR}$ with different functional forms for the outcome model, or different sets of covariates used for adjustment. We conjecture that rerandomization improves the coherence among a broader set of adjusted estimators (e.g., doubly robust and nonparametric estimators), and that this coherence gain extends beyond the restrictive case where $X^\rr = X^\adj$.

To investigate these conjectures and address the limitations of asymptotic theory, we conduct a simulation study in Section \ref{sec:simulations}. Based on the established rerandomization literature and our simulation study, we provide a practical guide in Section \ref{sec:practice} on implementing rerandomization, including practical challenges like computational concerns and scenarios with missing data.

%% file: text/simulations.tex
As explored in Section \ref{sec:theory}, rerandomization and linear covariate adjustment are asymptotically equivalent; however, it is unknown what this relationship is under small sample sizes. On top of that, the coherence between $\tauest{D}$ and $\tauest{L}$ is known when $X^\adj = X^\rr$, but no results explore $\tauest{DR}$ nor when $X^\adj \neq X^\rr$. Hence, to empirically investigate these open problems, we conduct a simulation study to assess the extent that rerandomization improves (i) the precision of adjusted estimators under finite samples, and (ii) the coherence among adjusted estimators (both parametric and nonparametric). We are particularly interested in how these properties behave when the amount of covariates in the design and analysis stages vary, as discussed in Section \ref{sec:theory:scenarios}.

In what follows, we first describe our simulation setup, and then we compare various estimators in terms of their precision and coherence. In addition to considering different estimators, we also consider the impact of the covariates one adjusts for, $X^\adj$, as well as the model used when implementing an estimator. Throughout, we focus on ATE estimators, because that is the estimand of interest in most of the rerandomization literature. We leave estimators for other estimands—such as quantile treatment effects and conditional average treatment effects—for future work, because such estimators under rerandomization are largely unexplored.

\subsection{Initial Simulation Design}\label{sec:simulations:design}

We follow previous simulation studies in the rerandomization literature, where first we simulate a single finite population, and then we generate many randomizations or rerandomizations with that one dataset \citep{li2018, branson2021, zhao2024b, lu2025rerandomization}. For each sample size $n \in \{100, 200, 500, 1000, 2000, 5000, 10000\}$ we simulate one dataset with $10$ covariates, and $X \sim N(\boldsymbol{0}, 0.3\boldsymbol{I} + 0.7 \boldsymbol{1}_{10}\boldsymbol{1}_{10}^\prime)$, where $\boldsymbol{I}$ is the identity matrix and  $\boldsymbol{1}_{10} = (1, \dots, 1)^\prime_{10\times 1}$.

We consider two settings for generating the potential outcomes:
\begin{align*}
\text{Linear setting} &
\begin{cases}
    Y_i(0) = \sum_{j=1}^{10} X_{ij} + \delta_i\\
    Y_i(1) = 1 + Y_i(0) \\
\end{cases} \\
\text{Non-linear setting} &
\begin{cases}
    Y_i(0) = \sum_{j=1}^{10} g(X_{ij}) + \delta_i\\
    Y_i(1) = 1 + Y_i(0) \\
\end{cases}
\end{align*}
with 
\begin{align*}
    g(x) &=  \begin{cases}
        x+0.4, \;\;\;\; x <-1 \\
        x^2-1.6, \;\;\;\; -1 \leq x < 1 \\
        \sin(x)-\sin(1)-0.6, \;\;\;\; \text{otherwise}
    \end{cases}
\end{align*}
for $i=1,\dots,n$,  where $\delta_i \sim \text{Normal}(0,\sigma^2)$, where $\sigma = \frac{1}{5} SD\left(\sum_{j=1}^{10}X_{ij}\right)$ for the linear setting, and for the non-linear setting we define $\sigma = \frac{1}{100} SD\left(\sum_{j=1}^{10}g(X_{ij})\right)$, where $SD$ is the standard deviation, and $g$ is a somewhat complex regression that is continuous but not smooth for all $x$. 

This choice of variance for the noise $\delta$ in the linear and non-linear settings is to enable fair comparison between our estimators across both settings. We choose the variance of $\delta$ so that the effect size, defined as $\tau/\{SD(Y_i(0))\sqrt{1-R^2}\}$, is approximately constant in both settings, where $R^2$ is the proportion of the variance of $Y_i(0)$ explained by a linear projection on the covariates. When the conditional mean of $Y_i(0)$ is linear in the covariates, as in the linear setting, this effect size equals $\tau/\sqrt{\var(Y_i(0)|X_i)}$. We view this as an approximate measure of signal-to-noise ratio. In the linear setting, $\tau/SD(Y_i(0))$ is approximately $0.1$ with $R^2 \approx 0.95$, while in the nonlinear setting, it is approximately $0.4$ with $R^2 \approx 0.35$, such that the effect size is approximately $0.6$
and $0.5$, respectively. This ensures comparable signal-to-noise ratios for hypothesis testing even for linearly-adjusted estimators. Furthermore, by considering linear and nonlinear scenarios, we can examine how rerandomization affects estimators when models are well-specified versus misspecified.



For each dataset, we compare two experimental designs. First, we consider complete randomization (CR), for which we assign $n_1 = n/2$ units to treatment and $n_0 = n/2$ to control completely at random. This corresponds to a known propensity score of $\pi = 0.5$; in Appendix~\ref{app:subsec:allocation} we also consider $\pi = 0.35$
and $\pi = 0.2$. Second, we consider rerandomization (RR), for which we use the Mahalanobis distance balance criterion with acceptance probability $p_a = 0.01$. To choose the threshold $a$, we choose the $p_a$ quantile of a $\chi^2_d$, where $d$ is the number of covariates used for rerandomization, because this is the asymptotic distribution of the Mahalanobis distance \citep{morgan2012}. Although this choice is based on asymptotics when we are studying finite samples, we found that results were similar when we instead picked the threshold using a finite-sample Monte Carlo approach, which we discuss in Section \ref{sec:practice:threshold}.


We examine four scenarios that are supposed to match the data availability scenarios defined in Section \ref{sec:theory:scenarios}. First, all 10 covariates are available in both design and analysis stages, $X^\rr = X^\adj$. Second, only the first 6 covariates are available in the design stage, $(X_1, \dots, X_6)' = X^\rr \subseteq X^\adj = X$. Third, only the last 6 covariates are available in the analysis stage, $X = X^\rr \supseteq X^\adj = (X_5, \dots, X_{10})'$. Fourth, only the first (last) 6 covariates are available in the design (analysis) stage, $X^\rr = (X_1, \dots, X_6)'$ and $X^\adj = (X_5,\dots, X_{10})'$. For each scenario, we generate 1000 randomizations under complete randomization and rerandomization.


We consider three estimators: the difference-in-means estimator $\tauest{D}$ from Equation (\ref{eq:theory:estimator-D}), the linearly-adjusted estimator with linear outcome models $\tauest{L}$ from Equation (\ref{eq:theory:estimator-L}), and the doubly-robust estimator $\tauest{DR}$ from Equation (\ref{eq:theory:estimator-DR}) where we set the outcome model estimator to be a random forest model from the \texttt{scikit-learn} package with
its default settings (100 trees). The outcome models of $\tauest{DR}$ are cross-fitted with $K=2$ folds (see Appendix~\ref{app:estimators}), while $\tauest{L}$ is fitted on the full sample.


We assess these estimators using two main metrics:
\begin{itemize}
    \item Variance ratio: We calculate the ratio of estimators' empirical variance under rerandomization and complete randomization
    \begin{equation*}
        \frac{\var(\tauest{*}|\A)}{\var(\tauest{*})},
    \end{equation*}
    for $* =$ D, L, DR. A ratio below 1 indicates that rerandomization makes the estimator more precise, compared to complete randomization.
    \item Pairwise coherence ratio: We measure the ratio of estimators' mean squared difference under rerandomization versus complete randomization:
    \begin{equation*}
        \frac{\E\left\{(\tauest{*} - \tauest{**} )^2|\A\right\}}{\E\left\{(\tauest{*} - \tauest{**})^2\right\}},
    \end{equation*}
    for $*, ** =$ D, L, DR. A ratio below 1 indicates that rerandomization makes the point estimates from different analytical approaches more similar to each other, compared to complete randomization. 
    We also consider an overall coherence ratio as the average pairwise coherence ratio.
    \item We also considered the power and coverage of the estimators under rerandomization and complete randomization. We leave these results in Appendix~\ref{app:simulations} since the estimators tend to have nominal coverage, and when the estimators are more precise under rerandomization, they also result in more power, following the results in the literature \citep{branson2024power}.
\end{itemize}

\subsection{Results}\label{sec:simulations:results}

\begin{figure}[ht]
\centering
\includegraphics[width=0.70\textwidth]{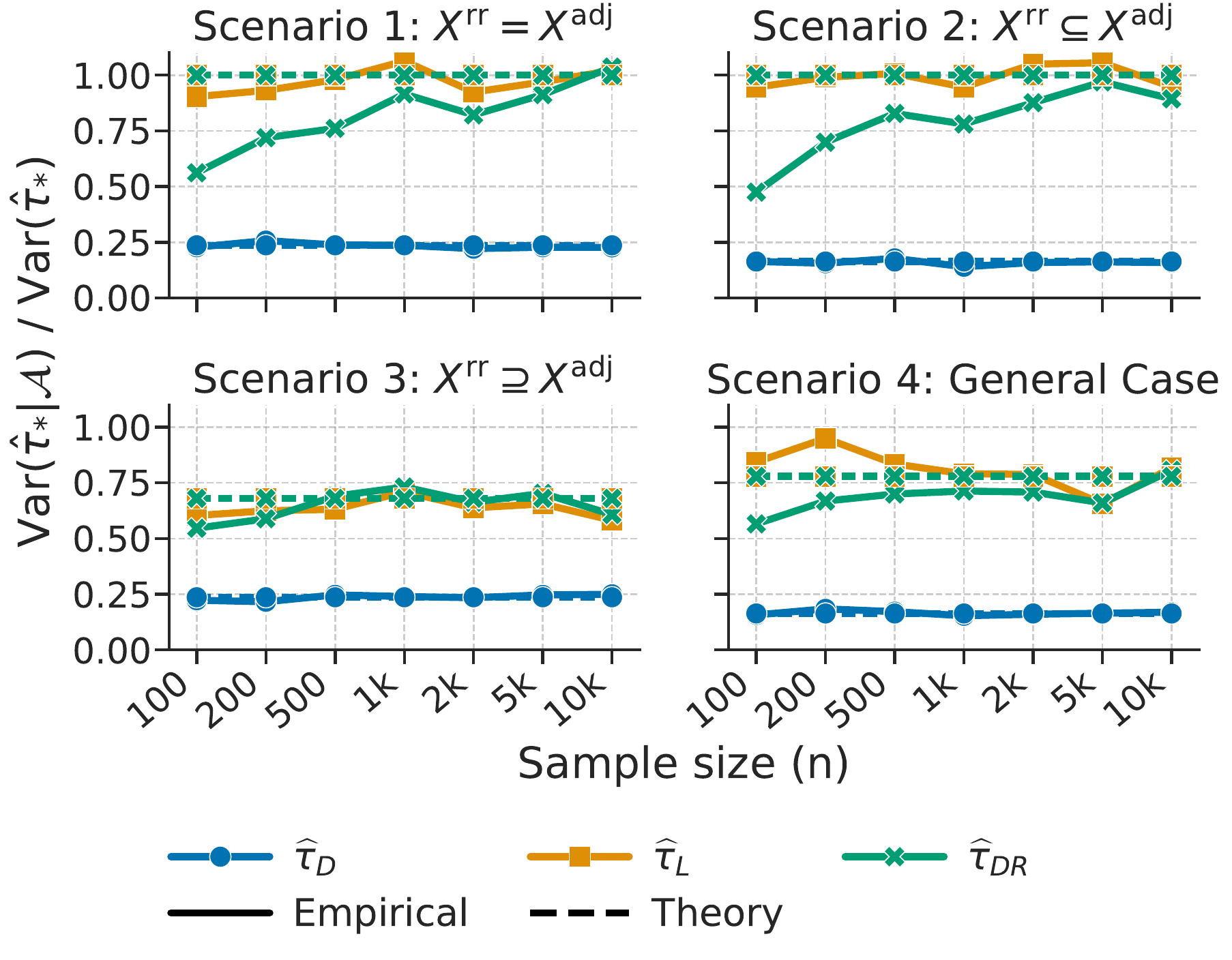}
\caption{Variance ratio in the linear setting for the difference-in-means estimator ($\tauest{D}$), the difference in linear outcome models ($\tauest{L}$), the doubly-robust estimator with random forest outcome model ($\tauest{DR}$). The sample sizes are on the x-axis, while the y-axis shows the variance ratios. The theoretical curve for $\tauest{DR}$ is derived under a super-population framework, while those for $\tauest{D}$ and $\tauest{L}$ are finite-population results. Each panel considers a different combination of covariates available in the design and analysis stages. Compared to asymptotic theory, $\tauest{DR}$ is benefited from rerandomization, especially for smaller samples in Scenarios 1 and 2. On the other hand, $\tauest{D}$ and $\tauest{L}$ tend to follow asymptotic theory.}
\label{fig:linear:precision}
\end{figure}

Each panel in Figure \ref{fig:linear:precision} corresponds to one of the four scenarios for the design stage and analysis stage covariates under the linear setting. The y-axis shows the variance ratio of $\tauest{D}, \tauest{L},$ and $\tauest{DR}$, where the closer the variance ratio is to zero, the more precise the estimator is under rerandomization. The x-axis represents the sample size. The dashed lines represent the known asymptotic theoretical result, while the solid lines show the empirical result achieved. Note that the theoretical result for $\tauest{DR}$ is derived under a super-population framework \citep{wang2024}, whereas our simulation considers a fixed finite population. In this case, the empirical results for both $\tauest{D}$ and $\tauest{L}$ follow the asymptotic results, where $\tauest{D}$ has the lowest variance ratio among the estimators, while $\tauest{L}$ has no precision gains in Scenarios 1 and 2. This is because, in these scenarios, all covariates used for rerandomization are also used for adjustment, and thus there can be no further asymptotic improvement via rerandomization. On the other hand, $\tauest{DR}$ is more precise under rerandomization in Scenarios 1 and 2, especially for smaller sample sizes, while the variance ratio tends to increase towards 1, which matches our conjecture in Section~\ref{sec:theory:debates} and illustrated in Figure~\ref{fig:precision}.

\begin{figure}[ht]
\centering
\includegraphics[width=0.70\textwidth]{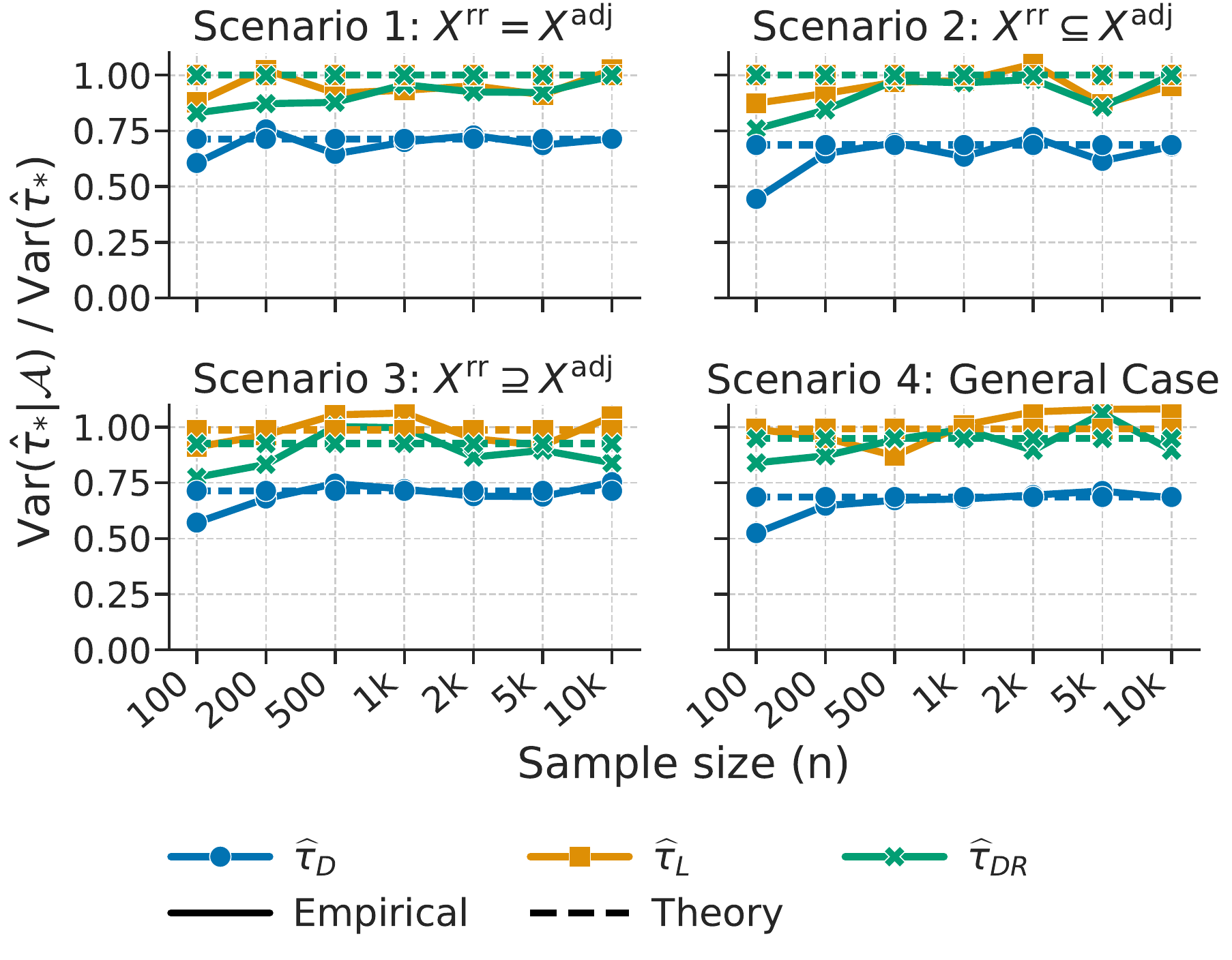}
\caption{Variance ratio in the non-linear setting for the difference-in-means estimator ($\tauest{D}$), the difference in linear outcome models ($\tauest{L}$), the doubly-robust estimator with random forest outcome model ($\tauest{DR}$). The sample sizes are on the x-axis, while the y-axis shows the variance ratios. The theoretical curve for $\tauest{DR}$ is derived under a super-population framework, while those for $\tauest{D}$ and $\tauest{L}$ are finite-population results. Each panel considers a different combination of covariates available in the design and analysis stages. Compared to asymptotic theory, $\tauest{DR}$ is somewhat benefited from rerandomization for smaller sample sizes. On the other hand, $\tauest{D}$ and $\tauest{L}$ tend to follow asymptotic theory.}
\label{fig:non-linear:precision}
\end{figure}

Figure \ref{fig:non-linear:precision} shows the same information as the previous figure, but for the non-linear setting. In this case, the empirical results $\tauest{D}$ and $\tauest{L}$ still follow the asymptotic theory, and since the $R^2_{D, X^\rr}$ is smaller under the non-linear setting, the precision benefits of rerandomization for $\tauest{D}$ diminish, while for $\tauest{L}$ there is almost no benefit. For $\tauest{DR}$ we still find some empirical precision improvements under smaller sample sizes in contrast to asymptotic theory, although the magnitude of these improvements are much more modest, compared to the linear setting.

Several of the curves in Figures~\ref{fig:linear:precision} and \ref{fig:non-linear:precision} are close to flat across the sample sizes, in particular those of $\tauest{D}$ and $\tauest{L}$, in all four scenarios and in both the linear and non-linear settings. This suggests that the finite-sample behavior of these two estimators is close to their asymptotic behavior in our simulations. Moreover, in the non-linear setting there is little variation across the panels of Figure~\ref{fig:non-linear:precision}: because the linear association between the outcomes and the covariates is weak, the choice of design and analysis covariates is less consequential.

\begin{figure}[ht]
\centering
\includegraphics[width=0.70\textwidth]{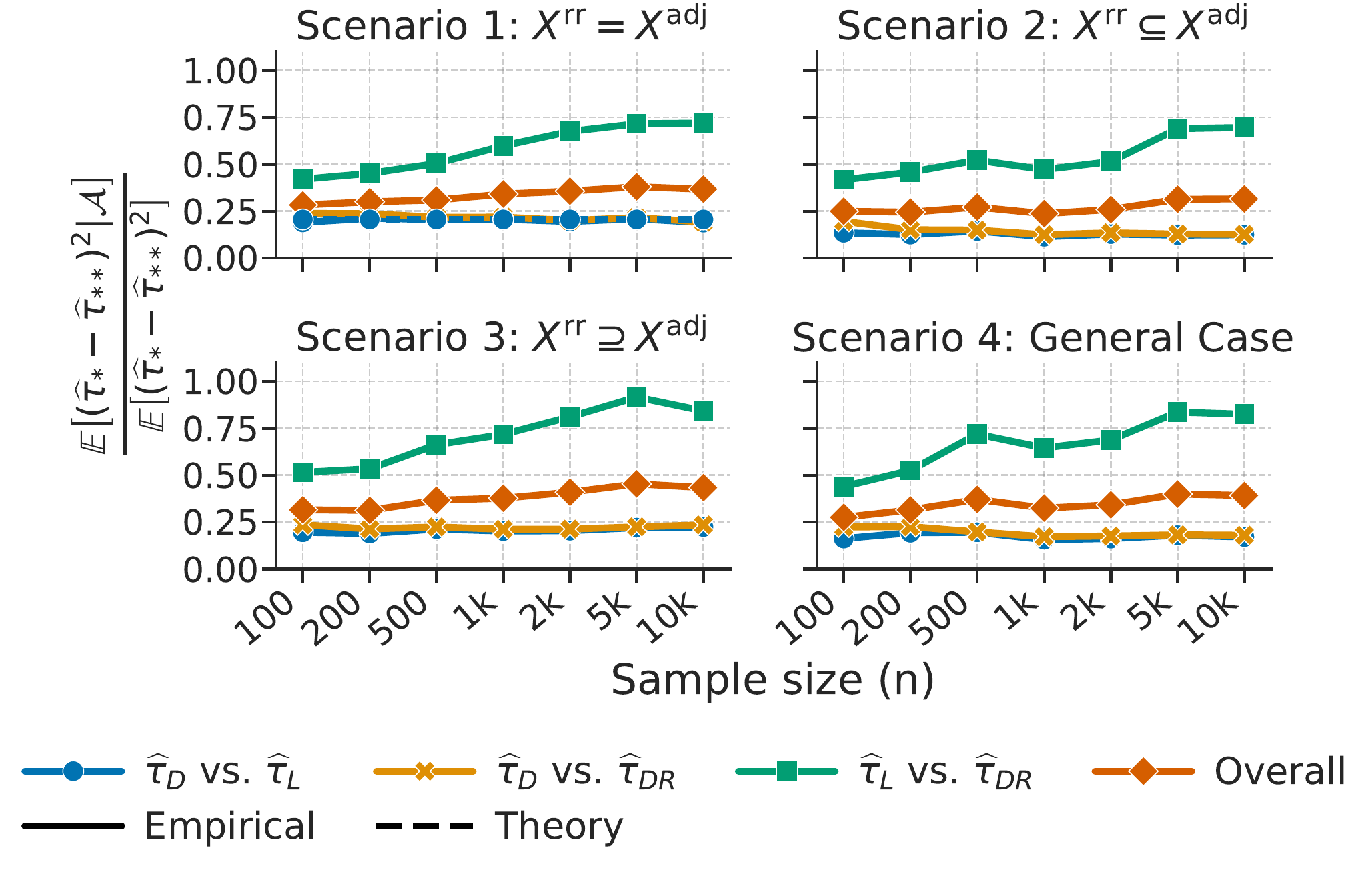}
\caption{Coherence ratios in the linear setting between the difference-in-means estimator ($\tauest{D}$), the difference in linear outcome models ($\tauest{L}$), and the doubly-robust estimator with random forest outcome model ($\tauest{DR}$). The overall coherence ratio is the average of the pairwise coherence ratios. The sample sizes are on the x-axis, while the y-axis shows the coherence ratios. Each panel considers a different combination of covariates available in the design and analysis stages. The only known asymptotic result for coherence is between $\tauest{D}$ and $\tauest{L}$ when $X^\adj = X^\rr$ (Scenario 1), which matches the empirical result we find. The lowest coherence ratios come when comparing $\tauest{D}$ with both $\tauest{L}$ and $\tauest{DR}$, which suggests that the unadjusted estimator gets more similar to the adjusted estimators. On top of that, since the coherence between all estimators increased, this suggests that rerandomization decreases the analysis' sensitivity to model specification.}
\label{fig:linear:coherence}
\end{figure}

Meanwhile, Figure \ref{fig:linear:coherence} displays analogous results for coherence in the linear setting. The y-axis shows the coherence ratios of the pairwise combination of $\tauest{D}, \tauest{L},$ and $\tauest{DR}$, and their overall coherence ratio, where the closer the coherence ratio is to zero, the more coherent the estimators are under rerandomization. The dashed line represents the known asymptotic theoretical result  while the solid lines show the empirical result achieved. Note that the only known asymptotic result for coherence is between $\tauest{D}$ and $\tauest{L}$ when $X^\adj = X^\rr$ (Scenario 1), which matches the empirical result we find. The lowest coherence ratios come when comparing $\tauest{D}$ with both $\tauest{L}$ and $\tauest{DR}$, which suggests that the unadjusted estimator is getting more similar to the adjusted estimators. On top of that, since the coherence between all estimators increased, this suggests that rerandomization decreases sensitivity to model specification, even asymptotically.

\begin{figure}[ht]
\centering
\includegraphics[width=0.70\textwidth]{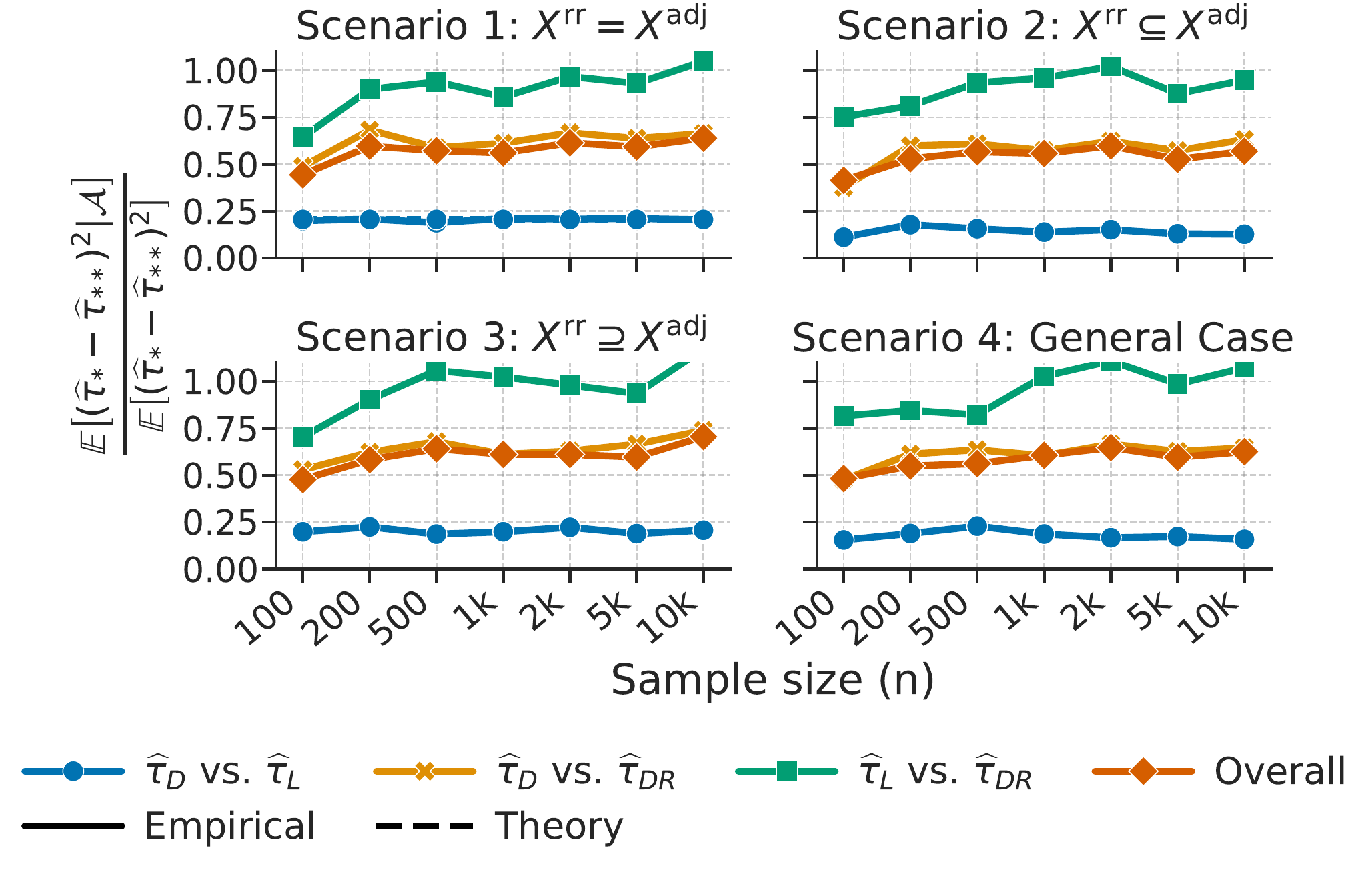}
\caption{Coherence ratios in the non-linear setting between the difference-in-means estimator ($\tauest{D}$), the difference in linear outcome models ($\tauest{L}$), and the doubly-robust estimator with random forest outcome model ($\tauest{DR}$). The overall coherence ratio is the average of the pairwise coherence ratios. The sample sizes are on the x-axis, while the y-axis shows the coherence ratios. Each panel considers a different combination of covariates available in the design and analysis stages. The only known asymptotic result for coherence is between $\tauest{D}$ and $\tauest{L}$ when $X^\adj = X^\rr$ (Scenario 1), which matches the empirical result we find. In general, the estimators are overall more coherent with one another, suggesting a decrease in sensitivity to model specification (albeit to a lesser degree than the linear case in Figure~\ref{fig:linear:coherence}).}
\label{fig:non-linear:coherence}
\end{figure}

Figure \ref{fig:non-linear:coherence} shows the same information as the previous figure, but for the non-linear setting. In this case, the empirical result of the coherence between $\tauest{D}$ and $\tauest{L}$ on Scenario 1 again match the asymptotic result. In general, the estimators are overall more coherent with one another, although the benefits are again much more modest than those in the linear setting. Notice that the coherence ratio of $\tauest{D}$ with $\tauest{L}$ and $\tauest{DR}$ is not the same anymore, with the coherence between $\tauest{D}$ and $\tauest{L}$ being the one with the most gains. This is due to the fact that $\tauest{DR}$ is able to explain more of the non-linearity between the covariates and potential outcomes, while $\tauest{D}$ after rerandomization and $\tauest{L}$ only capture the linear relationship between covariates and potential outcomes, hence $\tauest{DR}$ is more distant from the other estimators.

\subsection{Simulation extensions}\label{sec:simulations:extension}

In Section \ref{sec:theory:coherence} we connect the concept of coherence to p-hacking based on a single linear regression model that is used to adjust for different sets of covariates. In this case, rerandomization was shown to asymptotically improve the coherence between the estimators, reducing the sensitivity to the choice of covariates used. Beyond the choice of covariates, one can also consider the choice of model used for adjustment.

In order to understand the extent that rerandomization improves coherence among estimators that use different models for adjustment, we extend our analysis to consider two other doubly-robust estimators: $\tauest{Reg}$ with a linear outcome model, and $\tauest{Neural}$ with its outcome model defined by a neural network. For $\tauest{Neural}$ we use the \texttt{MLPRegressor} from the \texttt{scikit-learn} package with its default settings (one hidden layer with
100 units, trained for at most 200 iterations). In what follows, $\tauest{Forest}$ denotes the doubly-robust estimator using a random forest for the outcome model, as we had done previously. Also, note that $\tauest{Reg}$ is numerically the same as the linear regression estimator $\tauest{L}$ discussed earlier. However, we focus on doubly robust estimators because they commonly incorporate machine learning approaches for covariate adjustment \citep{wang2024, kennedy2024semiparametric}.

\begin{figure}[ht]
\centering
\includegraphics[width=0.70\textwidth]{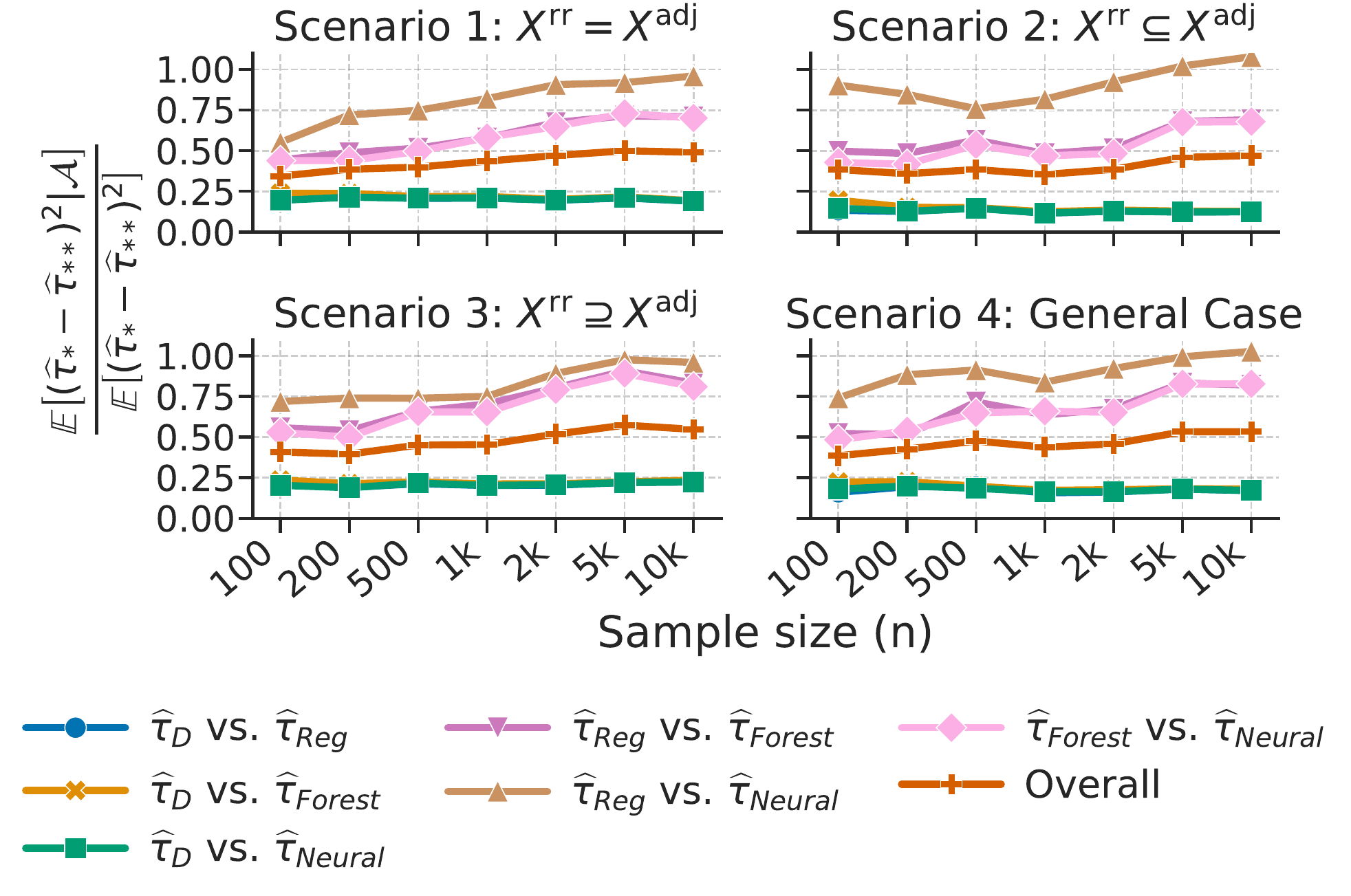}
\caption{Coherence ratios in the linear setting between the difference-in-means estimator ($\tauest{D}$), and the different doubly-robust estimators: linear outcome model ($\tauest{Reg}$), random forest outcome model ($\tauest{Forest}$), and neural network outcome model ($\tauest{Neural}$). The overall coherence ratio is the average of the pairwise coherence ratios. The sample sizes are on the x-axis, while the y-axis shows the coherence ratios. Each panel considers a different combination of covariates available in the design and analysis stages. No asymptotic results are known in this case. Since the coherence between all estimators increase, this suggests that rerandomization decreases the analysis' sensitivity to model specification.}
\label{fig:linear:DR_models:coherence}
\end{figure}

We find that all these doubly robust estimators and $\tauest{D}$ are more coherent under rerandomization in the linear setting, see Figure~\ref{fig:linear:DR_models:coherence}. Hence, this reinforces that rerandomization increases robustness to model specification. The coherence between these estimators also increase under the non-linear setting but to a lesser extent, see Appendix~\ref{app:subsec:MSNL10}.

It can also be the case that covariates for adjustment are selected in a data-driven fashion. Hence, we compare (i) when all 10 covariates are used for adjustment, which we denote as \texttt{Full}, and (ii) when a \texttt{Stepwise} selection procedure is used to select covariates for the aforementioned estimators. For simplicity and ease of computation, we used a forward-direction selection procedure within a linear regression model, where at each step we added the covariate that minimized cross-validated prediction error, and we continued until prediction error could no longer be lowered by adding an additional covariate. In our simulations, this procedure selected on average 5 covariates to use in the analysis stage. For simplicity, we consider the case where all 10 covariates were used for rerandomization.


\begin{figure}[ht]
\centering
\includegraphics[width=0.70\textwidth]{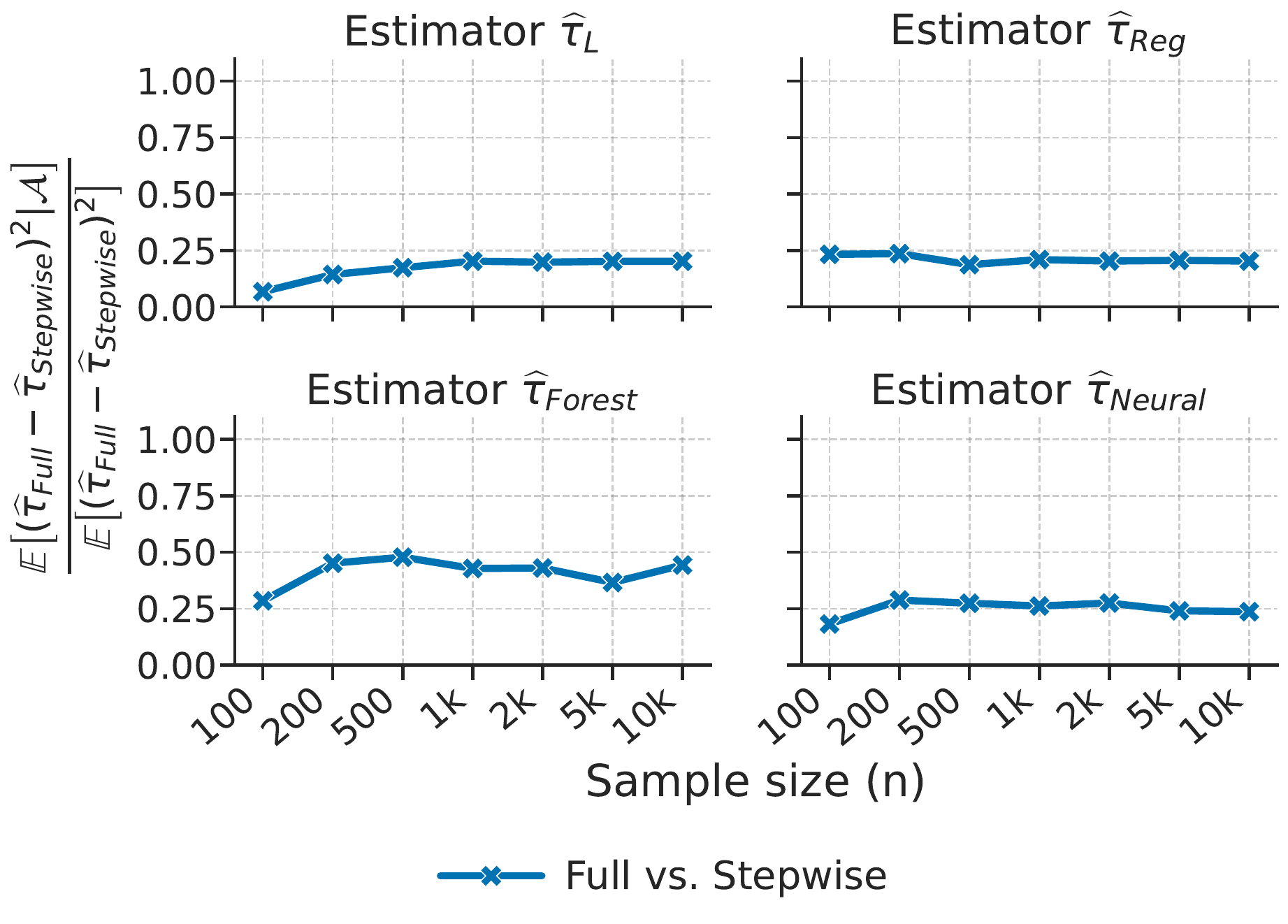}
\caption{Coherence ratios in the linear setting between different scenarios of data availability in the analysis stage. We consider the difference in linear outcome models $(\tauest{L})$, and the different doubly-robust estimators: linear outcome model ($\tauest{Reg}$), random forest outcome model ($\tauest{Forest}$), and neural network outcome model ($\tauest{Neural}$). While all covariates were used in the design stage, \texttt{Stepwise} is when covariate selection is done before conducting the analysis, and \texttt{Full} the scenario when all covariates are used for analysis. We consider the coherence within each estimator when estimated considering the two different scenarios. The sample sizes are on the x-axis, while the y-axis shows the coherence ratios. No asymptotic results are known in this case. Since the coherence between all estimators increased, this suggests that rerandomization decreases the analysis' sensitivity to covariate selection.}
\label{fig:linear:covariate_selection:coherence}
\end{figure}

Figure~\ref{fig:linear:covariate_selection:coherence} shows the coherence ratios in the linear setting for estimators that use all of the covariates (\texttt{Full}), and covariates chosen via the stepwise selection procedure (\texttt{Stepwise}). Similar to previous results, rerandomization increases the coherence among estimators for all sample sizes. In Appendix~\ref{app:subsec:CSNL10} we also consider the nonlinear setting---again there are coherence benefits, but they are more modest compared to the linear setting

In Appendix~\ref{app:simulations} we consider several other simulation scenarios, including (i) rerandomization based on Euclidean distance, and (ii) a high-dimensional (100 covariates) linear setting, (iii) varying the overlap between $X^\rr$ and $X^\adj$ while holding the number of covariates in each fixed, (iv) unequal treatment and control sample sizes, and (v) the pair-switching and BRAIN rerandomization algorithms discussed in Section~\ref{sec:practice:computational}. Again, the results also suggest that rerandomization increases precision and coherence of the estimators. For the Euclidean balance criterion, we find lower precision ratios for the difference-in-means estimator but similar ratios for adjusted estimators, compared to the Mahalanobis distance rerandomization, due to the high correlation between covariates. The Euclidean distance prioritizes balancing the top principal components of the covariates \citep{schindl2024}, which in our simulation include the direction along which the potential outcomes vary, whereas the Mahalanobis distance balances all directions equally; in Section \ref{sec:practice} we discuss why alternative balance metrics can be preferable in collinear settings. For the high-dimensional setting, we find variance ratios that match our conjecture in Figure \ref{fig:precision}; specifically, because the precision benefits for the difference-in-means estimator are less in the high-dimensional setting, the precision benefits are also less for adjusted estimators.


In summary, under finite-samples, these results suggest that rerandomization decreases sensitivity to analysis choices (e.g., covariate selection or model specification). Moreover, if the variance of the potential outcomes can be well explained by a linear projection on the covariates, rerandomization can also improve the estimators precision, thereby improving their statistical power while maintaining nominal coverage level. That said, as the linear association decreases or the number of covariates increases, these benefits decrease. Therefore, rerandomization should be viewed as a complementary ``safeguard" design strategy that enhances robustness and reliability of causal effect estimators while having the possibility to increase precision.


%% file: text/practice.tex
In this section we discuss practical considerations involved in rerandomized experiments, such as how to choose the balance criterion, how to conduct inference, how to deal with missing data, and the computational challenges that arise when implementing rerandomization. Although this section primarily acts as a guide for practitioners, we believe these details also illustrate interesting methodological problems that arise at the intersection of theory and practice for randomized experiments.

\subsection{Covariate balance criterion}\label{sec:practice:criterion}
\input{text/practice/criterion.tex}

\subsection{Inference}\label{sec:practice:inference}
\input{text/practice/inference.tex}

\subsection{Missing data}\label{sec:practice:missing}
\input{text/practice/missing-data.tex}

\subsection{Computational Burden}\label{sec:practice:computational}
\input{text/practice/computational.tex}

\subsection{Multi-arm experiments}\label{sec:practice:multi-arm}
\input{text/practice/multi-arm.tex}

%% file: text/practice/criterion.tex
\subsubsection{Metrics}





Rerandomization under the Mahalanobis distance provides an equal but diminishing balance improvement for each covariate as the number of covariates increases \citep{morgan2012}. If prior information on the relationship between covariates and potential outcomes is available, then it can be used to define a more specialized balance metric. 
For example, \cite{morgan2015} recommend organizing covariates into tiers of importance, and then specifying thresholds separately for each tier or one single threshold for a linear combination of the covariates in each tier. Alternatively, \cite{liu2025bayesian} demonstrate how prior information on the covariates' linear association with potential outcomes can be incorporated into a Bayesian criterion for rerandomization.



However, it can be difficult to specify the relative importance of a large number of covariates. Hence, \cite{branson2021} suggest to add a ridge parameter in the Mahalanobis distance, which emphasizes balance in the top principal components of the covariate space. In a similar vein, \cite{zhang2024} proposed running rerandomization using only the top principal components within a Mahalanobis distance

All of these metrics are based on the Mahalanobis distance or an adaptation of it. More generally, these are all special cases of quadratic forms \citep{lu2023, schindl2024}, formally $ \left(\overline{X}^\rr_1 - \overline{X}^\rr_0\right)^\prime A\left(\overline{X}^\rr_1 - \overline{X}^\rr_0\right)$ where $A$ is a positive semi-definite matrix chosen by the practitioner. While \cite{lu2023} consider quadratic forms for cluster rerandomization, \cite{schindl2024} derive general results for these rerandomization schemes and provides guidance on how to choose the appropriate matrix $A$ for the quadratic form given the covariates at hand. For example, \cite{schindl2024} demonstrate that the Euclidean distance is minimax optimal, in the sense that the variance reduction using the Euclidean distance is never "too far" from the reduction achieved by the optimal quadratic form


Outside of quadratic forms, \cite{gerber2012field} empirically recommended rerandomizing until all the p-values from significance tests of the difference between covariate means of the treatment groups are insignificant (e.g., p-values from two-samples t-tests or a logistic regression of treatment onto covariates).
\cite{zhao2024b} showed that rerandomizing with these $p$-value based metrics is asymptotically equivalent to Mahalanobis distance rerandomization, and \cite{lu2025rerandomization} used this rerandomization scheme to study how rerandomization can mitigate issues of $p$-hacking. 

All of the aforementioned balance criteria balance linear functions of the covariates. In order to balance non-linear functions of the covariates, one option is to incorporate such functions in the calculation of the Mahalanobis distance, at the cost of increasing the dimension size \citep{morgan2012}. Other options, for instance, would be to balance on the density of the covariates, or the estimated propensity score \citep{cytrynbaum2025}. Since the true propensity score is known in a randomized experiment, the estimated propensity score here refers to the fitted values of a model (e.g., a logistic regression) of the treatment assignment on the covariates, which are close to constant when the covariates are balanced. However, it is still unclear how to effectively balance non-linear functions of the covariates without significantly increasing the dimension size.

\subsubsection{Threshold definition} \label{sec:practice:threshold}

As defined in Section \ref{sec:criteria}, for a measure $\varphi$, the set of acceptable randomizations is $\A = \{Z:\varphi(X,Z) = 1\}$. For the metrics discussed in the previous section, we can rewrite the set of acceptable randomizations more specifically as $\A = \{Z: \phi(X,Z) \leq a\}$, where $a$ is the pre-specified threshold and $\phi(X,Z)$ is a non-negative metric of covariate imbalance such that $\phi(X,Z) = 0$ indicates perfect balance, and as $\phi(X,Z)$ increases the more imbalance there is.

The most common way to define the threshold $a$ is based on the proportion $p_a$ of randomizations deemed acceptable, i.e. $p_a = \Prob(\varphi(X,Z) = 1) = \Prob(\phi(X,Z) \leq a)$. As mentioned before, some common choices are between $p_a = 0.01$ and $p_a = 0.001$, however it has been shown that the marginal precision gains of causal estimators are decreasing in $a$. This means that there is not much additional benefit --- at least in terms of precision ---- in reducing $p_a$ even further once it is already small \citep{morgan2012}.

If the chosen covariate balance metric is the Mahalanobis distance, then asymptotically $M \sim \chi^2_d$, and $a$ is the $p_a$-quantile of the $\chi^2_d$ distribution. For a general $\phi$, its $p_a$-quantile can be estimated by Monte Carlo simulations. More specifically, one could generate $B \gg 0$ randomizations and for each randomization $Z^{(1)}, \dots, Z^{(B)}$ calculate $\phi\left(X, Z^{(1)}\right), \dots, \phi\left(X, Z^{(B)}\right)$, then take the $p_a$-quantile to be the value of $a$.

%% file: text/practice/inference.tex
\cite{morgan2012} first discussed randomization-based inference for rerandomized experiments, and later work developed the asymptotic distributions for some causal effect estimators. In this subsection, we outline how practitioners can analyze rerandomized experiments using either randomization-based inference (which does not require large samples for valid inference) or confidence intervals based on asymptotic results.

\subsubsection{Randomization-based inference}

Let the observed dataset be defined as $\{(X_i^{\obs}, Z_i^{\obs}, Y_i^{\obs})\}_{i=1}^{n}$. To conduct inference we proceed as follows:
\begin{enumerate}
    \item Define the sharp null hypothesis of no treatment effect $H_0 : Y_i(0) = Y_i(1)$ for all $i=1,\dots, n$.
    \item Calculate a test statistic with the observed data $T^{\obs}$. For instance $T^{\obs} = \tauest{D}^{\obs}$ is the $\tauest{D}$ estimator applied on the observed data.
    \item Generate $B \gg 0$ balanced randomizations \\ $(Z^{(1)}, \dots, Z^{(B)})$, meaning $Z^{(j)}$ is such that \\ $\varphi(X, Z^{(j)}) = 1,$ for $j =1,\dots, B$. 
    \item Define the datasets $\{(X_i^{\obs}, Z_i^{(j)}, Y_i^{\obs})\}_{i=1}^{n}$, for $j = 1,\dots, B$, and calculate the test statistics \\ $(T^{(1)}, \dots, T^{(B)})$ using each dataset, assuming the null hypothesis $H_0$ is true.
    \item Calculate the p-value as the proportion of test statistics in the rerandomized distribution that are at least as extreme as the observed statistic
    \begin{equation*}
        p = \frac{1}{B+1}\left[1+\sum_{j=1}^{B}\1\left(|T^{(j)}| > |T^{\obs}|\right)\right].
    \end{equation*}
\end{enumerate}

A key advantage of this approach is its finite-sample validity, however it has a few limitations. First, this only tests the sharp null hypothesis of no treatment effect for any subject. Second, to generate one acceptable randomization it takes an average of $1/p_a$ generated randomizations, hence to generate $B$ accepted randomizations it takes on average $B/p_a$. Constructing confidence intervals is even more computationally demanding, since it requires inverting the hypothesis test for a range of hypothesized treatment effects. In this case, it is often assumed that the treatment effect is additive. Extensions to allow for non-additive treatment effects have been studied in \cite{caughey2023randomisation}, who show the validity of randomization tests for bounded null hypotheses under rerandomization. In Section \ref{sec:practice:computational} we discuss recent tools to address the computational burden of generating acceptable randomizations.

\subsubsection{Confidence interval construction}\label{sec:practice:inference:ci}

An alternative is to use the asymptotic distribution of the causal effect estimators. Asymptotically, the estimators follow a convolution of a normal distribution and a component of a truncated multivariate normal distribution, as discussed in Section \ref{sec:theory:distributions}. Hence, $(1-\alpha)$-level confidence intervals would be given by
\begin{equation*}
    \tauest{*} \pm \widehat{v}_{1-\alpha/2}\left(\widehat{R}^2_{*, X^\rr}\right)\widehat{V}_{\tau\tau, *}^{1/2}n^{-1/2}
\end{equation*}
where $* =$ D,L,DR and $v_{1-\alpha/2}\left(\widehat{R}^2_{*, X^\rr}\right)$ is the $(1-\alpha/2)$-quantile of the asymptotic distribution of $\tauest{*}$, but using the estimate $\widehat{R}^2_{*, X^\rr}$. In Appendix \ref{app:estimators}  we provide the estimators for $V_{\tau \tau, *}$ and $R^2_{*, X^\rr}$.

To estimate $v_{1-\alpha/2}\left(\widehat{R}^2_{*, X^\rr}\right)$, one needs to generate $B \gg 0$ samples from the distribution
\begin{equation*}
    \sqrt{1-\widehat{R}_{*, X^\rr}^2}\epsilon + \sqrt{\widehat{R}_{*, X^\rr}^2}L_{d, a}.
\end{equation*}

Generating samples from $\epsilon \sim N(0,1)$ is straightforward. On the other hand, to generate samples from $L_{d,a}$ we use the following equivalence
\begin{equation*}
    L_{d,a} \sim \chi_{d,a}S\sqrt{\beta_d}
\end{equation*}
presented by \cite{li2018}, where $\chi_{d,a} \sim \chi_d|\chi^2_d \leq a$ is a truncated $\chi$ distribution, $S$ is a random sign taking $\pm 1$ with probability $1/2$, $\beta_d \sim \text{Beta}(1/2, (d-1)/2)$ is a Beta distribution degenerating to a point mass at 1 when $d=1$. 

To generate samples of $\chi_{d,a}$ one should generate samples from $\chi^2_d$, take the square root, and only accept the ones that are less than $a$. Hence, to generate samples of $L_{d,a}$, one just needs to generate samples from $\chi_{d,a}, S,$ and $\beta_d$, and multiply their values.  

Finally, after generating the $B$ samples $\left(\epsilon^{(1)}, \dots, \epsilon^{(B)}\right)$, $\left(L_{d,a}^{(1)}, \dots, L_{d,a}^{(B)}\right)$, and estimating $R_{*, X^\rr}^2$, one will have 
\begin{equation*}
    \sqrt{1-\widehat{R}_{*, X^\rr}^2}\epsilon^{(i)} + \sqrt{\widehat{R}_{*, X^\rr}^2}L_{d, a}^{(i)}
\end{equation*}
for $i = 1, \dots, B$, and take the $1-\alpha/2$ quantile of these values as an estimate for $v_{1-\alpha/2}\left(\widehat{R}^2_{*, X^\rr}\right)$.

%% file: text/practice/missing-data.tex
In practice, data is often subject to missingness due to various reasons such as attrition, nonresponse, or data recording errors. Thus, a recent literature addresses how to account for covariate imbalance in completely randomized experiments with missing data, where the missingness can be on the covariates and/or outcomes. Some of these results can be applied for rerandomization, and we explore them here.

First, one possibility is that all outcomes are observed but covariates are missing. This scenario was considered in \cite{zhao2024c}, and they studied five different strategies to account for missing covariates. Under these strategies, they prove analogous results for $\tauest{L}$ as the ones presented in Section \ref{sec:theory:distributions}, for complete randomization. From all the five strategies, the authors suggest the \textit{missingness-indicator} method. For this strategy, the covariate matrix is augmented with indicators denoting whether or not a covariate is missing. Then, missing values are imputed with some fixed value (e.g., zero); the choice of imputed value is not consequential, at least for linear models, because the missingness indicator accounts for the fact that these observations are missing. Once the updated covariate matrix is defined, all the rerandomization results explored in this paper hold. One caveat, however, is that augmenting by the missingness indicators can make the covariate matrix large, thus a balance metric that accounts for this high-dimensionality may be preferable.

Alternatively, it may be the case that all covariates are observed, but some outcomes are missing at random. In this case, to the best of our knowledge, the only known result in rerandomization is for $\tauest{L}$ and $\tauest{DR}$ \citep{wang2024}. While the rerandomization procedure remains the same, the results discussed in Section \ref{sec:theory} hold with minor changes on the estimator definition. For instance, we change $\widehat{\eta}_z(X_i)$ in $\tauest{DR}$ to be

\begin{align*}
    \widehat{\eta}_z\left(X_i\right) &= \frac{\1(Z_i = z)}{\pi^z(1-\pi)^{1-z}} \frac{R_i}{\widehat{\Prob}(R_i = 1 | X_i, Z_i = z)} \left(Y_i - \widehat{\mu}_z\left(X_i\right)\right) \\
    &+ \widehat{\mu}_z\left(X_i\right),
\end{align*}
for $z \in \{0,1\}$, where $R_i$ is an indicator variable that the outcome of unit $i$ is observed, and $\widehat{\Prob}(R_i = 1 | X_i, Z_i = z)$ is the estimated probability that the outcome is observed, conditional on covariates, in treatment group $z$. For example, this probability could be estimated from a logistic regression model or a machine learning approach; see \cite{wang2024} for details.

Finally, it may be the case that both outcomes and covariates are missing. Under this setting, one can combine the \textit{missingness-indicator} approach for handling missing covariates, and reweighting each observed unit by its probability of observing the outcome \citep{zhao2024a, zhao2024c}. However, no work in the rerandomization literature has explicitly studied this scenario.

%% file: text/practice/computational.tex
Inference on rerandomized experiments usually depends on rejective sampling to generate acceptable randomizations, thus making it computationally troublesome. Hence, some improvements and alternatives have recently been proposed in the literature.

Most covariate balance criteria are symmetric, i.e. $\varphi(X, Z) = \varphi(X, 1-Z)$, thus the number of generated treatment allocations can be halved. This is the first strategy defined in \cite{johansson2020}, named \textit{mirror allocations}. They extend this idea to a rank-based balance measure, based on the tiers of covariates defined in \cite{morgan2015}.

Motivated by the Metropolis-Hastings algorithm, \cite{zhu2023} propose a pair-switching rerandomization (PSRR) strategy that improves on the greedy pair-switching algorithm proposed by \cite{krieger2019nearly}. PSRR changes the treatment status of two randomly selected units---one from treatment and one from control---until an acceptable assignment is found. They find that PSRR is around 20x faster than rejective sampling. Finally, they extend PSRR to sequential experiments.

Although PSRR improves the efficiency of rejective sampling rerandomization, \cite{lu2025fast} show empirically that PSRR does not scale well with the dimension of the covariates, thus being inefficient for high-dimensional covariates. \cite{lu2025fast} consider the treatment allocation problem from a constrained optimization perspective rather than sampling, and proposes the balanced randomization via integer programming (BRAIN) approach that does not scale with sample size $n$, covariates dimension $d$, nor threshold $a$. Instead of using sampling, the BRAIN algorithm reformulates the problem of generating a balanced treatment assignment as a constrained optimization problem, thus being able to use combinatorial optimization techniques. Moreover, it is expandable for more structured experiments, such as stratified and clustered experiments.

On the other hand, the only available result for both BRAIN and PSRR is that, under these rerandomization approaches, $\tauest{D}$ is unbiased and is more precise than under complete randomization. These results assume (i) $n_1=n_0=n/2$, (ii) normality of $\tauest{D}$ and normality of the difference in covariates means under complete randomization, and (iii) additive treatment effect. No results on the asymptotic distribution, coherence, or statistical power of $\tauest{D}, \tauest{L},$ and $\tauest{DR}$ are available under the proposed rerandomization algorithms. Moreover, these algorithms do not necessarily sample uniformly from the set of acceptable randomizations, as acceptance-rejection sampling does. In Appendix~\ref{app:subsec:design} we compare these two algorithms against acceptance-rejection rerandomization empirically, in the setting of Section~\ref{sec:simulations} with $X^\rr = X^\adj$, on precision, coherence and power for all three estimators. The three designs produce essentially the same results. This suggests that, even though these approaches do not sample uniformly from the set of acceptable randomizations, that does not drastically affect the finite-sample or asymptotic behavior of the estimators, at least for this setting.

Another option that avoids rejection sampling would be to use importance sampling. Although not developed for rerandomization, \cite{branson2019randomization} derive an importance-sampling approach for computing randomization-based $p$-values for Bernoulli trial experiments that condition on summary statistics (e.g., the number of treated units). Thus, this approach could be extended to rerandomization, where covariate balance is used as the conditioning statistic. Although the resulting randomization test $p$-values are biased, the size of this bias is inversely related to the number of complete randomization draws, which are computationally cheap to generate.

We leave a list of available software for conducting rerandomized experiments in Appendix \ref{app:software}.

%% file: text/practice/multi-arm.tex
While the preceding discussion focuses on two-arm experiments, rerandomization can be extended to multi-arm settings. The two-arm setup with a single treatment assignment vector $Z \in \{0,1\}^n$ generalizes directly: In a multi-arm experiment with $K$ treatments, the assignment vector takes values in $\{0,1,\dots,K-1\}^n$. The balance criterion can be applied by checking balance across all $\binom{K}{2}$ groups, though computational costs increase with the number of pairwise comparisons.

Another option would be to balance only the most scientifically important treatment comparisons or use an omnibus balance criterion across all arms. However, general results for multi-arm rerandomized experiments remain underdeveloped, and further research is needed to determine which omnibus balance metrics would be most promising for a multi-arm experiment. That said, some work has been done on multi-arm experiments with specific structures, such as $2^K$ factorial experiments.

In the $2^K$ factorial design, $K$ binary factors are studied simultaneously. \cite{brason2016improving} proposed a computationally efficient approach that exploits the factorial structure: rather than balancing all $2^K$ treatment combinations jointly, they balance covariates with respect to each factor marginally. Specifically, for each factor $k = 1,\dots, K$, compute the Mahalanobis distance $M_k$ between units receiving levels $0$ and $1$ of that factor, and accept the randomization only if $M_k < a$ for all $k$. This requires only $2^K-1$ balance checks rather than $\binom{2^K}{2}$. \cite{li2020factorial} extend the asymptotic theory to this setting, showing that factorial rerandomization preserves unbiasedness while reducing variance of main effect and interaction estimators.

%% file: text/conclusion.tex
Rerandomization is a simple experimental design strategy that ensures covariate balance prior to conducting experiments. Given its simplicity, rerandomization has been used for many years by practitioners \citep{bruhn2009}, but statistical inference after rerandomization has only recently been studied \citep{morgan2012}. Rerandomization improves the precision of the difference-in-means estimator, thereby acting as a design stage analog to linear covariate adjustment. That said, theoretical results have found that rerandomization does not improve the asymptotic precision of adjusted estimators, as long as the covariates used for rerandomization are also used for adjustment \citep{li2020rerandomization, wang2024}. This has raised doubts about whether rerandomization is needed if covariates are adjusted for in the analysis stage. That said, these results come with several important caveats. In particular, these results are asymptotic and focus on the marginal distribution of estimators after rerandomization. Although some works have conjectured that rerandomization may still improve the precision of adjusted estimators in finite samples \citep{yang2021rejective, zhu2024design, wang2024} and considered rerandomization's ability to improve estimators' joint behavior, such as their coherence \citep{zhao2024b}, these topics have received less attention in the literature.

In this paper, we first provided a comprehensive review of rerandomization, including properties of the difference-in-means, linear regression, and nonparametric doubly robust estimators after rerandomization using the Mahalanobis distance. If more covariates are used in the design stage than the analysis stage, then the asymptotic distribution of these estimators follows a convolution of a normal and a component of a
truncated multivariate normal, resulting in greater precision compared to complete randomization. On the other hand, if the same or more covariates are used in the analysis stage, then rerandomization does not alter the asymptotic distribution of these estimators. Nonetheless, estimators' joint behavior is affected: For example, the difference-in-means and linear regression estimators become more coherent with each other after rerandomization.


Second, we conduct several simulation studies to assess the performance and relationship among parametric and nonparametric estimators under rerandomization in finite samples. We found that rerandomization improves the precision of adjusted estimators in finite samples---aligned with empirical evidence from the constrained randomization literature \citep{li2016evaluation, li2017evaluation, zhou2022constrained}---especially when the number of covariates is small, the covariates have a linear association with the potential outcomes, and a stringent rerandomization threshold is used. We also found that rerandomization decreases sensitivity to model choices (e.g., parametric versus nonparametric specifications, or the use of covariate selection procedures) in the analysis stage by increasing the coherence among estimators. To complement our review and simulation study, we provided a practical guide for practitioners who want to run rerandomized experiments. For example, we discussed how researchers can choose balance criteria, how to handle missing data, and how to construct confidence intervals for the unique asymptotic distribution of estimators after rerandomization.

That said, several important limitations and avenues for future research remain. Although we discussed the idea that rerandomization can mitigate $p$-hacking---i.e., the analyst has less ability to alter the treatment effect $p$-value via model specification \citep{lu2025rerandomization}---our simulation study focused on precision and coherence, hence a more direct empirical investigation on p-hacking after rerandomization is warranted. Furthermore, we focused on binary treatments. While rerandomization can be extended to multi-valued treatments, the theoretical properties and practical considerations in this setting, particularly concerning the choice of balance metrics, deserve deeper investigation. Future work should also explore whether rerandomization can improve the estimation of other estimands, such as quantile treatment effects or conditional average treatment effects, or how to implement rerandomization when common causal assumptions are violated, such as no interference.

%% file: text/appendix.tex
\section{Finite-Population Quantities Needed to Characterize the Estimators}\label{app:statistics}
\input{text/appendix/statistics}

\section{Estimators of \texorpdfstring{$V_{\tau\tau, *}$}{V} and \texorpdfstring{$R^2_{*, X^\rr}$}{R\texttwosuperior}}\label{app:estimators}

\input{text/appendix/estimators}

\section{Available software}\label{app:software}
\input{text/appendix/software}

\section{Additional simulation results}\label{app:simulations}
\input{text/appendix/simulations}

%% file: text/appendix/statistics.tex
Here we define several summary statistics of covariates and potential outcomes within a finite-population framework, which are needed to characterize the distributions of estimators after complete randomization or rerandomization. Furthermore, we formalize the assumptions needed so that asymptotic distributions of estimators after rerandomization are well-defined.

The behavior of estimators after complete randomization or rerandomization depends on the finite-population of covariates and potential outcomes. Recall that $X^{\text{rr}}$ denotes the $n \times d$ covariate matrix used for rerandomization, and $X_i^{\text{rr}}$ denotes the $d$-length vector of covariates for subject $i$. Meanwhile, $Y_i(z)$ denotes the potential outcomes for subject $i$, where $z \in \{0, 1\}$. Given this notation, the finite-population mean, variances and covariances of the covariates and potential outcomes are defined as:

\begin{equation*}
    \mean{X}^\rr = \frac{1}{n}\sum_{i=1}^{n}X^\rr_i, \;\;\;\;\;\;\;\;\;\;\;\; \mean{Y}(z) = \frac{1}{n}\sum_{i=1}^{n}Y_i(z),
\end{equation*}
\begin{equation*}
    S^2_{X^\rr} = \frac{1}{n-1}\sum_{i=1}^{n}(X^\rr_i - \mean{X}^\rr)(X^\rr_i - \mean{X}^\rr)',
\end{equation*}
\begin{equation*}
    S^2_{Y(z)} = \frac{1}{n-1}\sum_{i=1}^{n}\left(Y_i(z)-\mean{Y}(z)\right)^2,
\end{equation*}
\begin{equation*}
     S_{Y(z), X^\rr} = \frac{1}{n-1}\sum_{i=1}^{n}\left(Y_i(z)-\mean{Y}(z)\right)\left(X^\rr_i - \mean{X}^\rr\right)',
\end{equation*}
with $S_{X^\rr, Y(z)} = S'_{Y(z), X^\rr}$. Meanwhile, recall that $\tau_i = Y_i(1) - Y_i(0)$ denotes the individual treatment effect for subject $i$, and $\tau$ denotes the average treatment effect across the $n$ subjects. Then, the finite-population variance of the individual treatment effects is defined as
\begin{equation*}
    S^2_{\tau} = \frac{1}{n-1}\sum_{i=1}^{n}(\tau_i - \tau)^2.
\end{equation*}

Moreover, the finite-population variance of the linear projection of the potential outcomes $Y(z)$ onto the covariates $X^{\text{rr}}$ is defined as
\begin{equation*}
    S^2_{Y(z)|X^\rr} \equiv S_{Y(z), X^\rr}\left(S^2_{X^\rr}\right)^{-1}S_{X^\rr, Y(z)}
\end{equation*}
and similarly define $S^2_{\tau | X^\rr}$ as the finite-population variance of the linear projection of the individual treatment effects onto covariates. Note that these quantities do not assume a linear regression model in order to be defined.

Furthermore, we can define the finite-population covariance between $X^\rr$ and $X^\adj$ as
\begin{equation*}
    S_{X^\rr, X^\adj} = \frac{1}{n-1}\sum_{i=1}^{n}(X^\rr_i - \mean{X}^\rr)(X^\adj_i - \mean{X}^\adj)',
\end{equation*}
and similarly define all finite-population variances and covariances with respect to the covariates used in the analysis stage, $S^2_{X^\adj}, S_{Y(z), X^\adj}$, $S^2_{Y(z)|X^\adj}, S^2_{\tau|X^\adj}$.

These are all finite-population quantities, which can be estimated by their respective sample quantities. We denote the sample quantities for the variances and covariances with a lower-case $s$.

Note that since we do not observe individual treatment effects in our sample, $S^2_\tau$ can typically not be consistently estimated. That said, a consistent estimator for $S^2_{\tau | X^{rr}}$ that is commonly used in the literature to estimate $S^2_{\tau}$ is
\begin{equation*}
    s^2_{\tau|X^\rr} = (s_{Y(1), X^\rr} - s_{Y(0), X^\rr})(S^2_{X^\rr})^{-1}(s_{X^\rr, Y(1)} - s_{X^\rr, Y(0)})
\end{equation*}

Although $s^2_{\tau | X^{rr}}$ is a consistent estimator for $S^2_{\tau | X^{rr}}$, it consistently under-estimates $S^2_{\tau}$, in the sense that its asymptotic limit is less than $S^2_{\tau}$ \citep{li2018, li2020rerandomization}. Thus, using $s^2_{\tau | X^{rr}}$ as an estimator for $S^2_{\tau}$ yields conservative inference, which we discuss in the next section.

For the asymptotic results in Section \ref{sec:theory}, we need that as $n\to \infty$, for $z \in \{0,1\}$,
\begin{itemize}
    \item $n_z/n$, the proportion of units under treatment and control, have a positive limit;
    \item the finite population variances and covariances $S^2_{Y(z)}, S^2_{\tau}, S^2_{X^\rr}, S_{Y(z), X^\rr}, S^2_{X^\adj}, S_{Y(z), X^\adj}$ are well-defined, have finite limiting values, and $S^2_{X^\rr}, S^2_{X^\adj}$ need to be nonsingular;
    \item control over the behavior of the tail of the distribution of the potential outcomes and covariates, meaning $\frac{1}{n}\max_{1 \leq i \leq n} |Y_i(z) - \mean{Y}(z)|^2 \to 0$, $\frac{1}{n}\max_{1 \leq i \leq n} ||X^\rr_i - \mean{X}^\rr||^2 \to 0$, and \\ $\frac{1}{n}\max_{1 \leq i \leq n} ||X^\adj_i - \mean{X}^\adj||^2 \to 0$.
\end{itemize}

%% file: text/appendix/estimators.tex
In Section \ref{sec:theory} we defined the difference-in-means estimator ($\tauest{D}$) in Equation (\ref{eq:theory:estimator-D}), the linearly adjusted estimator ($\tauest{L}$) in Equation (\ref{eq:theory:estimator-L}), and the doubly-robust estimator ($\tauest{DR}$) in Equation (\ref{eq:theory:estimator-DR}). For each estimator, we provided their asymptotic distribution under complete randomization and rerandomization in Section \ref{sec:theory:distributions}. These distributions depend on a variance term $V_{\tau\tau, *}$, and a linear projection term $R^2_{*, X^\rr}$, which we define and provide estimators for below.

\subsection*{Estimator $\tauest{D}$}

The variance of $\tauest{D}$ is given by
\begin{equation*}
    V_{\tau\tau, \text{D}} = \frac{n}{n_1}S^2_{Y(1)} + \frac{n}{n_0}S^2_{Y(0)} - S^2_{\tau}
\end{equation*}
To estimate $V_{\tau\tau, \text{D}}$, two options are 
\begin{equation*}
    \widehat{V}_{\tau\tau, \text{D}} = \frac{n}{n_1}s^2_{Y(1)} + \frac{n}{n_0}s^2_{Y(0)}
\end{equation*}
and
\begin{equation*}
    \widehat{V}_{\tau\tau, \text{D}} = \frac{n}{n_1}s^2_{Y(1)} + \frac{n}{n_0}s^2_{Y(0)} - s^2_{\tau|X^\rr}.
\end{equation*}

The estimator $s^2_{\tau | X^\rr}$ asymptotically under-estimates $S^2_{\tau}$, in the sense that $\lim_{n \rightarrow \infty} s^2_{\tau | X^\rr} \leq S^2_{\tau}$. Thus, both estimators above asymptotically over-estimate the true variance $V_{\tau\tau, \text{D}}$, thereby yielding conservative inference. However, the second estimator can yield more precise inference, compared to the first, and hence is preferred in the literature \citep{li2018, li2020rerandomization}.

Furthermore, $R^2_{\text{D}, X^\rr}$ is defined as
\begin{equation*}
    R^2_{\text{D}, X^\rr} = \frac{\frac{n}{n_1}S^2_{Y(1)|X^\rr} + \frac{n}{n_0}S^2_{Y(0)|X^\rr} - S^2_{\tau|X^\rr} }{ \frac{n}{n_1}S^2_{Y(1)} + \frac{n}{n_0}S^2_{Y(0)} - S^2_{\tau} }
\end{equation*}
which can be estimated by
\begin{equation*}
    \widehat{R}^2_{\text{D}, X^\rr} = \frac{\frac{n}{n_1}s^2_{Y(1)|X^\rr} + \frac{n}{n_0}s^2_{Y(0)|X^\rr} - s^2_{\tau|X^\rr} }{ \frac{n}{n_1}s^2_{Y(1)} + \frac{n}{n_0}s^2_{Y(0)} - s^2_{\tau|X^\rr} }.
\end{equation*}

Because the denominator for $\widehat{R}^2_{\text{D}, X^\rr}$ asymptotically over-estimates the true denominator in $R^2_{\text{D}, X^\rr}$, the estimator $\widehat{R}^2_{\text{D}, X^\rr}$ asymptotically under-estimates the true $R^2_{\text{D}, X^\rr}$. This will in turn yield conservative inference.

\subsection*{Estimator $\tauest{L}$}

Considering the linear model $\mu_z\left(X_i\right) = \E\left(Y_i|X_i,Z_i=z\right) = \beta_{0z} + \beta_{1z}X_i$, we define $Y_i(z;\beta) \equiv Y_i(z) - \mu_z\left(X_i^\adj\right)$ for $z=0,1$ to be the linear residual of the potential outcomes, $\tau_i(\beta) \equiv \tau_i - \left[\mu_1\left(X_i^\adj\right) - \mu_0\left(X_i^\adj\right)\right]$ the linear residual of the individual treatment effect, and $\tau(\beta) \equiv \frac{1}{n}\sum_{i=1}^{n}\tau_i(\beta)$ the average linear residual treatment effect.


Thereby, we can define the finite-population variances of $Y_i(z;\beta)$ and its linear projection on $X^\rr_i$ and $X^\adj_i$ as $S^2_{Y(z;\beta)}$, $S^2_{Y(z;\beta)|X^\rr}$, and $S^2_{Y(z;\beta)|X^\adj}$, respectively. Moreover, we similarly define the finite-population variances of $\tau_i(\beta)$ and its linear projection on $X^\rr_i$ and $X^\adj_i$ as $S^2_{\tau(\beta)}$, $S^2_{\tau(\beta)|X^\rr}$, and $S^2_{\tau(\beta)|X^\adj}$, respectively. As before, we define their respective sample estimates with lower case $s$.


The variance of $\tauest{L}$ is given by
\begin{equation*}
    V_{\tau\tau, \text{L}} = \frac{n}{n_1}S^2_{Y(1;
    \beta)} + \frac{n}{n_0}S^2_{Y(0;
    \beta)} - S^2_{\tau(\beta)}
\end{equation*}
and its estimator following Equation (20) of \cite{li2020rerandomization},
\begin{equation*}
    \widehat{V}_{\tau\tau, \text{L}} = \frac{n}{n_1}s^2_{Y(1;
    \beta)} + \frac{n}{n_0}s^2_{Y(0;
    \beta)} - s^2_{\tau(\beta)|X^\adj}
\end{equation*}
where
\begin{align*}
    &s^2_{\tau(\beta)|X^\adj} = \\
    &\quad (s_{Y(1; \beta), X^\adj} - s_{Y(0; \beta), X^\adj})(S^2_{X^\adj})^{-1}(s_{X^\adj, Y(1; \beta)} - s_{X^\adj, Y(0; \beta)})
\end{align*}

while $R^2_{\text{L}, X^\rr}$ is
\begin{equation*}
    R^2_{\text{L}, X^\rr} = \frac{\frac{n}{n_1}S^2_{Y(1;\beta)|X^\rr} + \frac{n}{n_0}S^2_{Y(0;\beta)|X^\rr} - S^2_{\tau(\beta)|X^\rr} }{ \frac{n}{n_1}S^2_{Y(1;\beta)} + \frac{n}{n_0}S^2_{Y(0;\beta)} - S^2_{\tau(\beta)} }
\end{equation*}
and one estimator follows directly as
\begin{equation*}
    \widehat{R}^2_{\text{L}, X^\rr} = \frac{\frac{n}{n_1}s^2_{Y(1;\beta)|X^\rr} + \frac{n}{n_0}s^2_{Y(0;\beta)|X^\rr} - s^2_{\tau(\beta)|X^\rr} }{ \frac{n}{n_1}s^2_{Y(1;\beta)} + \frac{n}{n_0}s^2_{Y(0;\beta)} - s^2_{\tau(\beta)|X^\adj} },
\end{equation*}
where
\begin{align*}
    &s^2_{\tau(\beta)|X^\rr} = \\
    &\quad (s_{Y(1; \beta), X^\rr} - s_{Y(0; \beta), X^\rr})(S^2_{X^\rr})^{-1}(s_{X^\rr, Y(1; \beta)} - s_{X^\rr, Y(0; \beta)})
\end{align*}

which is the estimator proposed by \cite{li2020rerandomization} in Equation (21).

Again, because $ s^2_{\tau(\beta)|X^\adj}$ asymptotically over-estimates $S^2_{\tau(\beta)}$, the estimators $\widehat{V}_{\tau\tau, \text{L}}$ and $\widehat{R}^2_{\text{L}, X^\rr}$ over-estimate and under-estimate their true estimands, respectively. This yields conservative inference.

\subsection*{Estimator $\tauest{DR}$}

The doubly-robust estimator is considered in \cite{wang2024}, where the authors assume a super-population setting. Thereby, the formulations of $V_{\tau\tau, \text{DR}}$ and $R^2_{\text{DR}, X^\rr}$ change slightly from $V_{\tau\tau, *}$ and $R^2_{*, X^\rr}$, $*=$ D, L, but the intuition is analogous. The formulations in this case are based on efficient influence functions (EIF). For general models $\mu_z(X_i) = \E(Y_i|X_i,Z_i=z)$, the true EIF values for subject $i$ in treatment group $z \in \{0, 1\}$ are
\begin{align*}
     \eif_i(z) &= \frac{\1(Z_i = z)}{\pi^z(1-\pi)^{1-z}}\left(Y_i - \mu_z\left(X^\adj_i\right)\right) + \mu_z\left(X^\adj_i\right) \\
     & - \mean{\eta}_z
\end{align*}
where $\mean{\eta}_z$ is
\begin{equation*}
    \mean{\eta}_z = \frac{1}{n} \sum_{i=1}^{n} \left[\frac{\1(Z_i = z)}{\pi^z(1-\pi)^{1-z}}\left(Y_i - \mu_z\left(X^\adj_i\right)\right) + \mu_z\left(X^\adj_i\right)\right],
\end{equation*}
$\eif_i = \eif_i(1) - \eif_i(0)$, and $\pi = n_1/n$.

To implement $\tauest{DR}$, cross-fitting is typically employed to decouple the estimation of nuisance parameters from treatment effect estimation via EIFs, thereby mitigating bias induced by overfitting.
Specifically, it is assumed that, when implementing $\tauest{DR}$, the sample is split in $K$ folds, where $K$-1 folds are used to estimate $\mu_z\left(X_i^\adj\right)$ for $z \in \{0, 1\}$, and the remaining fold is used to compute $\widehat{EIF}_i(z)$. Then, the procedure is repeated swapping the folds until $\widehat{EIF}_i(z)$ is computed across all folds, and the resulting estimates are averaged to obtain $\tauest{DR}$. Commonly, $K$ is chosen to be 2, 5, or 10.

Hence, the variance of $\tauest{DR}$ is given by
\begin{equation*}
    V_{\tau\tau, \text{DR}} = \frac{1}{K}\sum_{k=1}^{K}\frac{1}{|\Ical_k|}\sum_{i \in \Ical_k}\eif_i^2
\end{equation*}
and
\begin{equation*}
    R^2_{\text{DR}, X^\rr} = \frac{1}{V}C'\left(n\Sigma\right)^{-1}C
\end{equation*}
where $\Sigma = \cov(\tauestrr)$, $\Ical_k$ denotes the index set for the $k$-th fold from the cross-fitting procedure,
and $C = \frac{1}{K}\sum_{k=1}^{K}\frac{1}{|\Ical_k|} \sum_{i \in \Ical_k} \frac{Z_i - \pi}{\pi(1-\pi)}\eif_i\left(X_i^\rr - \mean{X^\rr}\right)$.

Meanwhile the estimators are defined in the Appendix B.3 of \cite{wang2024} in Equations (12) and (13),
\begin{equation*}
    \widehat{V}_{\tau\tau, \text{DR}} = \frac{1}{K}\sum_{k=1}^{K}\frac{1}{|\Ical_k|}\sum_{i \in \Ical_k}\widehat{\eif}_i^2
\end{equation*}
and
\begin{equation*}
    \widehat{R}^2_{\text{DR}, X^\rr} = \frac{1}{V}\widehat{C}'\left(n\widehat{\Sigma}\right)^{-1}\widehat{C}
\end{equation*}
where $\widehat{C} = \frac{1}{K}\sum_{k=1}^{K}\frac{1}{|\Ical_k|} \sum_{i \in \Ical_k} \frac{Z_i - \pi}{\pi(1-\pi)}\widehat{\eif}_i\left(X_i^\rr - \mean{X^\rr}\right)$, with
\begin{equation*}
    \widehat{\eif}_i(z) = \frac{\1(Z_i = z)}{\pi^z(1-\pi)^{1-z}}\left(Y_i - \widehat{\mu}_z\left(X^\adj_i\right)\right) + \widehat{\mu}_z\left(X^\adj_i\right) 
     - \widehat{\mean{\eta}}_z,
\end{equation*}
\begin{equation*}
    \widehat{\mean{\eta}}_z = \frac{1}{n} \sum_{i=1}^{n} \left[\frac{\1(Z_i = z)}{\pi^z(1-\pi)^{1-z}}\left(Y_i - \widehat{\mu}_z\left(X^\adj_i\right)\right) + \widehat{\mu}_z\left(X^\adj_i\right)\right],
\end{equation*}
and
$\widehat{\eif}_i = \widehat{\eif}_i(1) - \widehat{\eif}_i(0)$.

%% file: text/appendix/software.tex
There is still a lack of tailored software for rerandomized experiments. To our knowledge, there are only a few publicly available packages for specific applications of rerandomized experiments, all of which are in R. We briefly describe these packages below



\textbf{fastrerandomize}: \href{https://github.com/cjerzak/fastrerandomize-software}{github}

Optimizes hardware usage to increase computational speed and better allocate memory to find balanced randomizations. Based on \cite{jerzak2023degrees, goldstein2025}.

\textbf{rerandPower}: \href{https://cran.r-universe.dev/rerandPower}{CRAN}

Implements power and sample size calculations for $\tauest{D}$ after Mahalanobis distance rerandomization. Based on \cite{branson2024power}.


%% file: text/appendix/simulations.tex
In addition to the simulation results in Section~\ref{sec:simulations}, we considered two other metrics:
\begin{itemize}
    \item Coverage: We measure the coverage of each estimator under  complete randomization and rerandomization:
    \begin{equation*}
        \Prob(\tau \in \text{CI}(\tauest{*};\alpha)) \;\;\; \text{and} \;\;\;  \Prob(\tau \in \text{CI}(\tauest{*};\alpha)|\A)
    \end{equation*}
    for $* =$ D, L, DR, and significance level $\alpha$.
    \item Power: We measure the power of each estimator under complete randomization and rerandomization:
    \begin{equation*}
        \Prob(\text{reject } H_0 | \tau \neq 0) \;\;\; \text{and} \;\;\;  \Prob(\text{reject } H_0 | \tau \neq 0, \A)
    \end{equation*}
    where the null hypothesis is $H_0 : \tau = 0$.
\end{itemize}

Furthermore, we considered many other settings (e.g., estimators based on covariate selection, and estimators with different model specifications). In this section, we describe additional results for the following cases:
\begin{itemize}
    \item Linear setting with 10 covariates (\ref{app:subsec:L10}), when varying model specification of the doubly robust estimator (\ref{app:subsec:MSL10}) or covariate selection (\ref{app:subsec:CSL10});
    \item Non-linear setting with 10 covariates (\ref{app:subsec:NL10}), when varying model specification of the doubly robust estimator (\ref{app:subsec:MSNL10}) or covariate selection (\ref{app:subsec:CSNL10});
    \item Linear setting with 100 covariates (\ref{app:subsec:L100}), when varying model specification of the doubly robust estimator (\ref{app:subsec:MSL100}) or covariate selection (\ref{app:subsec:CSL100});
    \item Linear setting with 10 covariates but performing rerandomization based on the Euclidean distance (\ref{app:subsec:EL10});
    \item Linear and non-linear settings with 10 covariates, where we vary the overlap between the covariates used in the design and analysis stages (\ref{app:subsec:overlap});
    \item Linear and non-linear settings with 10 covariates, where we consider unequal treatment allocations (\ref{app:subsec:allocation});
    \item Linear and non-linear settings with 10 covariates, where we use pair-switching rerandomization (PSRR) or BRAIN to find acceptable randomizations (\ref{app:subsec:design}).
\end{itemize}

\subsection{Linear setting with 10 covariates}\label{app:subsec:L10}

\begin{figure}[ht]
\centering
\includegraphics[width=0.70\textwidth]{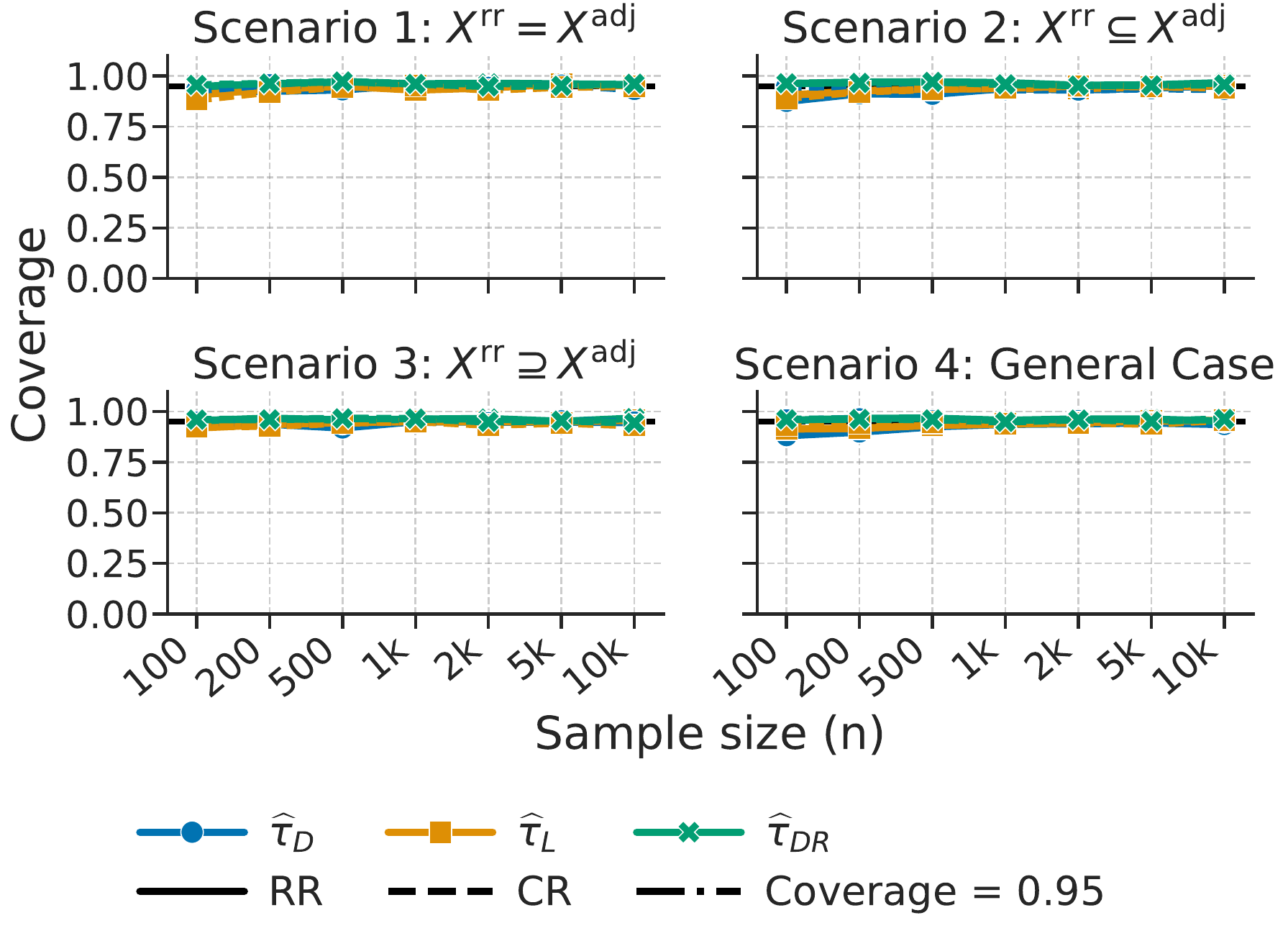}
\caption{Coverage comparison between the difference-in-means ($\tauest{D}$), difference-in-linear outcome models ($\tauest{L}$) and doubly-robust estimator with random forest outcome model ($\tauest{DR}$) for different sample sizes and covariates available for design and analysis under complete randomization (CR) and Mahalanobis-based rerandomization (RR) for the linear setting with 10 covariates. For all data availability scenarios both adjusted estimators achieve nominal coverage level. However, the difference-in-means estimator has undercoverage for the smaller
sample sizes, especially when balancing on less covariates in the design stage (scenarios 2 and 4). See \cite{li2018} for a detailed discussion on this property. Under complete randomization, the unadjusted estimator achieves nominal coverage level.}
\label{fig:linear:coverage}
\end{figure}

\begin{figure}[ht]
\centering
\includegraphics[width=0.70\textwidth]{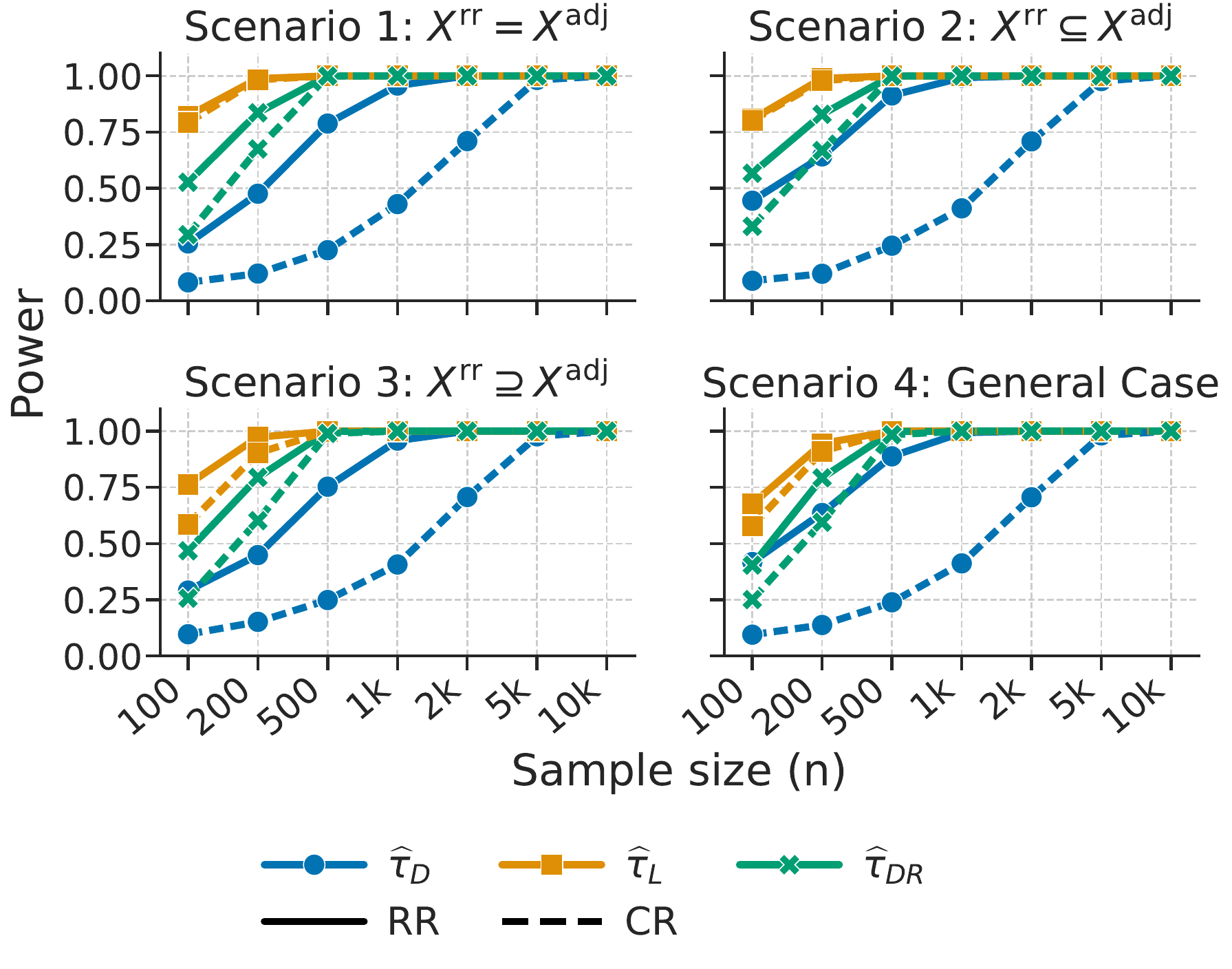}
\caption{Power comparison between the difference-in-means ($\tauest{D}$), difference-in-linear outcome models ($\tauest{L}$) and doubly-robust estimator with random forest outcome model ($\tauest{DR}$) for different sample sizes and covariates available for design and analysis under complete randomization (CR) and Mahalanobis-based rerandomization (RR) for the linear setting with 10 covariates. All estimators are at least as powerful under rerandomization in comparison to complete randomization for all scenarios of data availability.}
\label{fig:linear:power}
\end{figure}

In Section~\ref{sec:simulations:results} we discussed the precision and coherence results of the linear setting with 10 covariates. Now we discuss coverage and power. Figures~\ref{fig:linear:coverage} and \ref{fig:linear:power} show the results on coverage and power, respectively. All estimators are at least as powerful under rerandomization compared to complete randomization. Moreover, both adjusted estimators achieve nominal coverage level under rerandomization, except for a slight undercoverage of $\tauest{L}$ at $n=100$. On the other hand, the difference-in-means estimator has undercoverage for the smaller sample sizes, especially when balancing on fewer covariates in the design stage (scenarios 2 and 4), and achieves nominal coverage level as the sample size increases. The intuition is based on the interplay between $R^2_{D, X^\rr}$ and $v_{d,a}$ on the variance of $\tauest{D}$ under rerandomization. For this data generating process, the decrease in the number of covariates used for rerandomization from scenarios 1 and 3 to scenarios 2 and 4 does not change $R^2_{D, X^\rr}$ significantly, in fact it changes from approximately 0.99 to 0.95. However, the value of $v_{d,a}$ changes significantly from approximately 0.21 to 0.11, thereby causing $\tauest{D}$ to have undercoverage in some scenarios.


\subsection{Doubly robust estimator model specification on the linear setting with 10 covariates} \label{app:subsec:MSL10}

\begin{figure}[ht]
\centering
\includegraphics[width=0.70\textwidth]{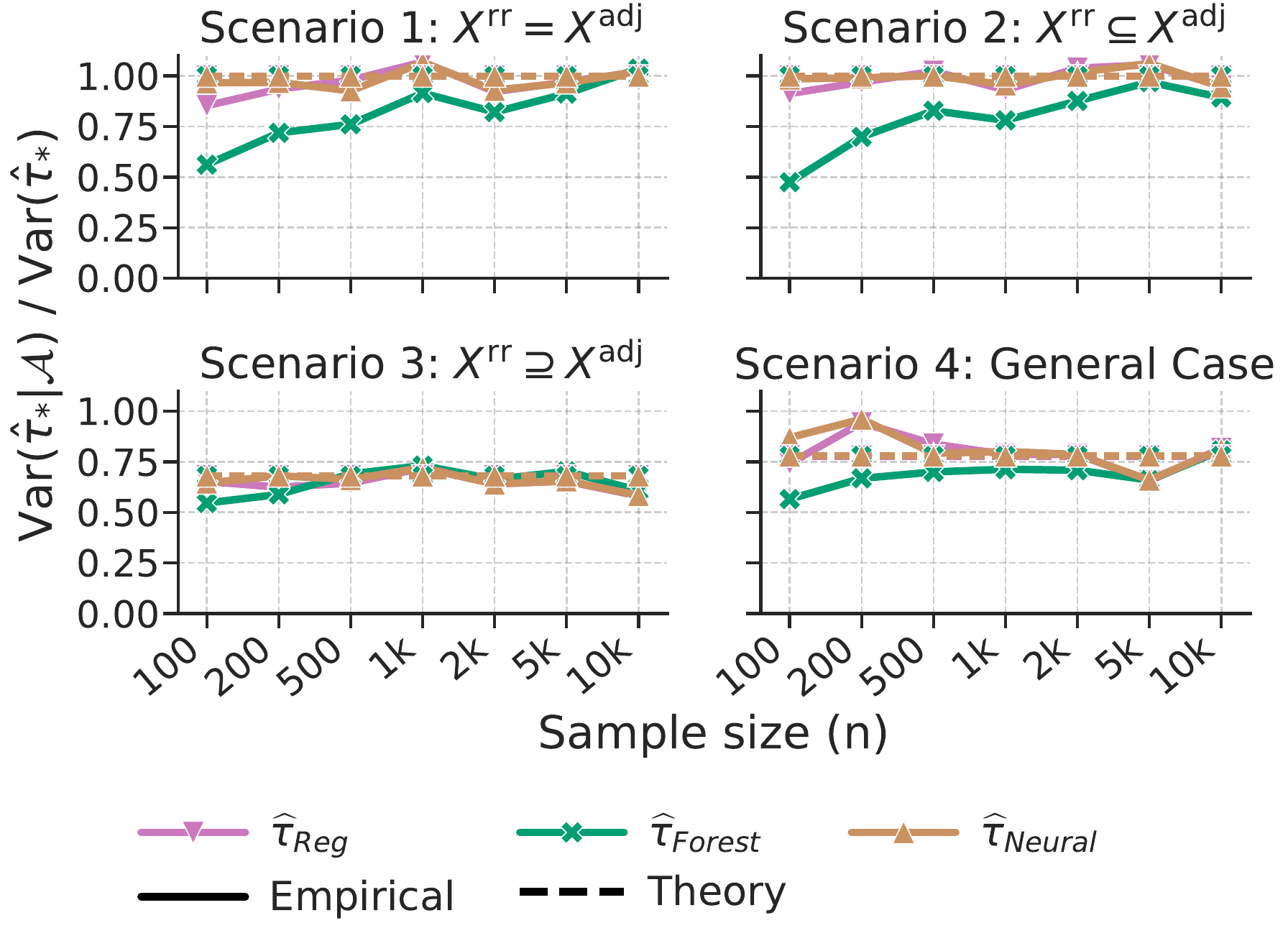}
\caption{Precision gain comparison between the doubly-robust estimator with linear outcome model ($\tauest{Reg}$), the doubly-robust estimator with random forest outcome model ($\tauest{Forest}$), the doubly-robust estimator with neural network outcome model ($\tauest{Neural}$) for different sample sizes and covariates available for design and analysis for the linear setting with 10 covariates. There is significant finite-sample precision gains for $\tauest{Forest}$ under all scenarios. On the other hand, the variance ratios of $\tauest{Neural}$ and $\tauest{Reg}$ follow the asymptotic theory.}
\label{fig:linear:DR_models:precision}
\end{figure}

\begin{figure}[ht]
\centering
\includegraphics[width=0.70\textwidth]{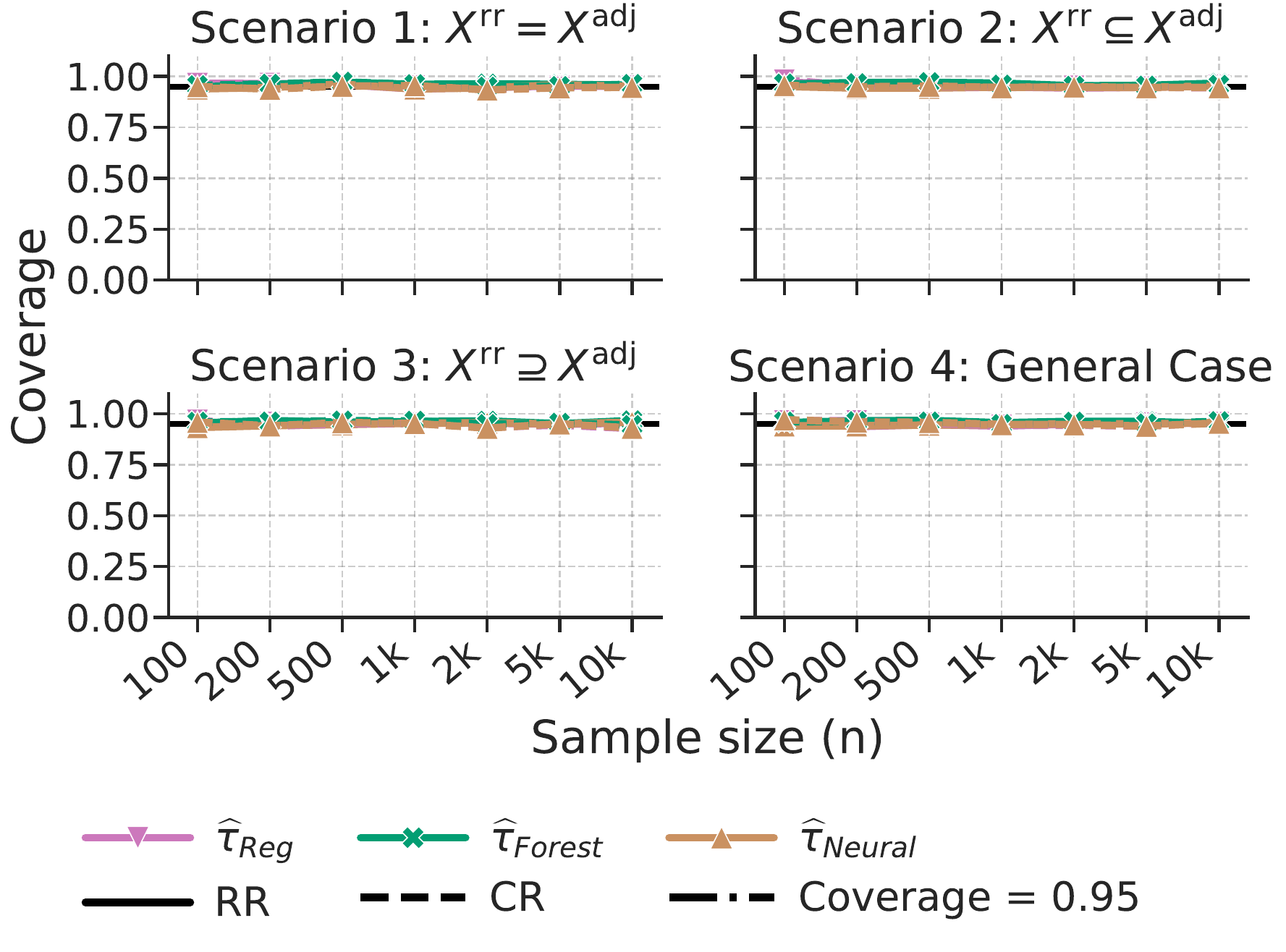}
\caption{Coverage comparison between the doubly-robust estimator with linear outcome model ($\tauest{Reg}$), the doubly-robust estimator with random forest outcome model ($\tauest{Forest}$), the doubly-robust estimator with neural network outcome model ($\tauest{Neural}$) for different sample sizes and covariates available for design and analysis under complete randomization (CR) and Mahalanobis-based rerandomization (RR) for the linear setting with 10 covariates. For all data availability scenarios all estimators achieve nominal coverage level.}
\label{fig:linear:DR_models:coverage}
\end{figure}

\begin{figure}[ht]
\centering
\includegraphics[width=0.70\textwidth]{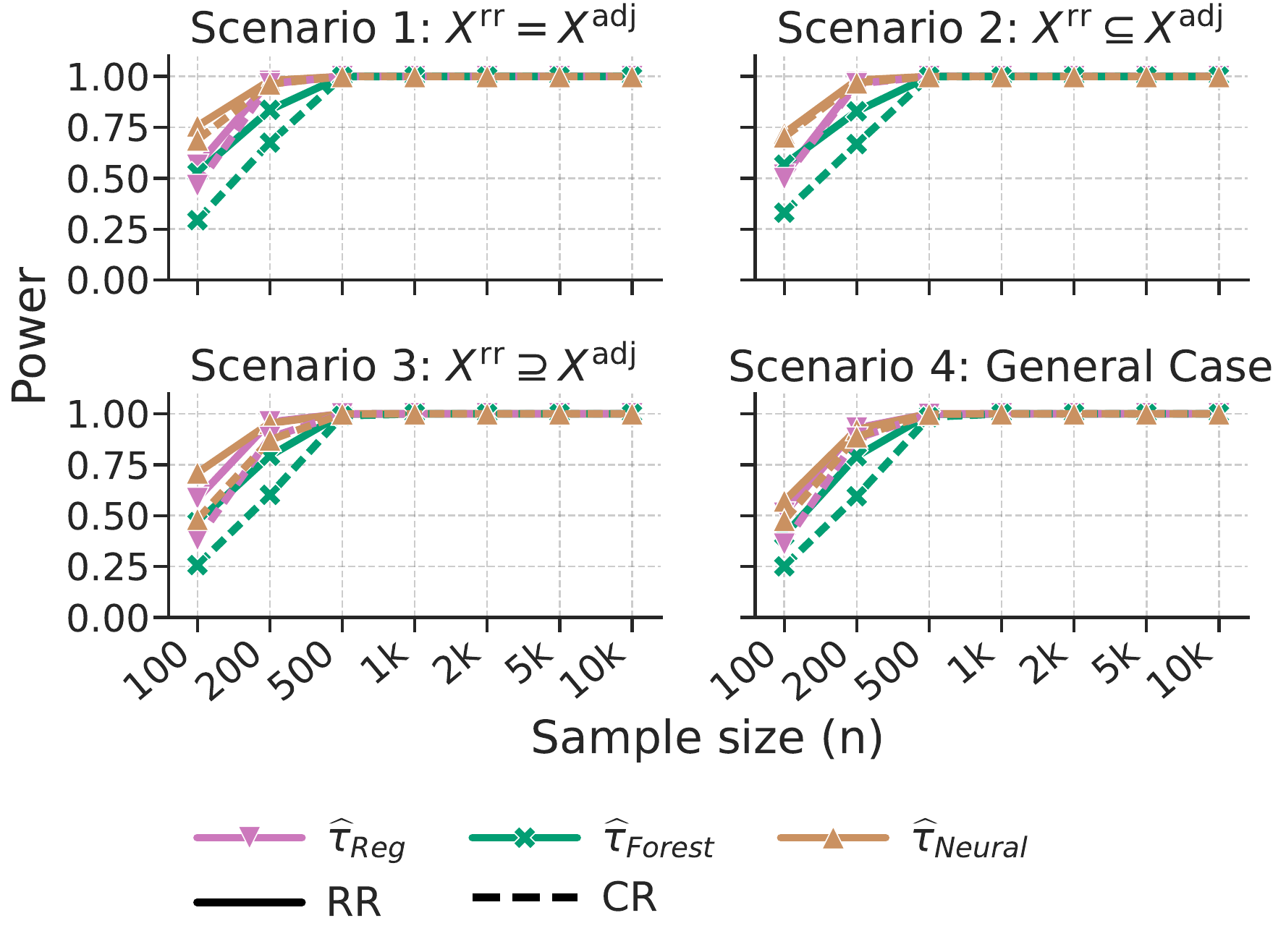}
\caption{Power comparison between the doubly-robust estimator with linear outcome model ($\tauest{Reg}$), the doubly-robust estimator with random forest outcome model ($\tauest{Forest}$), the doubly-robust estimator with neural network outcome model ($\tauest{Neural}$) for different sample sizes and covariates available for design and analysis under complete randomization (CR) and Mahalanobis-based rerandomization (RR) for the linear setting with 10 covariates. All estimators are at least as powerful under rerandomization in comparison to complete randomization for all scenarios of data availability.}
\label{fig:linear:DR_models:power}
\end{figure}

In Section~\ref{sec:simulations:extension} we discussed the coherence results when varying the specification of $\mu_z$, for $z \in \{0,1\}$, within the doubly robust estimator in a linear setting with 10 covariates. We considered the following the doubly robust estimators: $\tauest{Reg}$ with a linear outcome model, $\tauest{Neural}$ with its outcome model defined by a neural network, $\tauest{Forest}$ with a random forest on the outcome model. Figures~\ref{fig:linear:DR_models:precision}, \ref{fig:linear:DR_models:coverage} and \ref{fig:linear:DR_models:power} show the results on precision, coverage, and power, respectively. We find that $\tauest{Forest}$ have significant finite-sample precision gains, while $\tauest{Neural}$ and $\tauest{Reg}$ follow the asymptotic theory. All estimators achieve nominal coverage and they are at least as powerful under rerandomization compared to complete randomization.

\subsection{Covariate selection on the linear setting with 10 covariates}\label{app:subsec:CSL10}

\begin{figure}[ht]
\centering
\includegraphics[width=0.70\textwidth]{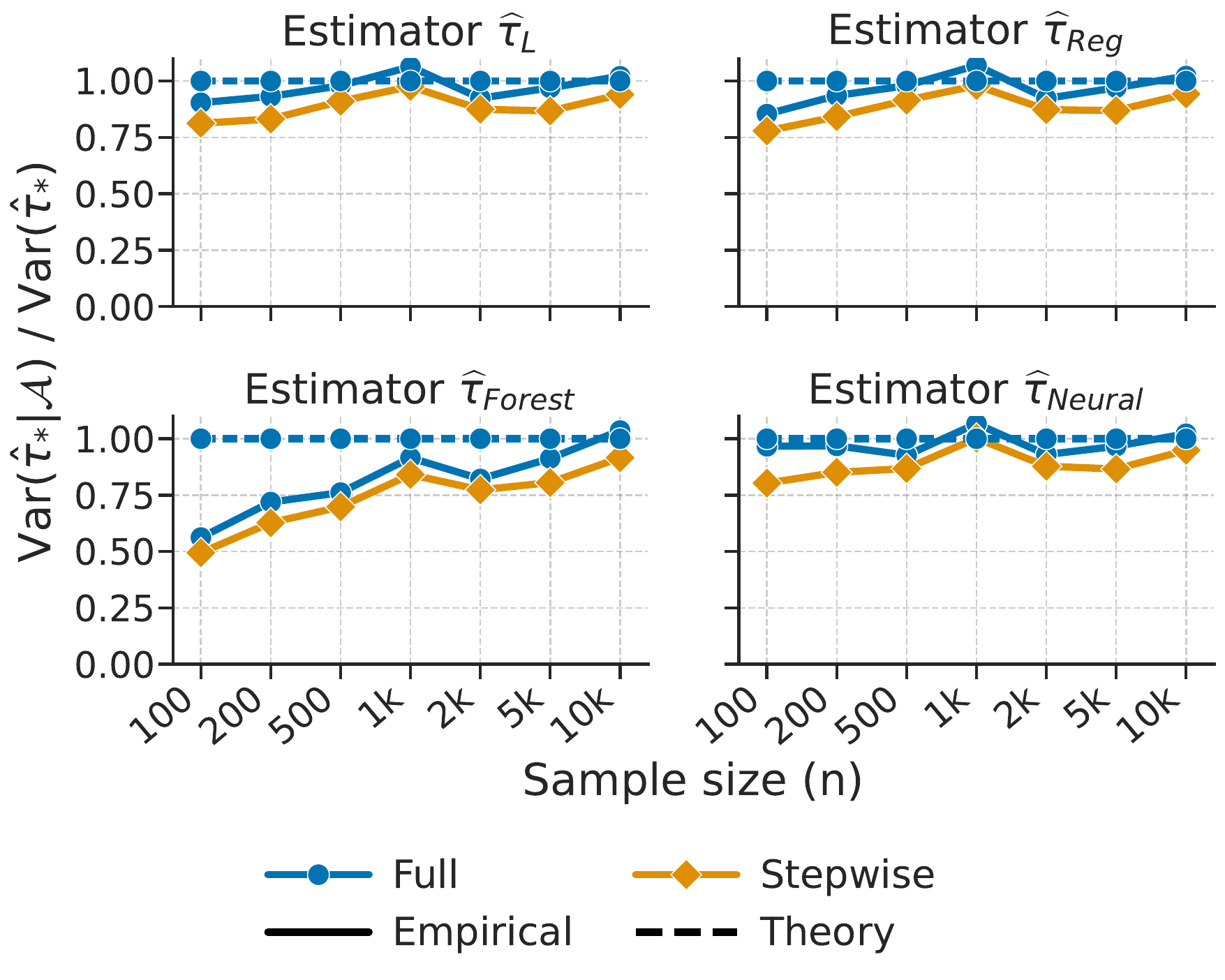}
\caption{Variance ratio comparison for the difference in linear outcome model ($\tauest{L}$), the doubly-robust estimator with linear outcome model ($\tauest{Reg}$), the doubly-robust estimator with random forest outcome model ($\tauest{Forest}$), the doubly-robust estimator with neural network outcome model ($\tauest{Neural}$) for different sample sizes in the linear setting with 10 covariates when different amounts of covariates are considered for adjustment in the analysis stage. \texttt{Stepwise} refers to when the covariate selection is done in the analysis stage prior to estimation, and \texttt{Full} refers to when the analysis is conducted using all covariates. When the covariate selection is done in the analysis stage (\texttt{Stepwise}), there is significant precision benefits when doing rerandomization for small sample sizes on all estimators. Moreover, $\tauest{Forest}$ are benefited by rerandomization on small sample sizes for both scenarios while $\tauest{Neural}$ is benefited by rerandomization on small sample sizes under \texttt{Stepwise}.}
\label{fig:linear:covariate_selection:precision}
\end{figure}

\begin{figure}[ht]
\centering
\includegraphics[width=0.70\textwidth]{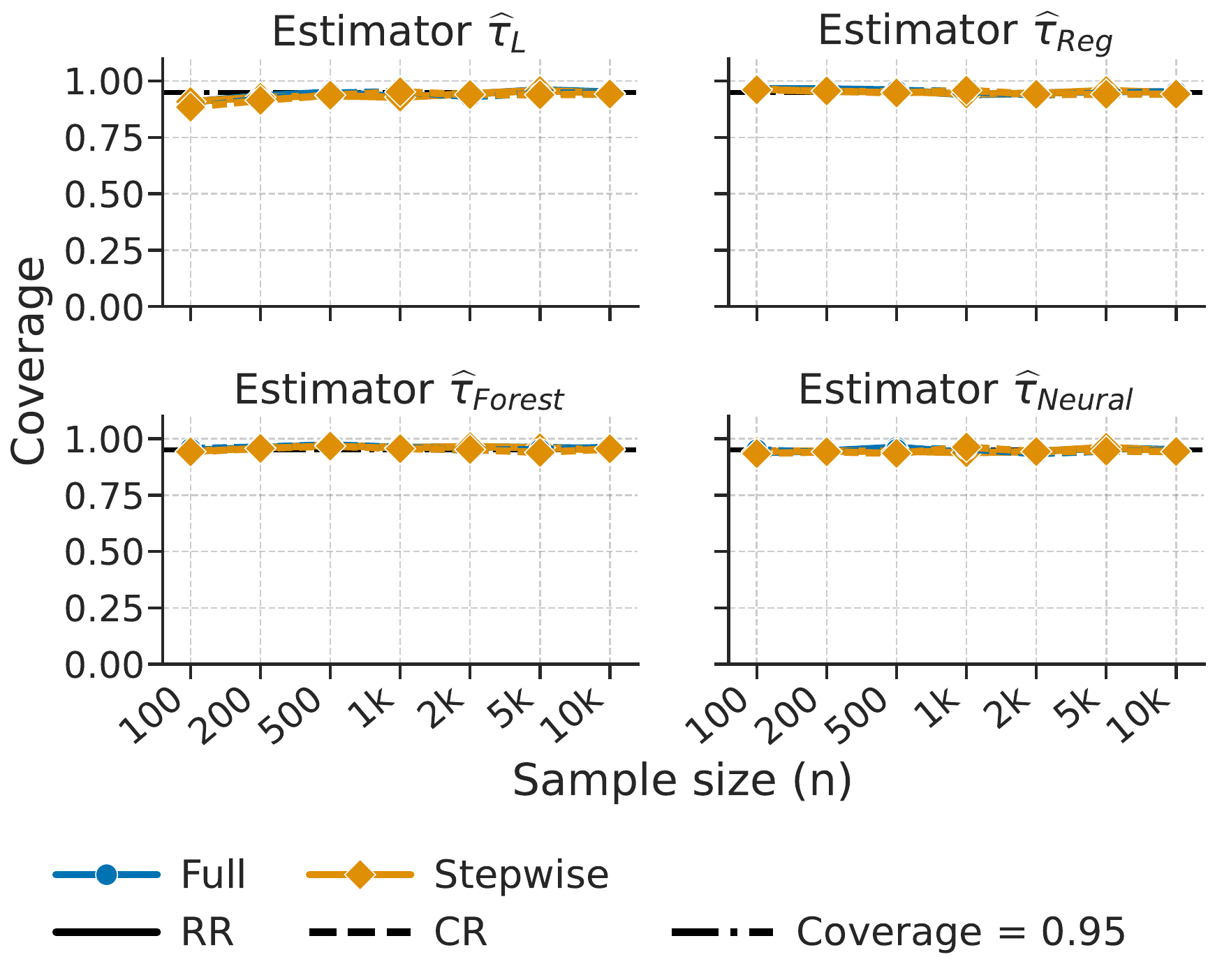}
\caption{Coverage for the difference in linear outcome model ($\tauest{L}$), the doubly-robust estimator with linear outcome model ($\tauest{Reg}$), the doubly-robust estimator with random forest outcome model ($\tauest{Forest}$), the doubly-robust estimator with neural network outcome model ($\tauest{Neural}$) for different sample sizes in the linear setting with 10 covariates when different amounts of covariates are considered for adjustment in the analysis stage. \texttt{Stepwise} refers to when the covariate selection is done in the analysis stage prior to estimation, and \texttt{Full} refers to when the analysis is conducted using all covariates.  The comparison is between under complete randomization (CR) and Mahalanobis-based rerandomization (RR). All estimators achieve nominal coverage under both CR and RR.}
\label{fig:linear:covariate_selection:coverage}
\end{figure}

\begin{figure}[ht]
\centering
\includegraphics[width=0.70\textwidth]{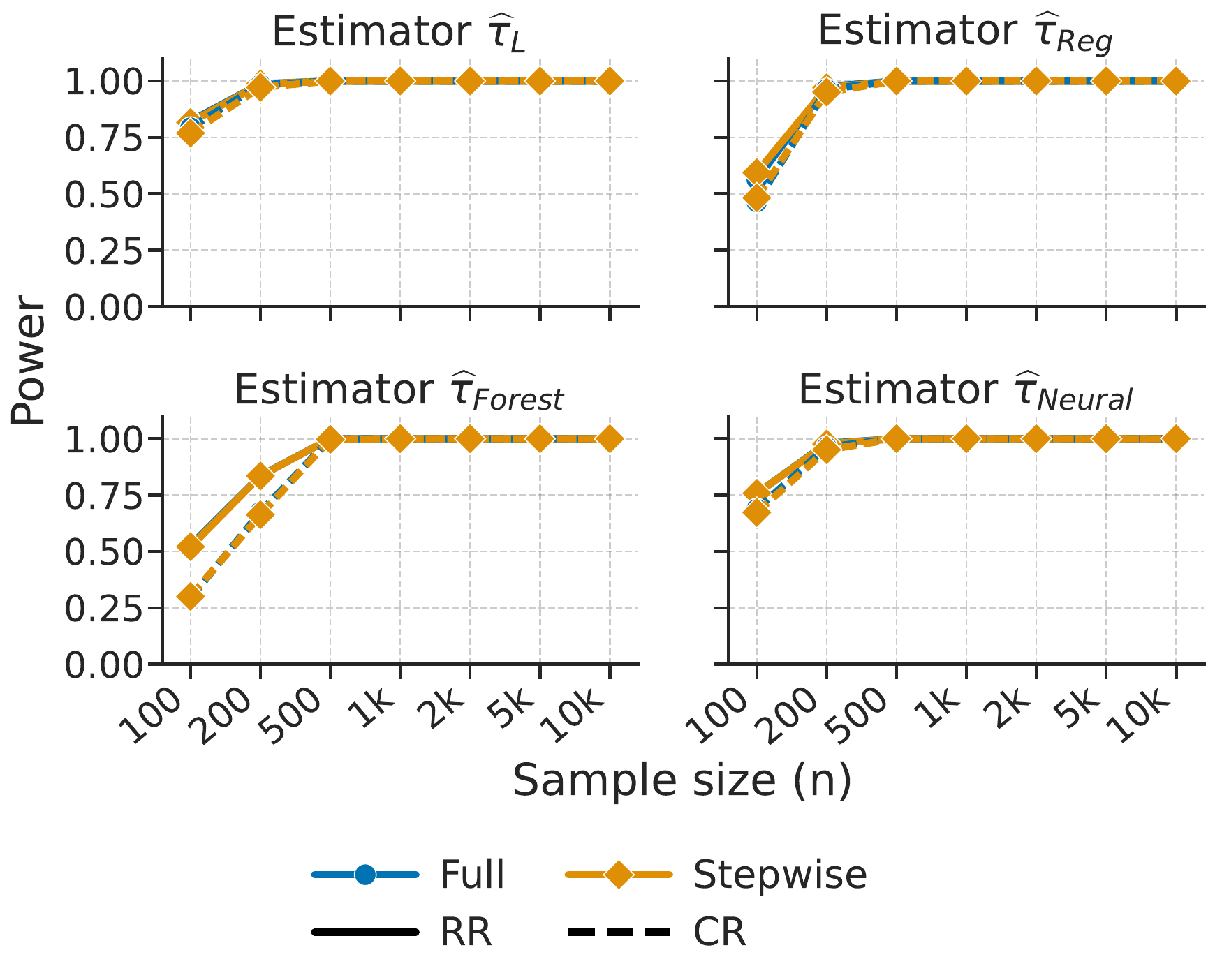}
\caption{Power for the difference in linear outcome model ($\tauest{L}$), the doubly-robust estimator with linear outcome model ($\tauest{Reg}$), the doubly-robust estimator with random forest outcome model ($\tauest{Forest}$), the doubly-robust estimator with neural network outcome model ($\tauest{Neural}$) for different sample sizes in the linear setting with 10 covariates when different amounts of covariates are considered for adjustment in the analysis stage. \texttt{Stepwise} refers to when the covariate selection is done in the analysis stage prior to estimation, and \texttt{Full} refers to when the analysis is conducted using all covariates.  The comparison is between under complete randomization (CR) and Mahalanobis-based rerandomization (RR). All estimators are at least as powerful under RR in comparison to CR for all scenarios.}
\label{fig:linear:covariate_selection:power}
\end{figure}

In Section~\ref{sec:simulations:extension} we discussed the coherence results when covariate selection is performed in the linear setting with 10 covariates. We compared (i) when all 10 covariates are used for adjustment, \texttt{Full}, and (ii) when a \texttt{Stepwise} selection procedure is used to select covariates. For both cases, we considered that all covariates were used for rerandomization. Figures~\ref{fig:linear:covariate_selection:precision}, \ref{fig:linear:covariate_selection:coverage} and \ref{fig:linear:covariate_selection:power} show the results on precision, coverage, and power, respectively. We find that there are significant precision benefits when doing rerandomization for small sample sizes on all estimators. Moreover, all estimators achieve nominal coverage and they are at least as powerful under rerandomization compared to complete randomization.

\subsection{Non-linear setting with 10 covariates}\label{app:subsec:NL10}

\begin{figure}[ht]
\centering
\includegraphics[width=0.70\textwidth]{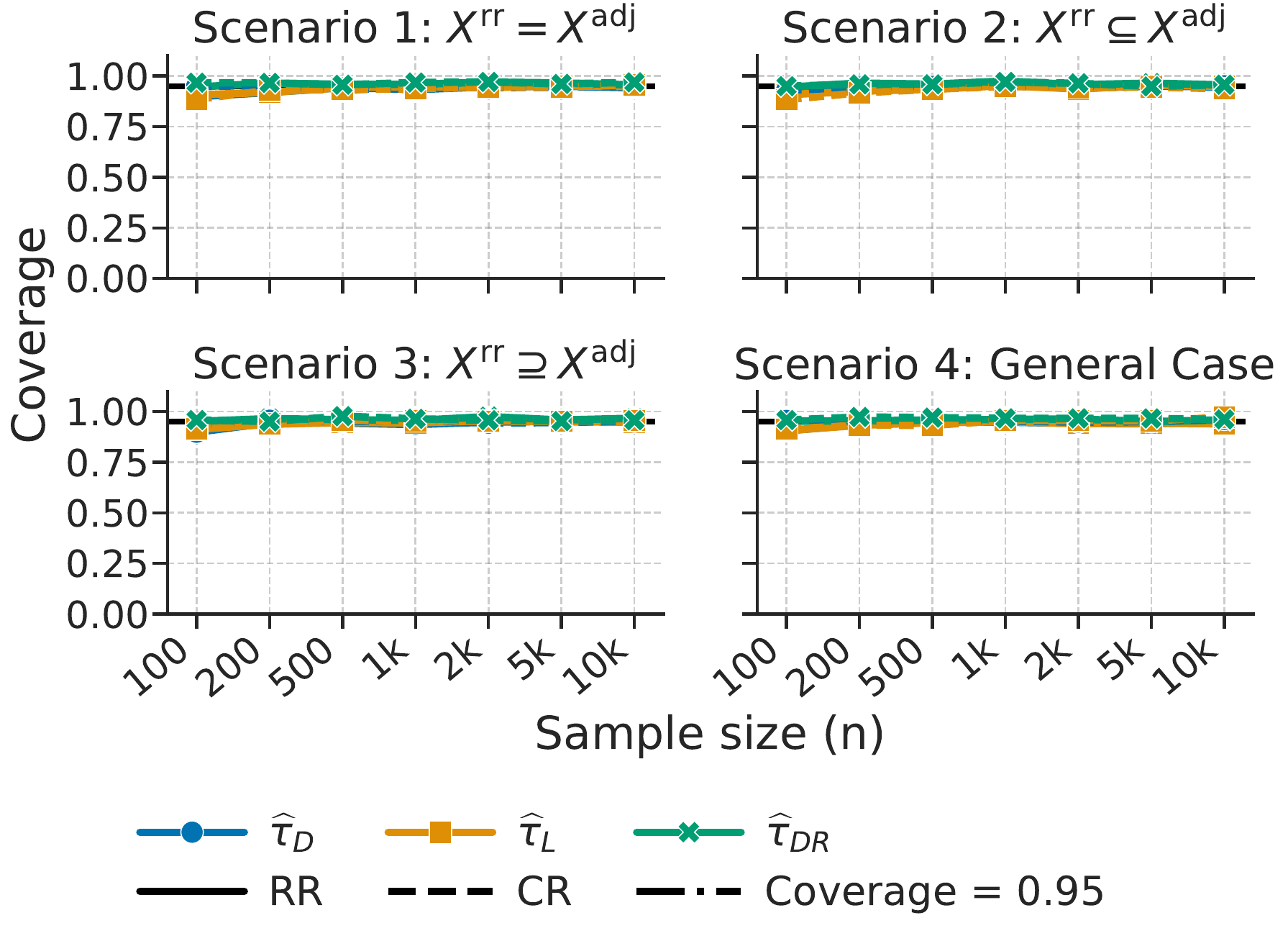}
\caption{Coverage comparison between the difference-in-means ($\tauest{D}$), difference-in-linear outcome models ($\tauest{L}$) and doubly-robust estimator with random forest outcome model ($\tauest{DR}$) for different sample sizes and covariates available for design and analysis under complete randomization (CR) and Mahalanobis-based rerandomization (RR) for the non-linear setting with 10 covariates. All estimators achieve nominal coverage under both CR and RR except for a slight undercoverage of $\tauest{D}$ and $\tauest{L}$ at $n=100$.}
\label{fig:non-linear:coverage}
\end{figure}

\begin{figure}[ht]
\centering
\includegraphics[width=0.70\textwidth]{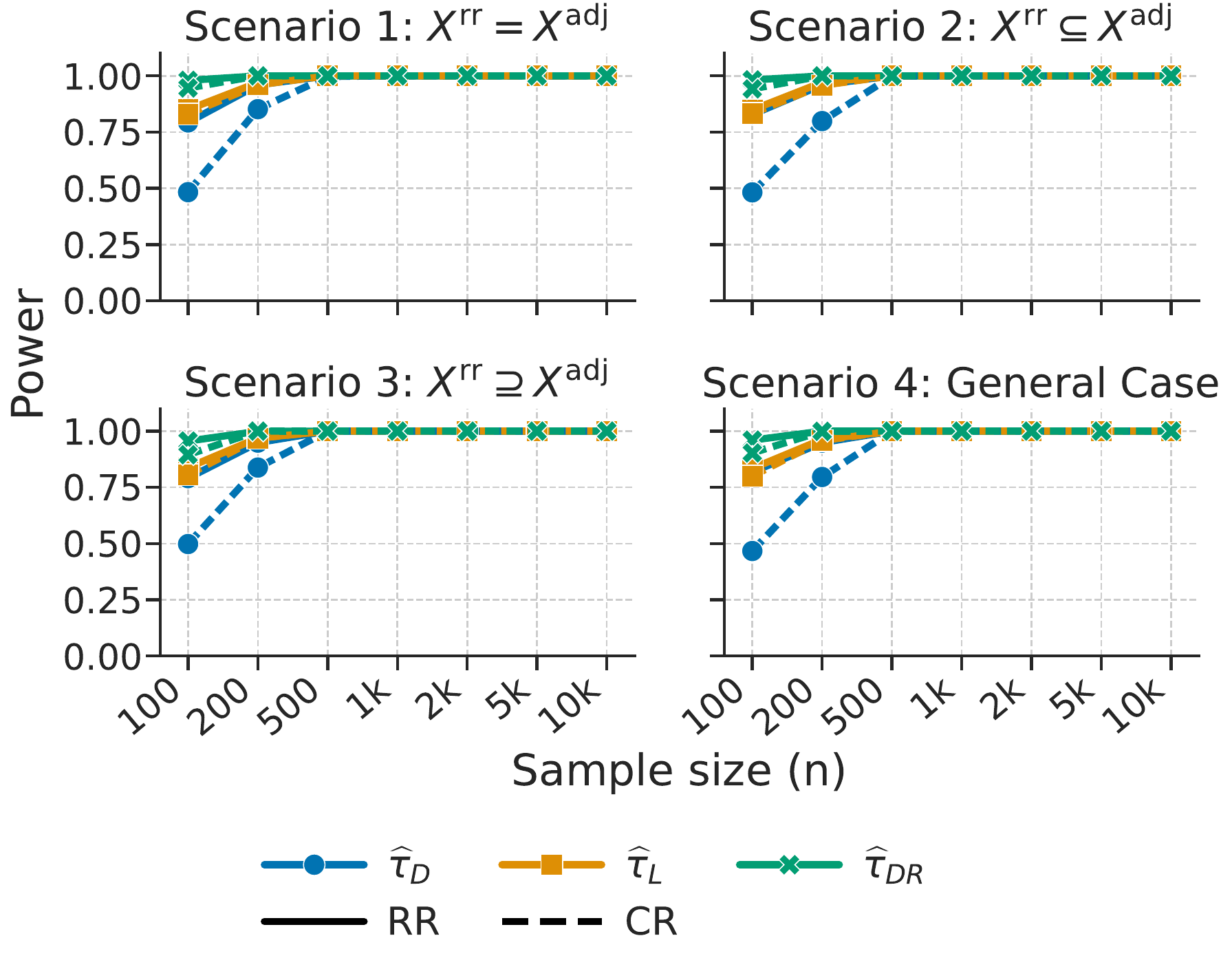}
\caption{Power comparison between the difference-in-means ($\tauest{D}$), difference-in-linear outcome models ($\tauest{L}$) and doubly-robust estimator with random forest outcome model ($\tauest{DR}$) for different sample sizes and covariates available for design and analysis under complete randomization (CR) and Mahalanobis-based rerandomization (RR) for the non-linear setting with 10 covariates. All estimators are at least as powerful under RR in comparison to CR for all scenarios of data availability.}
\label{fig:non-linear:power}
\end{figure}

In Section~\ref{sec:simulations:results} we discussed the precision and coherence results of the non-linear setting with 10 covariates. Now we discuss coverage and power. Figures~\ref{fig:non-linear:coverage} and \ref{fig:non-linear:power} show the results on coverage and power, respectively. All estimators are at least as powerful under rerandomization compared to complete randomization and they achieve nominal coverage, except for a slight undercoverage
of $\tauest{D}$ and $\tauest{L}$ at $n=100$.

\subsection{Doubly robust estimator model specification on the non-linear setting with 10 covariates}\label{app:subsec:MSNL10}

\begin{figure}[ht]
\centering
\includegraphics[width=0.70\textwidth]{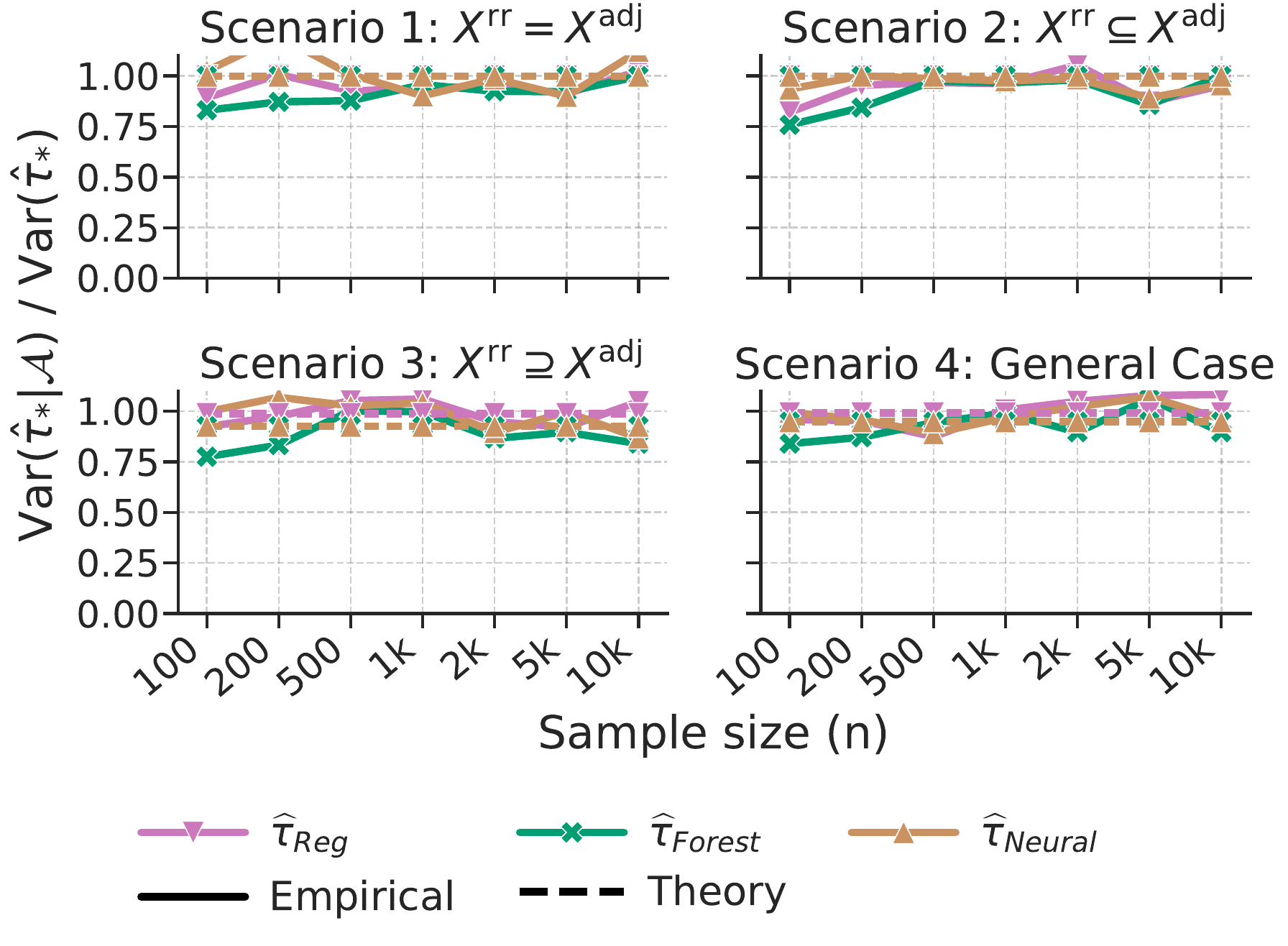}
\caption{Variance ratio comparison between the doubly-robust estimator with linear outcome model ($\tauest{Reg}$), the doubly-robust estimator with random forest outcome model ($\tauest{Forest}$), the doubly-robust estimator with neural network outcome model ($\tauest{Neural}$) for different sample sizes and covariates available for design and analysis for the non-linear setting with 10 covariates. There is significant finite-sample precision gains for $\tauest{Forest}$ under all scenarios. On the other hand, the variance ratios of $\tauest{Reg}$ and $\tauest{Neural}$ follow the asymptotic theory.}
\label{fig:non-linear:DR_models:precision}
\end{figure}

\begin{figure}[ht]
\centering
\includegraphics[width=0.70\textwidth]{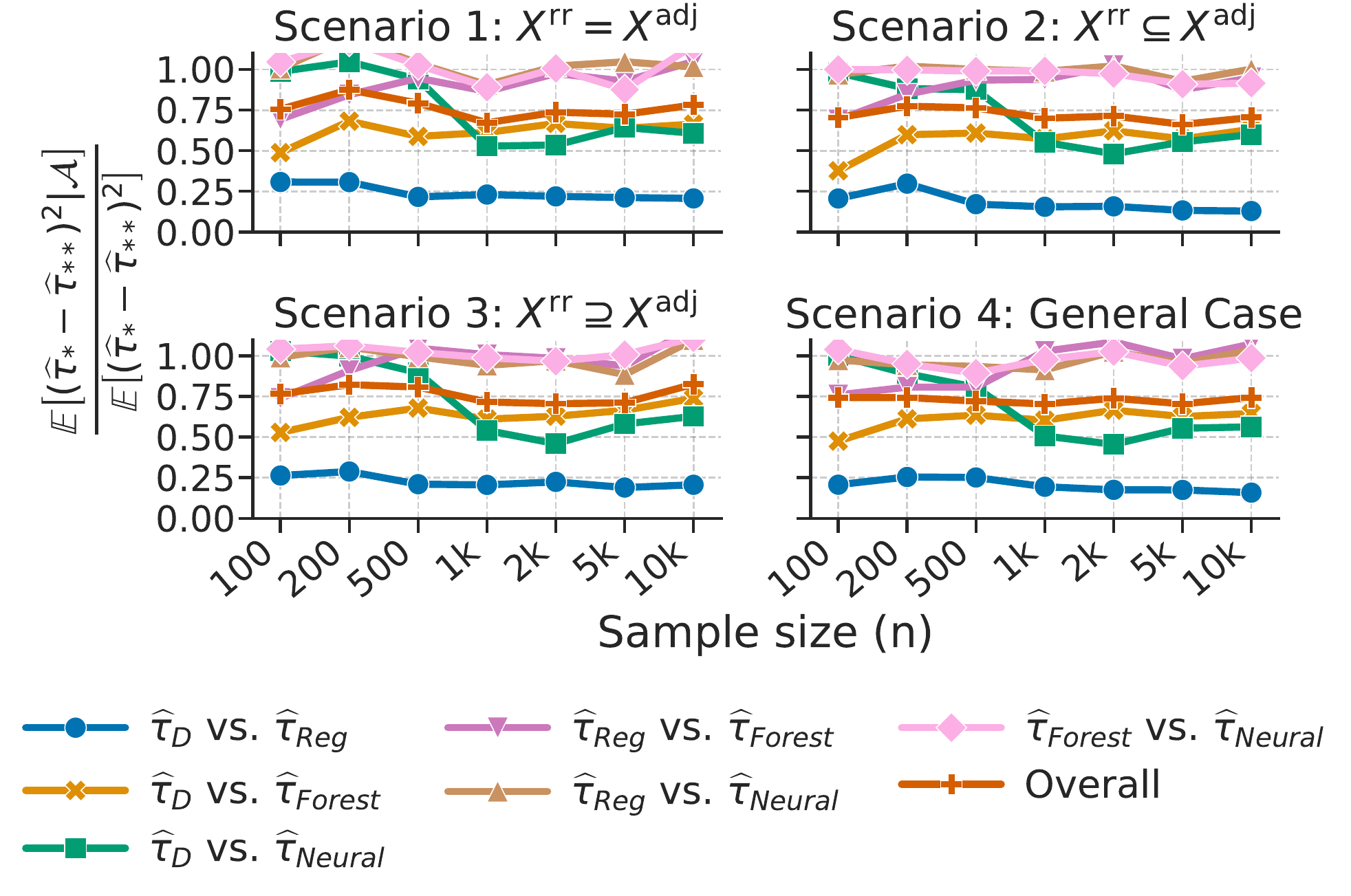}
\caption{Coherence ratio comparison between the difference-in-means estimator ($\tauest{D}$), the doubly-robust estimator with linear outcome model ($\tauest{Reg}$), the doubly-robust estimator with random forest outcome model ($\tauest{Forest}$), the doubly-robust estimator with neural network outcome model ($\tauest{Neural}$) for different sample sizes and covariates available for design and analysis for the non-linear setting with 10 covariates. There is improvement in the overall coherence across all data availability scenarios. The biggest increases in coherence is for the $\tauest{D}$ estimator, since it doesn’t adjust for covariates. Hence, it gets more similar to the other estimators under rerandomization.}
\label{fig:non-linear:DR_models:coherence}
\end{figure}

\begin{figure}[ht]
\centering
\includegraphics[width=0.70\textwidth]{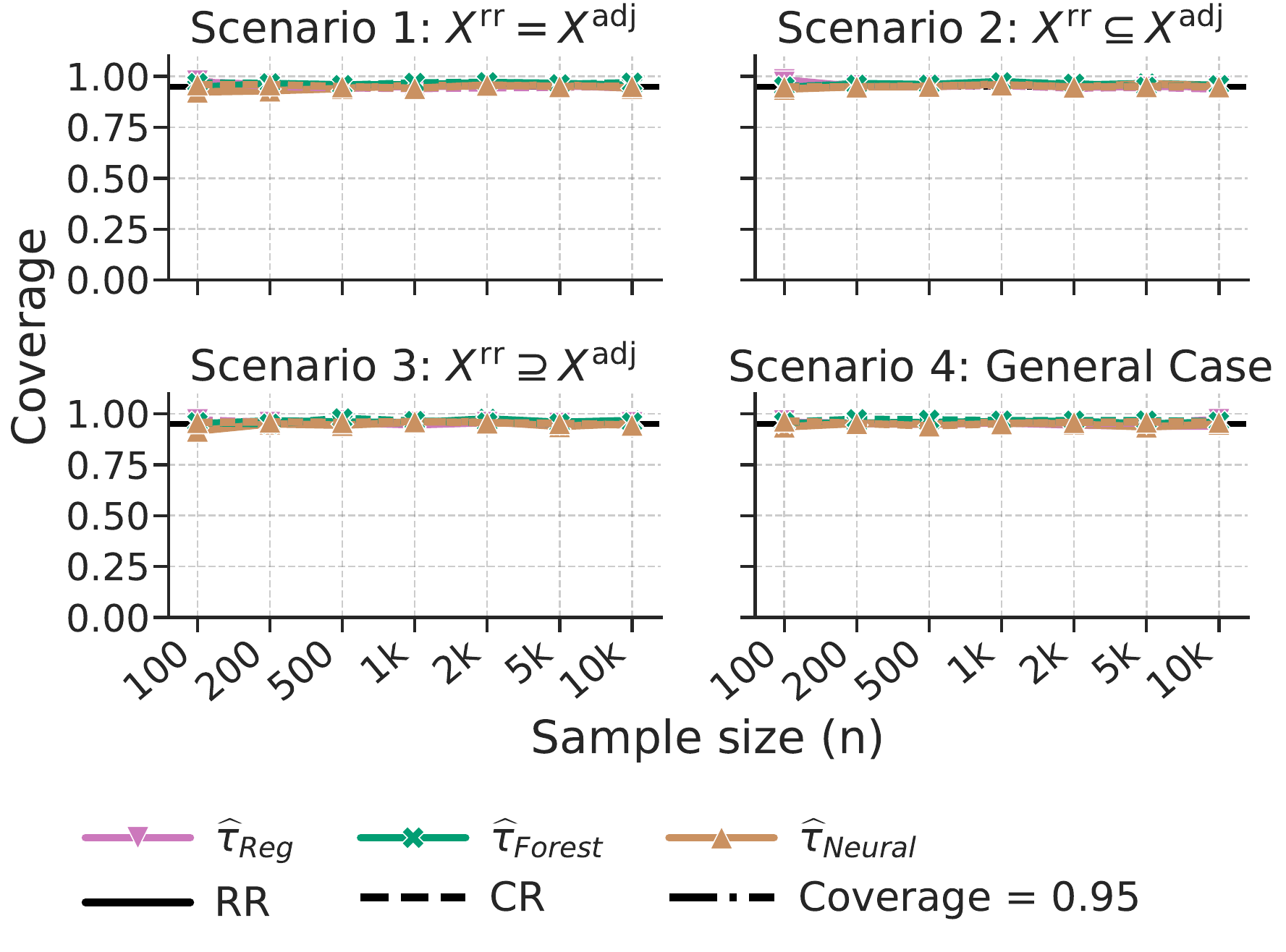}
\caption{Coverage comparison between the doubly-robust estimator with linear outcome model ($\tauest{Reg}$), the doubly-robust estimator with random forest outcome model ($\tauest{Forest}$), the doubly-robust estimator with neural network outcome model ($\tauest{Neural}$) for different sample sizes and covariates available for design and analysis under complete randomization (CR) and Mahalanobis-based rerandomization (RR) for the non-linear setting with 10 covariates. For all data availability scenarios all estimators achieve nominal coverage level.}
\label{fig:non-linear:DR_models:coverage}
\end{figure}

\begin{figure}[ht]
\centering
\includegraphics[width=0.70\textwidth]{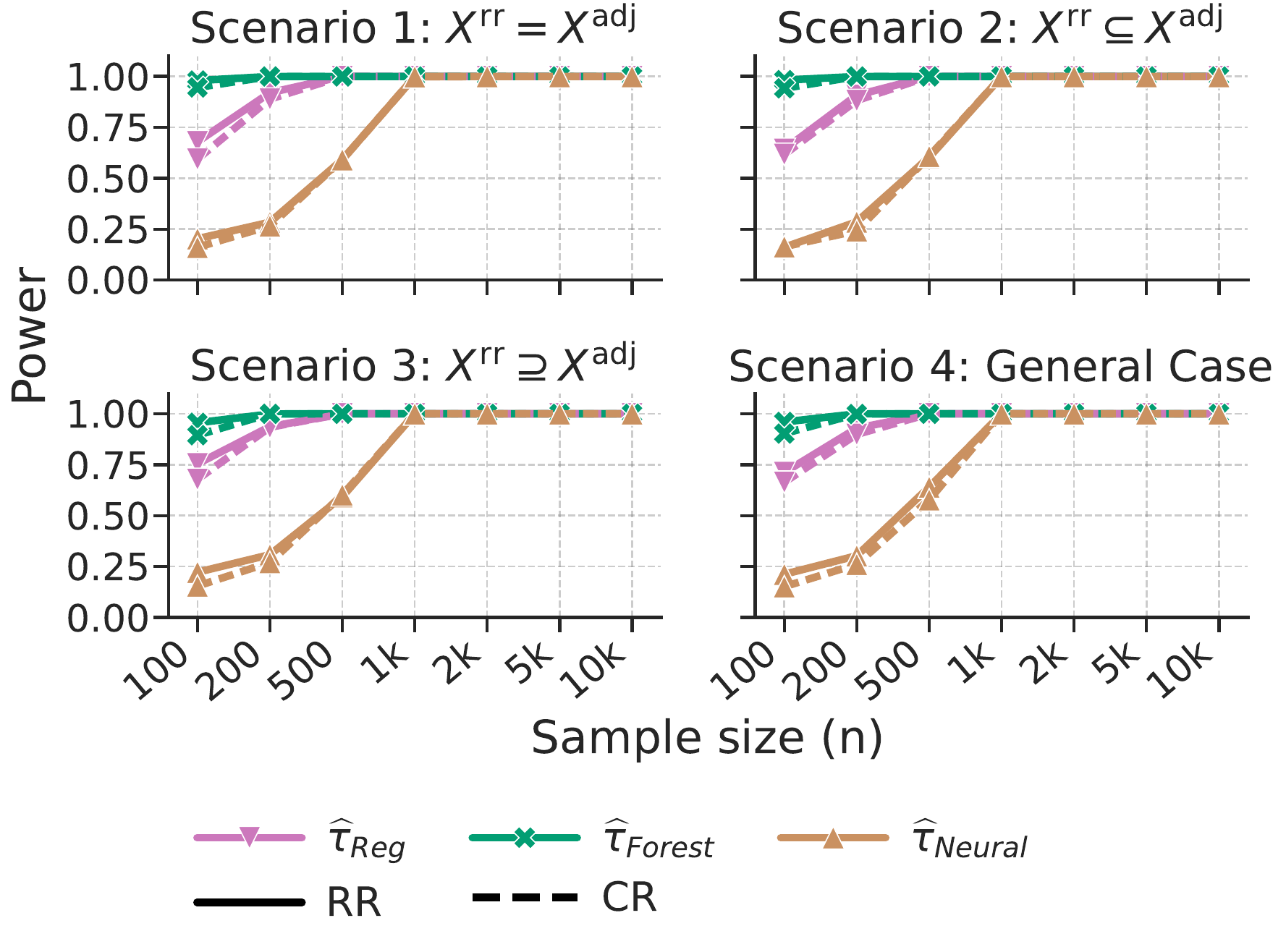}
\caption{Power comparison between the doubly-robust estimator with linear outcome model ($\tauest{Reg}$), the doubly-robust estimator with random forest outcome model ($\tauest{Forest}$), the doubly-robust estimator with neural network outcome model ($\tauest{Neural}$) for different sample sizes and covariates available for design and analysis under complete randomization (CR) and Mahalanobis-based rerandomization (RR) for the non-linear setting with 10 covariates. All estimators are at least as powerful under rerandomization in comparison to complete randomization for all scenarios of data availability. $\tauest{Neural}$ has small power under RR and CR for the smaller sample sizes, indicating wide confidence intervals because of underfitting.}
\label{fig:non-linear:DR_models:power}
\end{figure}

Figures~\ref{fig:non-linear:DR_models:precision}, \ref{fig:non-linear:DR_models:coherence}, \ref{fig:non-linear:DR_models:coverage} and \ref{fig:non-linear:DR_models:power} show the results on precision, coherence coverage, and power when varying the specification of $\mu_z$, for $z \in \{0,1\}$, within the doubly robust estimator in a non-linear setting with 10 covariates. We find that there are modest finite-sample precision gains for $\tauest{Forest}$ in all scenarios, and these gains are less than in the linear setting. On the other hand, the variance ratios of $\tauest{Reg}$ and $\tauest{Neural}$ follow the asymptotic theory. In terms of coherence, there is significant improvement in the overall coherence among estimators, although it is less than under the linear setting. All estimators achieve nominal coverage and are at least as powerful under rerandomization compared to complete randomization. $\tauest{Neural}$, on the other hand, has low power for the smaller sample sizes, indicating wide confidence intervals because of underfitting.

\subsection{Covariate selection on the non-linear setting with 10 covariates}\label{app:subsec:CSNL10}

\begin{figure}[ht]
\centering
\includegraphics[width=0.70\textwidth]{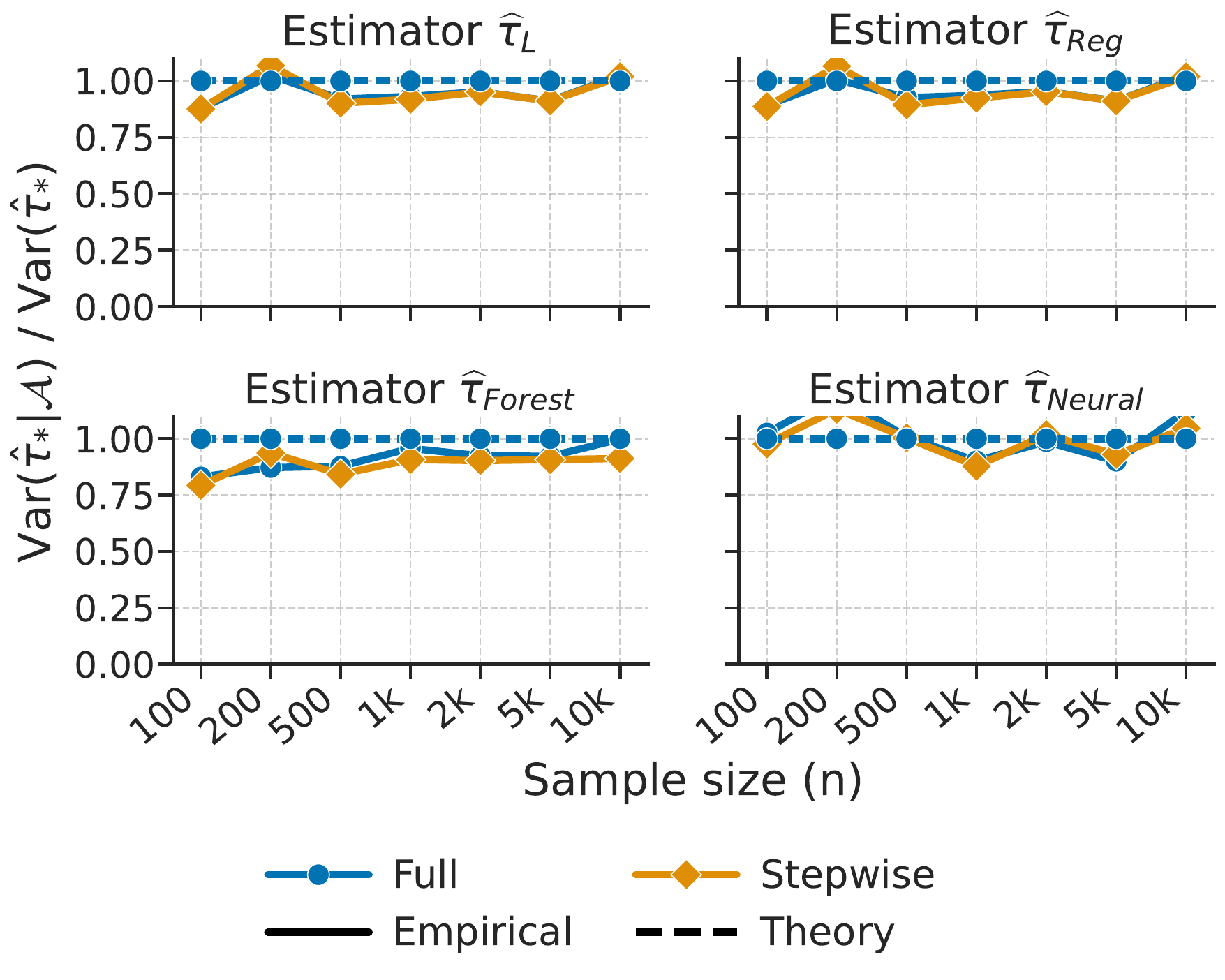}
\caption{Variance ratio comparison for the difference in linear outcome model ($\tauest{L}$), the doubly-robust estimator with linear outcome model ($\tauest{Reg}$), the doubly-robust estimator with random forest outcome model ($\tauest{Forest}$), the doubly-robust estimator with neural network outcome model ($\tauest{Neural}$) for different sample sizes in the non-linear setting with 10 covariates when different amounts of covariates are considered for adjustment in the analysis stage. \texttt{Stepwise} refers to when the covariate selection is done in the analysis stage prior to estimation, and \texttt{Full} refers to when the analysis is conducted using all covariates. $\tauest{Forest}$ is the only estimator benefited from rerandomization under finite samples in this case. On the other hand, the precision of all other estimators is not improved under rerandomization.}
\label{fig:non-linear:covariate_selection:precision}
\end{figure}

\begin{figure}[ht]
\centering
\includegraphics[width=0.70\textwidth]{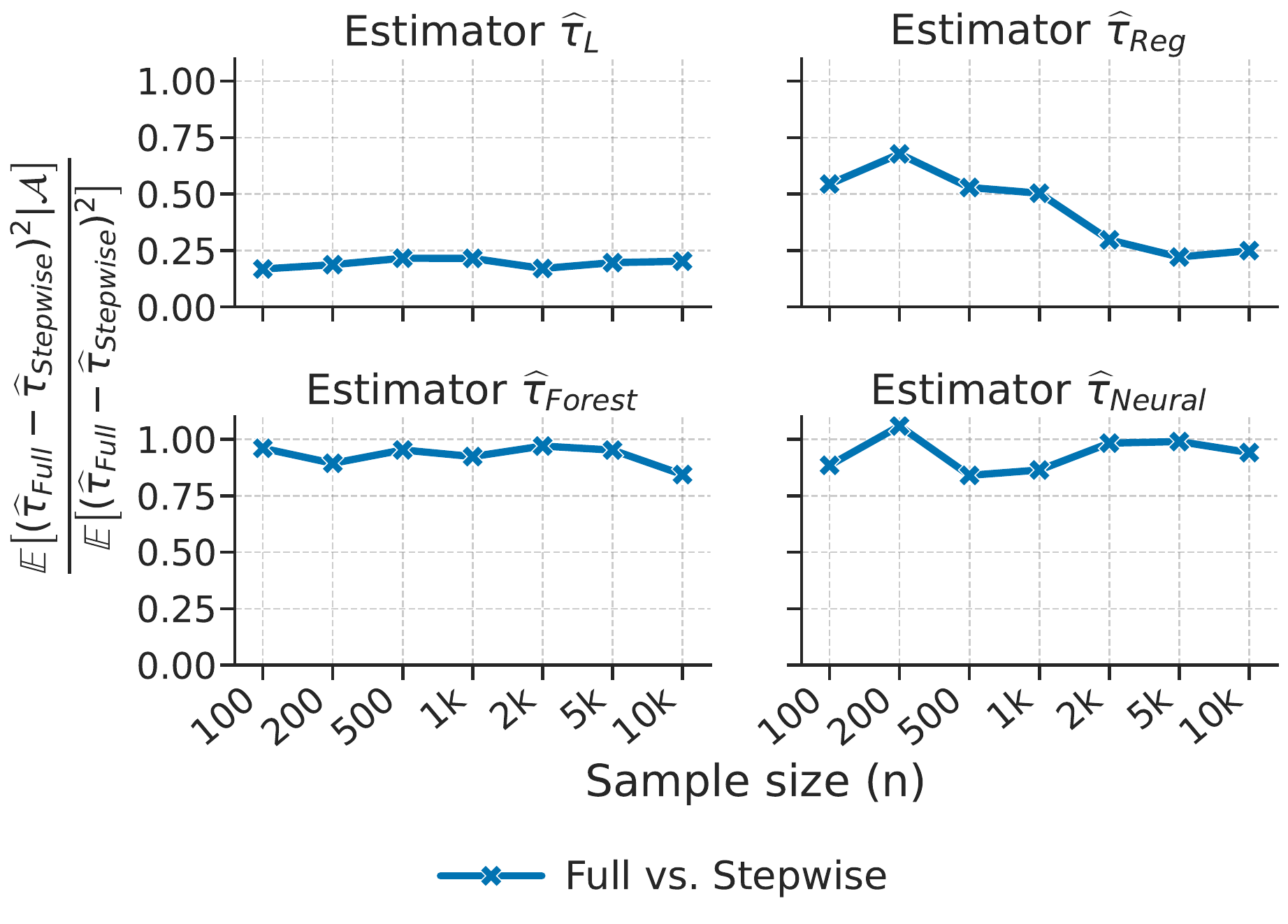}
\caption{Coherence ratio comparison for the difference in linear outcome model ($\tauest{L}$), the doubly-robust estimator with linear outcome model ($\tauest{Reg}$), the doubly-robust estimator with random forest outcome model ($\tauest{Forest}$), the doubly-robust estimator with neural network outcome model ($\tauest{Neural}$) for different sample sizes in the non-linear setting with 10 covariates when different amounts of covariates are considered for adjustment in the analysis stage. \texttt{Stepwise} refers to when the covariate selection is done in the analysis stage prior to estimation, and \texttt{Full} refers to when the analysis is conducted using all covariates. $\tauest{L}$ and $\tauest{Reg}$ are the estimators that benefit the most from rerandomization in this case.}
\label{fig:non-linear:covariate_selection:coherence}
\end{figure}

\begin{figure}[ht]
\centering
\includegraphics[width=0.70\textwidth]{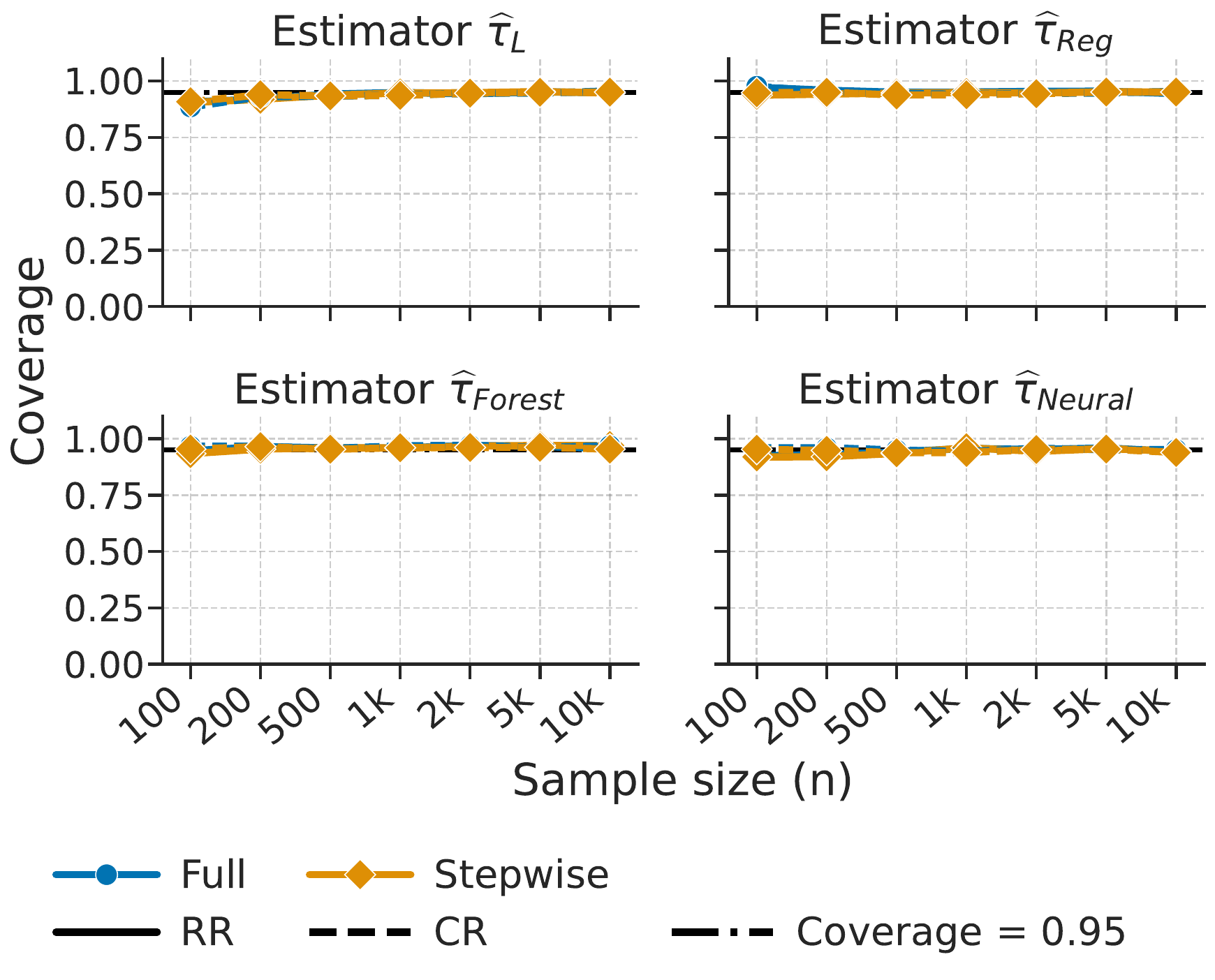}
\caption{Coverage for the difference in linear outcome model ($\tauest{L}$), the doubly-robust estimator with linear outcome model ($\tauest{Reg}$), the doubly-robust estimator with random forest outcome model ($\tauest{Forest}$), the doubly-robust estimator with neural network outcome model ($\tauest{Neural}$) for different sample sizes in the linear setting with 10 covariates when different amounts of covariates are considered for adjustment in the analysis stage. \texttt{Stepwise} refers to when the covariate selection is done in the analysis stage prior to estimation, and \texttt{Full} refers to when the analysis is conducted using all covariates.  The comparison is between under complete randomization (CR) and Mahalanobis-based rerandomization (RR). All estimators achieve nominal coverage under both CR and RR.}
\label{fig:non-linear:covariate_selection:coverage}
\end{figure}

\begin{figure}[ht]
\centering
\includegraphics[width=0.70\textwidth]{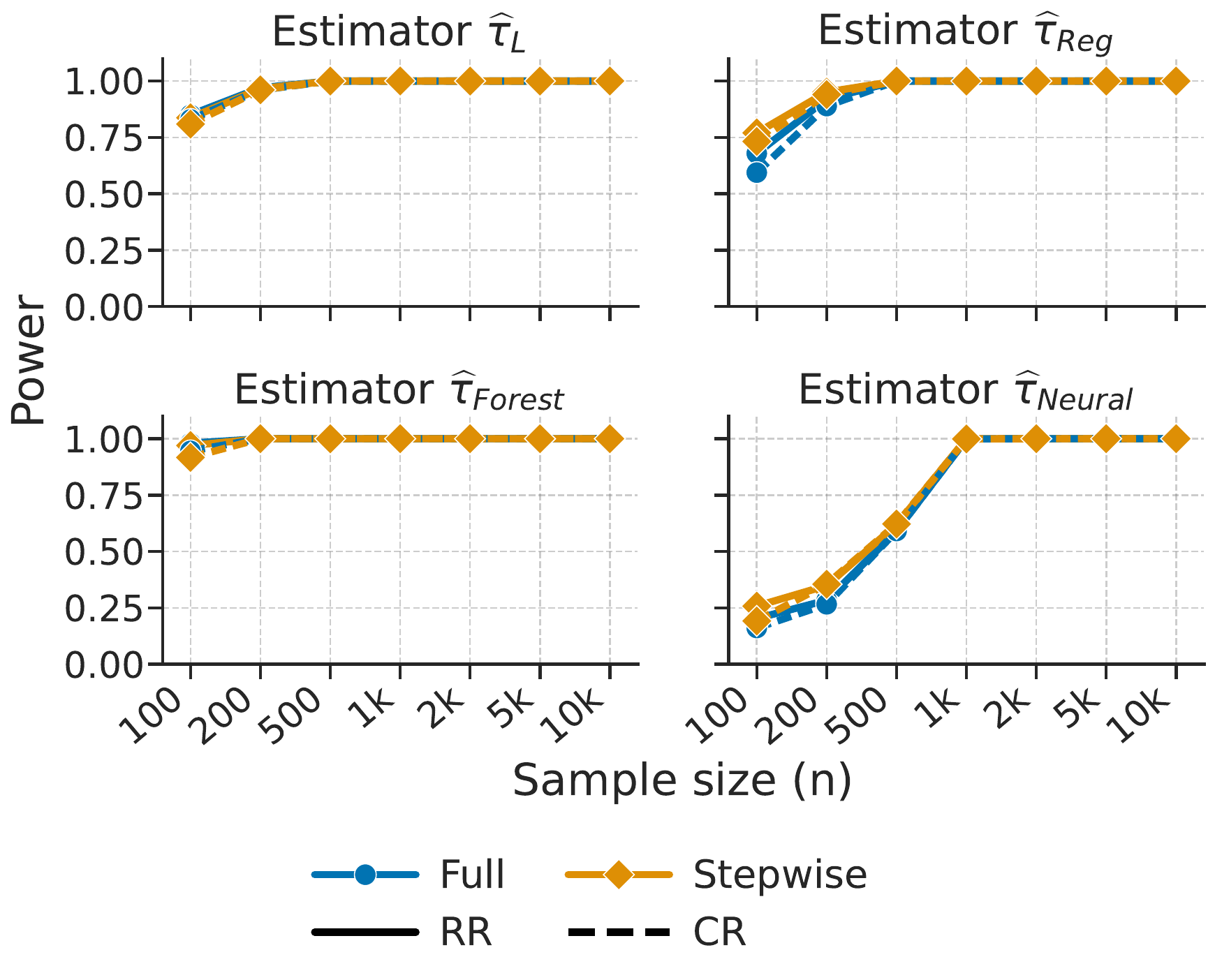}
\caption{Power for the difference in linear outcome model ($\tauest{L}$), the doubly-robust estimator with linear outcome model ($\tauest{Reg}$), the doubly-robust estimator with random forest outcome model ($\tauest{Forest}$), the doubly-robust estimator with neural network outcome model ($\tauest{Neural}$) for different sample sizes in the linear setting with 10 covariates when different amounts of covariates are considered for adjustment in the analysis stage. \texttt{Stepwise} refers to when the covariate selection is done in the analysis stage prior to estimation, and \texttt{Full} refers to when the analysis is conducted using all covariates.  The comparison is between under complete randomization (CR) and Mahalanobis-based rerandomization (RR). Although more significant in $\tauest{Neural}$, all estimators show some low power for the smaller sample sizes, indicating wide confidence intervals. At the same time, all estimators are at least as powerful under RR in comparison to CR.}
\label{fig:non-linear:covariate_selection:power}
\end{figure}

Figures~\ref{fig:non-linear:covariate_selection:precision}, \ref{fig:non-linear:covariate_selection:coherence}, \ref{fig:non-linear:covariate_selection:coverage} and \ref{fig:non-linear:covariate_selection:power} show the results on precision, coherence coverage, and power when covariate selection is performed in a non-linear setting with 10 covariates. We find that there are modest finite-sample precision gains for $\tauest{Forest}$, and these gains are less than in the linear setting. On the other hand, the precision of all other estimators is not improved under rerandomization. Analogously for coherence, $\tauest{L}$ and $\tauest{Reg}$ are the estimator that benefit the most from rerandomization under finite samples in this case. Finally, all estimators achieve nominal coverage and are at least as powerful under rerandomization compared to complete randomization. Again, $\tauest{Neural}$ has low power for the smaller sample sizes, indicating wide confidence intervals because of underfitting.

\subsection{Linear setting with 100 covariates}\label{app:subsec:L100}

\begin{figure}[ht]
\centering
\includegraphics[width=0.70\textwidth]{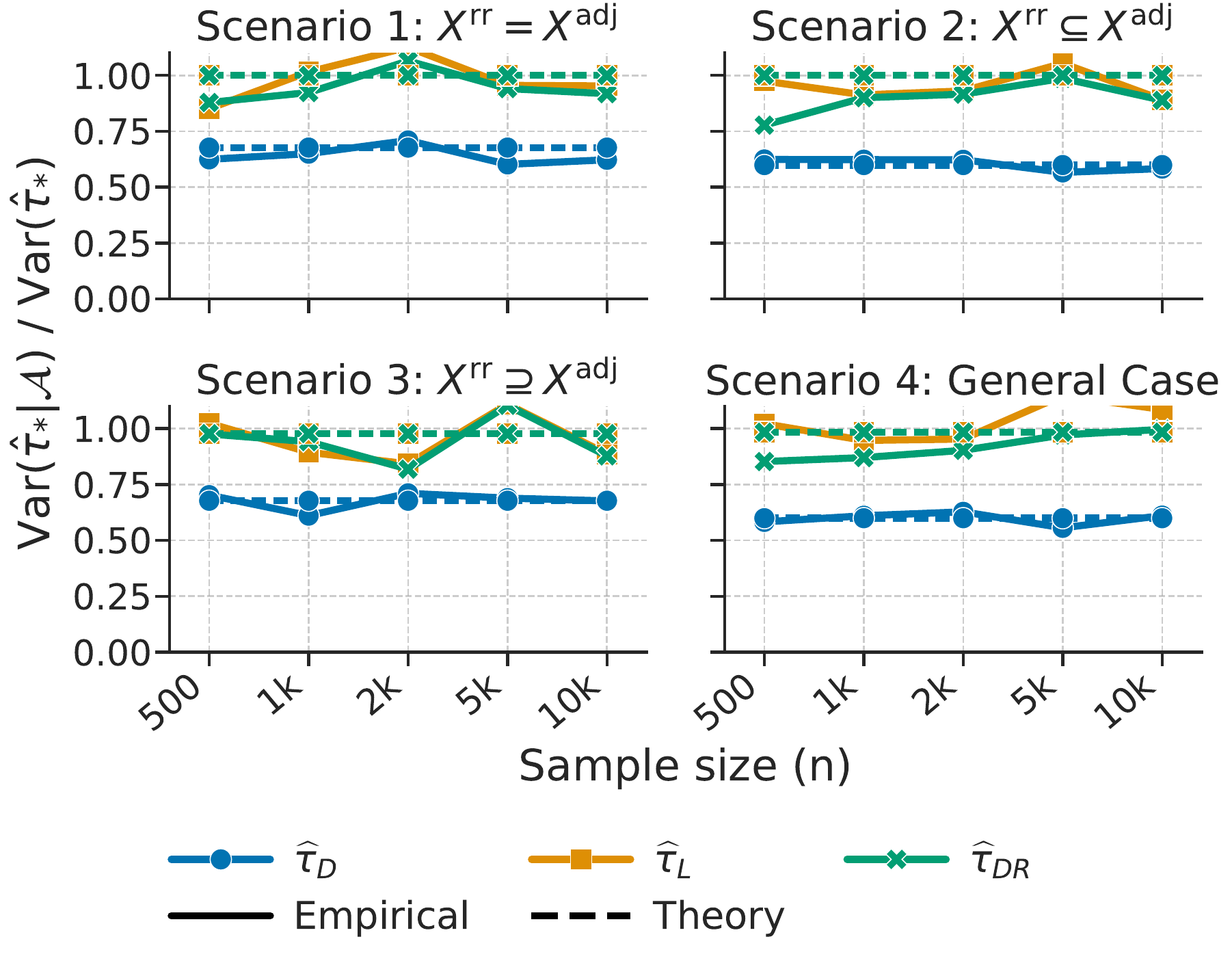}
\caption{Variance ratio comparison between the difference-in-means estimator ($\tauest{D}$), the difference in linear outcome models ($\tauest{L}$), the doubly-robust estimator with random forest outcome model ($\tauest{DR}$) for different sample sizes and covariates available for design and analysis for the linear setting with 100 covariates. Under small sample sizes the adjusted estimators are unstable due to the high-dimensionality of the estimation problem, and the linear model is well-specified. Regardless, $\tauest{DR}$ is benefited from rerandomization under small sample sizes for Scenarios 2 and 4 since its empirical precision gains are bigger than the asymptotic theory. On the other hand, $\tauest{D}$ and $\tauest{L}$ follow asymptotic theory.}
\label{fig:dgp3:precision}
\end{figure}

\begin{figure}[ht]
\centering
\includegraphics[width=0.70\textwidth]{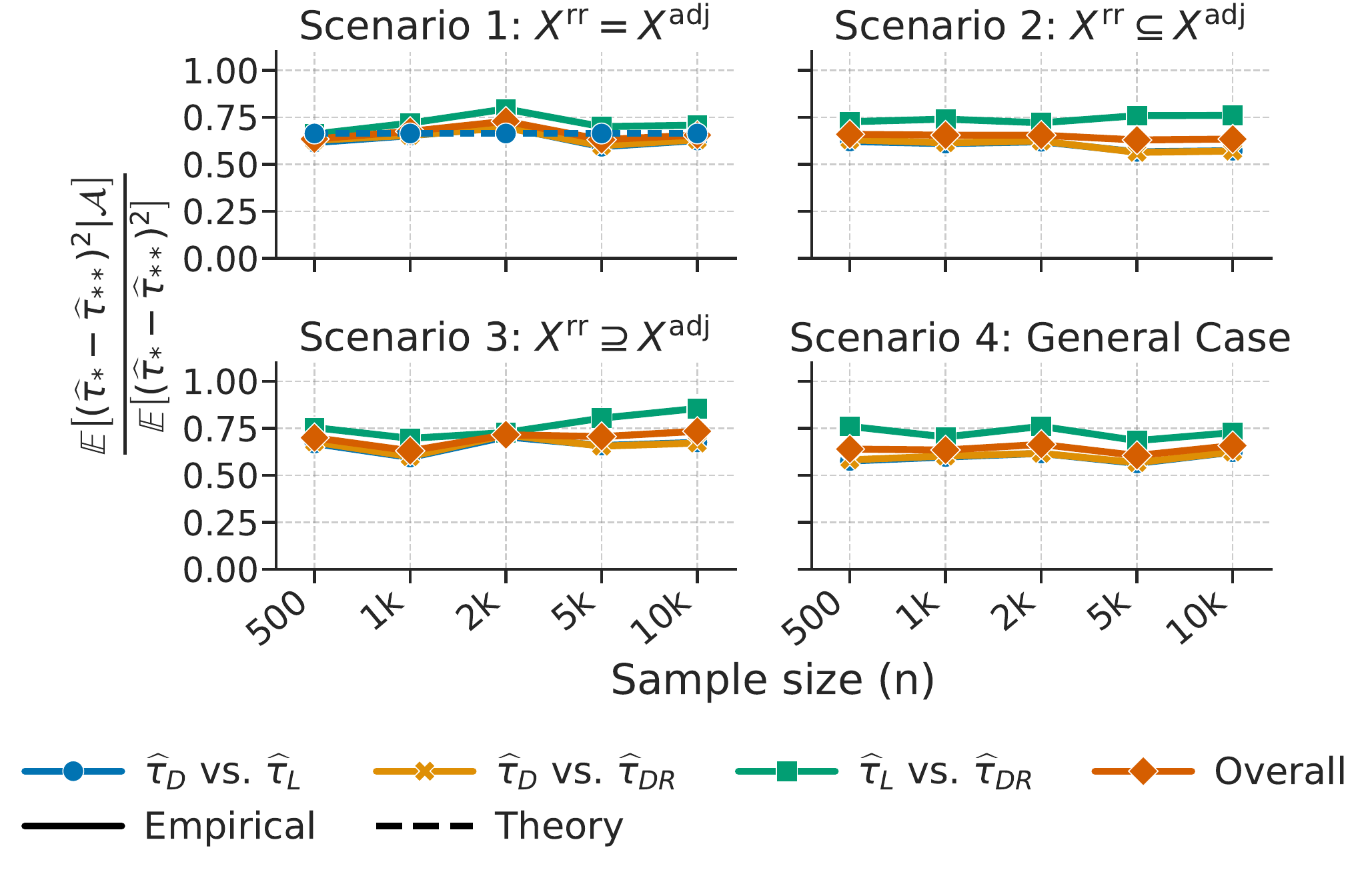}
\caption{Coherence ratio comparison between the difference-in-means estimator ($\tauest{D}$), the difference in linear outcome models ($\tauest{L}$), the doubly-robust estimator with random forest outcome model ($\tauest{DR}$) for different sample sizes and covariates available for design and analysis for the linear setting with 100 covariates. There is significant improvement in the overall coherence across all data availability scenarios.}
\label{fig:dgp3:coherence}
\end{figure}

\begin{figure}[ht]
\centering
\includegraphics[width=0.70\textwidth]{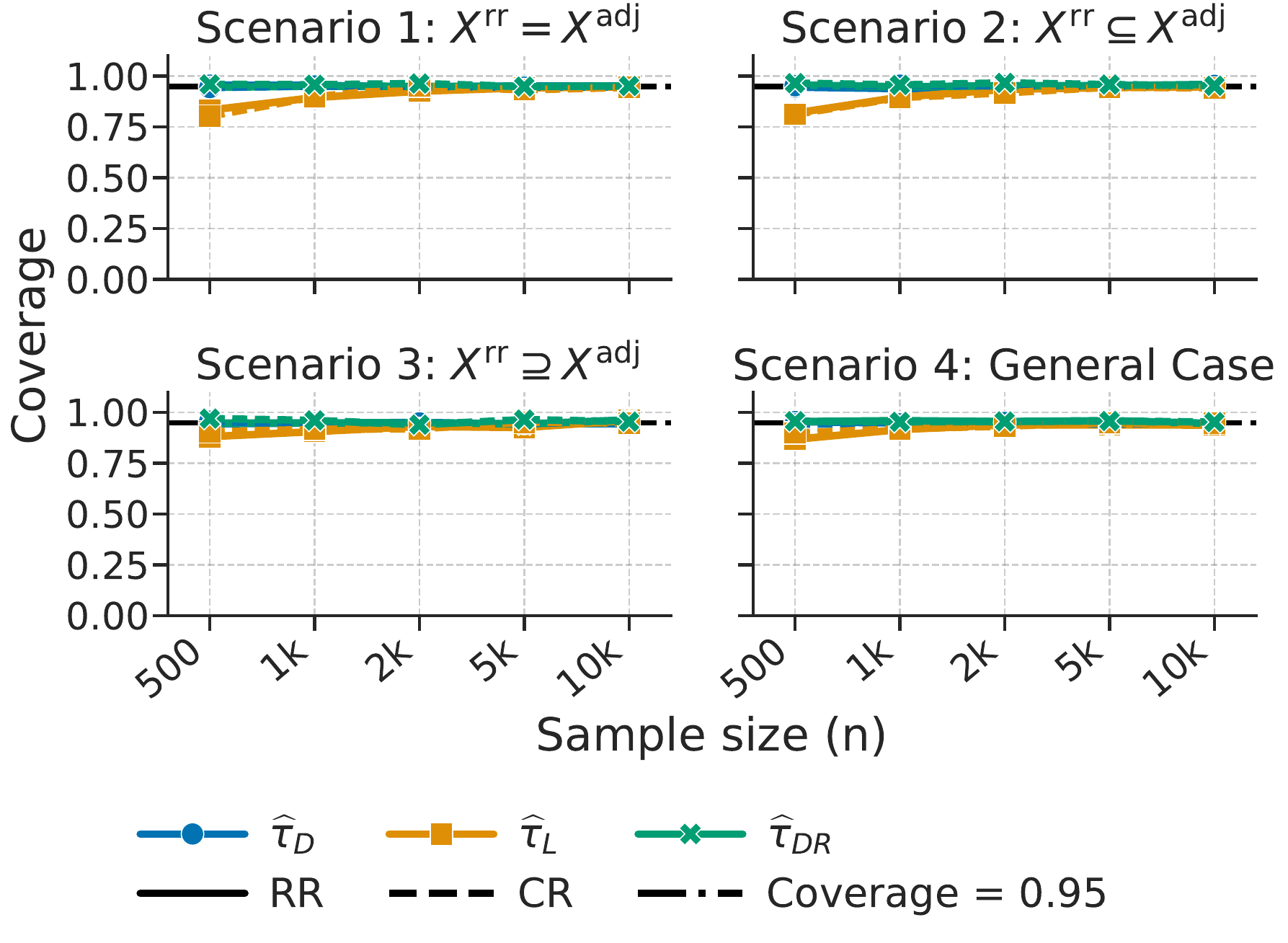}
\caption{Coverage comparison between the difference-in-means ($\tauest{D}$), difference-in-linear outcome models ($\tauest{L}$) and doubly-robust estimator with random forest outcome model ($\tauest{DR}$) for different sample sizes and covariates available for design and analysis under complete randomization (CR) and Mahalanobis-based rerandomization (RR) for the linear setting with 100 covariates.
Under small sample sizes $\tauest{L}$ show a slight undercoverage, while the other estimators achieve nominal coverage.}
\label{fig:dgp3:coverage}
\end{figure}

\begin{figure}[ht]
\centering
\includegraphics[width=0.70\textwidth]{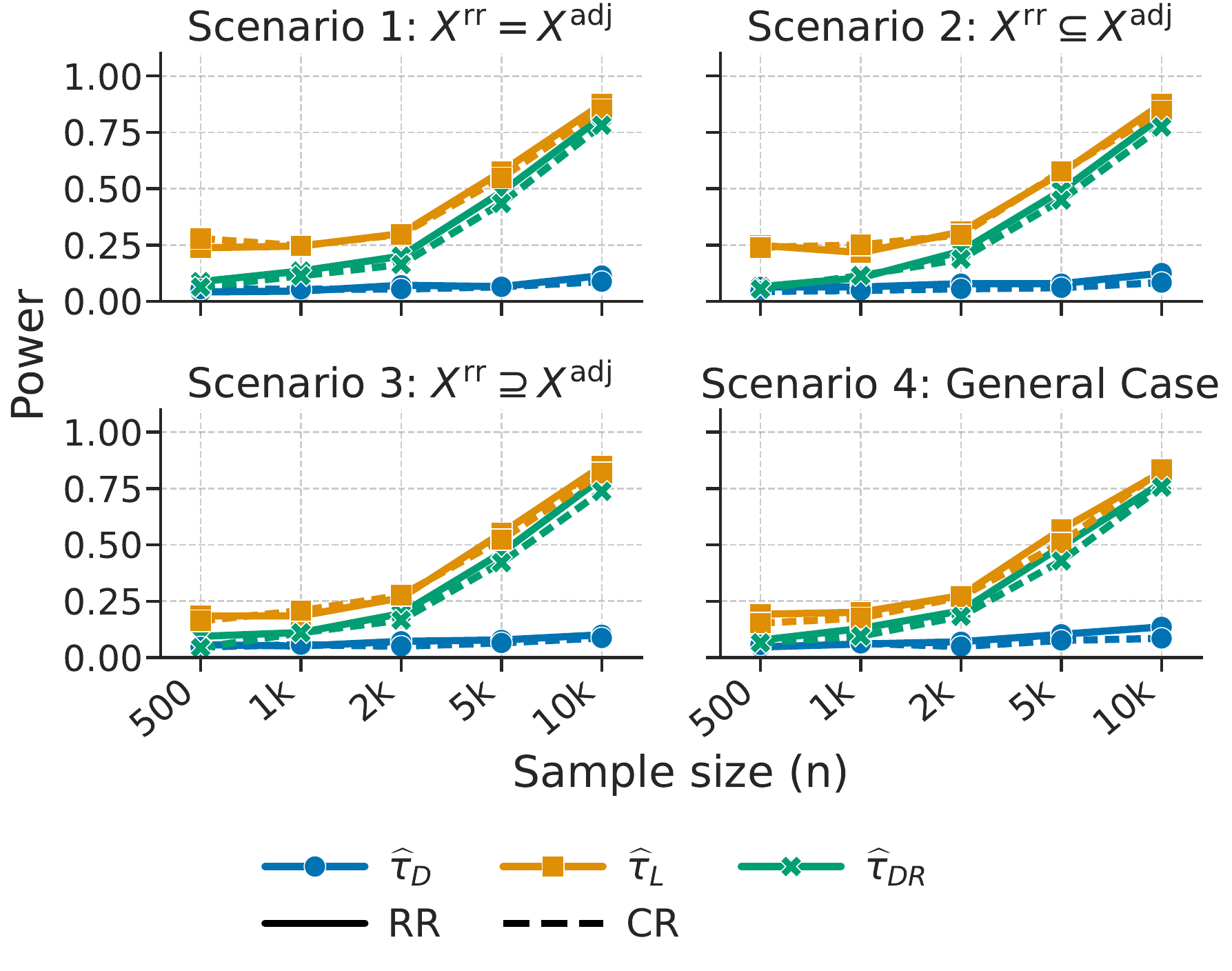}
\caption{Power comparison between the difference-in-means ($\tauest{D}$), difference-in-linear outcome models ($\tauest{L}$) and doubly-robust estimator with random forest outcome model ($\tauest{DR}$) for different sample sizes and covariates available for design and analysis under complete randomization (CR) and Mahalanobis-based rerandomization (RR) for the linear setting with 100 covariates. The power for all the estimators increase as the sample size increases and they are at least as powerful under RR compared to CR.}
\label{fig:dgp3:power}
\end{figure}

Figures~\ref{fig:dgp3:precision}, \ref{fig:dgp3:coherence}, \ref{fig:dgp3:coverage} and \ref{fig:dgp3:power} show the results on precision, coherence, coverage, and power for the linear setting with 100 covariates. Under small sample sizes, the adjusted estimators are unstable due to the high-dimensionality of the estimation problem. However, $\tauest{DR}$ benefits from rerandomization in Scenarios 2 and 4 under small sample sizes since its empirical precision gains are greater than the asymptotic theory. Furthermore, $\tauest{D}$ and $\tauest{L}$ follow asymptotic theory. In terms of coherence, there is significant improvement in overall coherence. Under small samples we find undercoverage for $\tauest{L}$ while the other estimators achieve nominal coverage. The power for all the estimators increase as the sample size increases and they are at least as powerful under rerandomization, compared to complete randomization.

\subsection{Doubly robust estimator model specification in the linear setting with 100 covariates}\label{app:subsec:MSL100}

\begin{figure}[ht]
\centering
\includegraphics[width=0.70\textwidth]{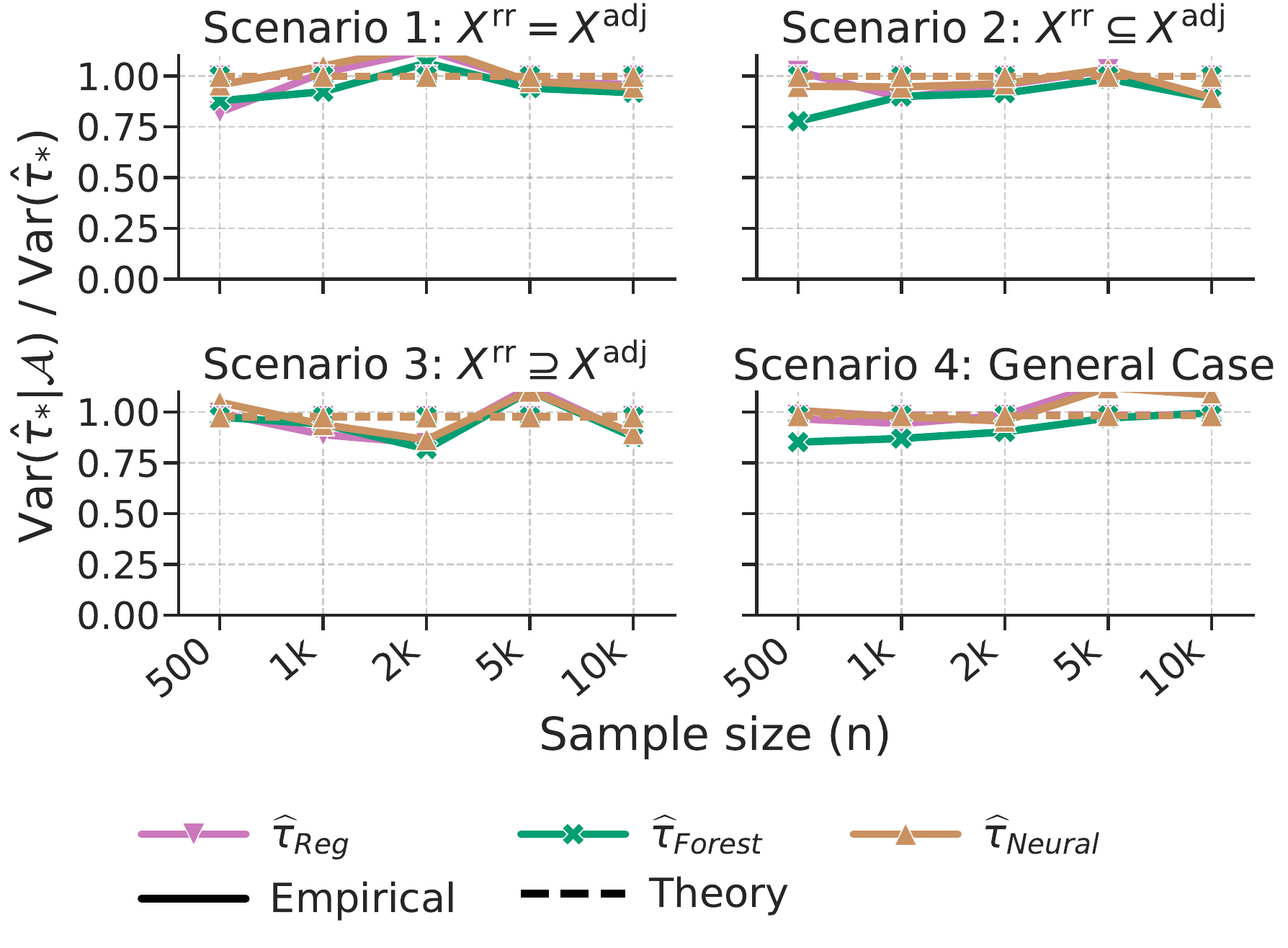}
\caption{Variance ratio comparison between the doubly-robust estimator with linear outcome model ($\tauest{Reg}$), the doubly-robust estimator with random forest outcome model ($\tauest{Forest}$), the doubly-robust estimator with neural network outcome model ($\tauest{Neural}$) for different sample sizes and covariates available for design and analysis for the linear setting with 100 covariates. Although there is some instability on the estimators, we find finite-sample precision benefits for $\tauest{Forest}$.}
\label{fig:dgp3:DR_models:precision}
\end{figure}

\begin{figure}[ht]
\centering
\includegraphics[width=0.70\textwidth]{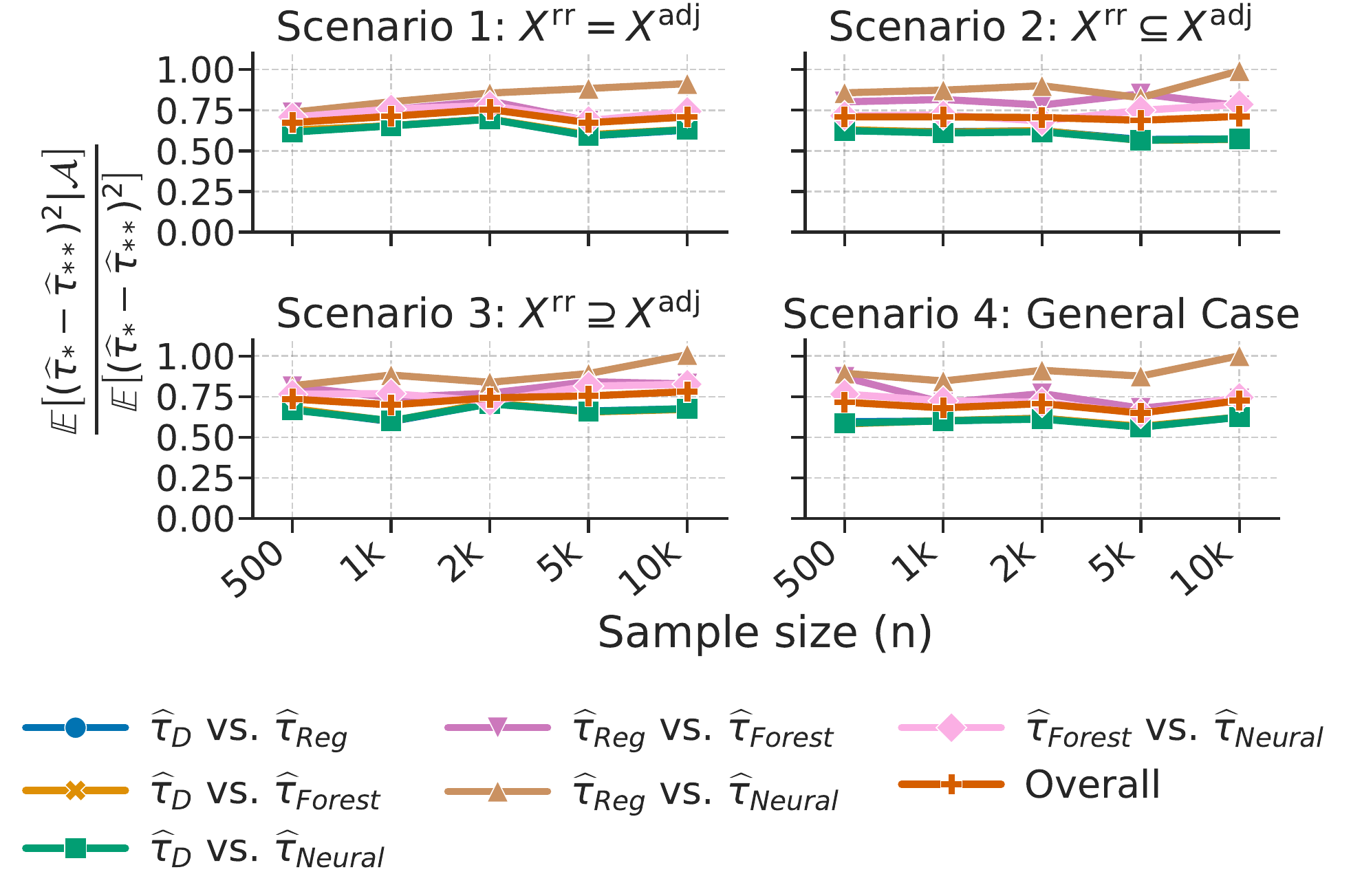}
\caption{Coherence ratio comparison between the difference-in-means estimator ($\tauest{D}$), the doubly-robust estimator with linear outcome model ($\tauest{Reg}$), the doubly-robust estimator with random forest outcome model ($\tauest{Forest}$), the doubly-robust estimator with neural network outcome model ($\tauest{Neural}$) for different sample sizes and covariates available for design and analysis for the linear setting with 100 covariates. There is significant improvement in the overall coherence across all data availability scenarios.}
\label{fig:dgp3:DR_models:coherence}
\end{figure}

\begin{figure}[ht]
\centering
\includegraphics[width=0.70\textwidth]{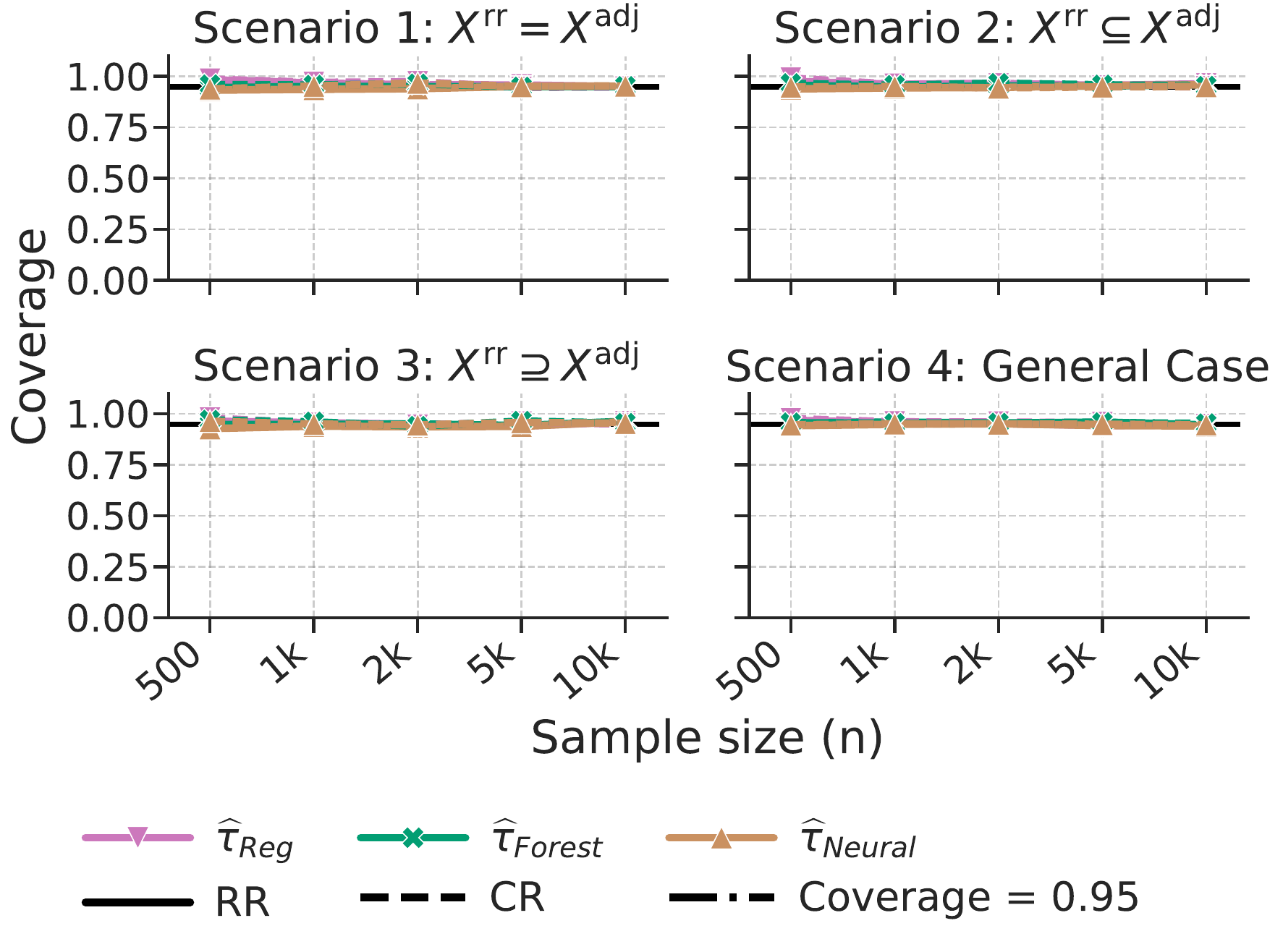}
\caption{Coverage comparison between the doubly-robust estimator with linear outcome model ($\tauest{Reg}$), the doubly-robust estimator with random forest outcome model ($\tauest{Forest}$), the doubly-robust estimator with neural network outcome model ($\tauest{Neural}$) for different sample sizes and covariates available for design and analysis under complete randomization (CR) and Mahalanobis-based rerandomization (RR) for the linear setting with 100 covariates. All estimators achieve nominal coverage}
\label{fig:dgp3:DR_models:coverage}
\end{figure}

\begin{figure}[ht]
\centering
\includegraphics[width=0.70\textwidth]{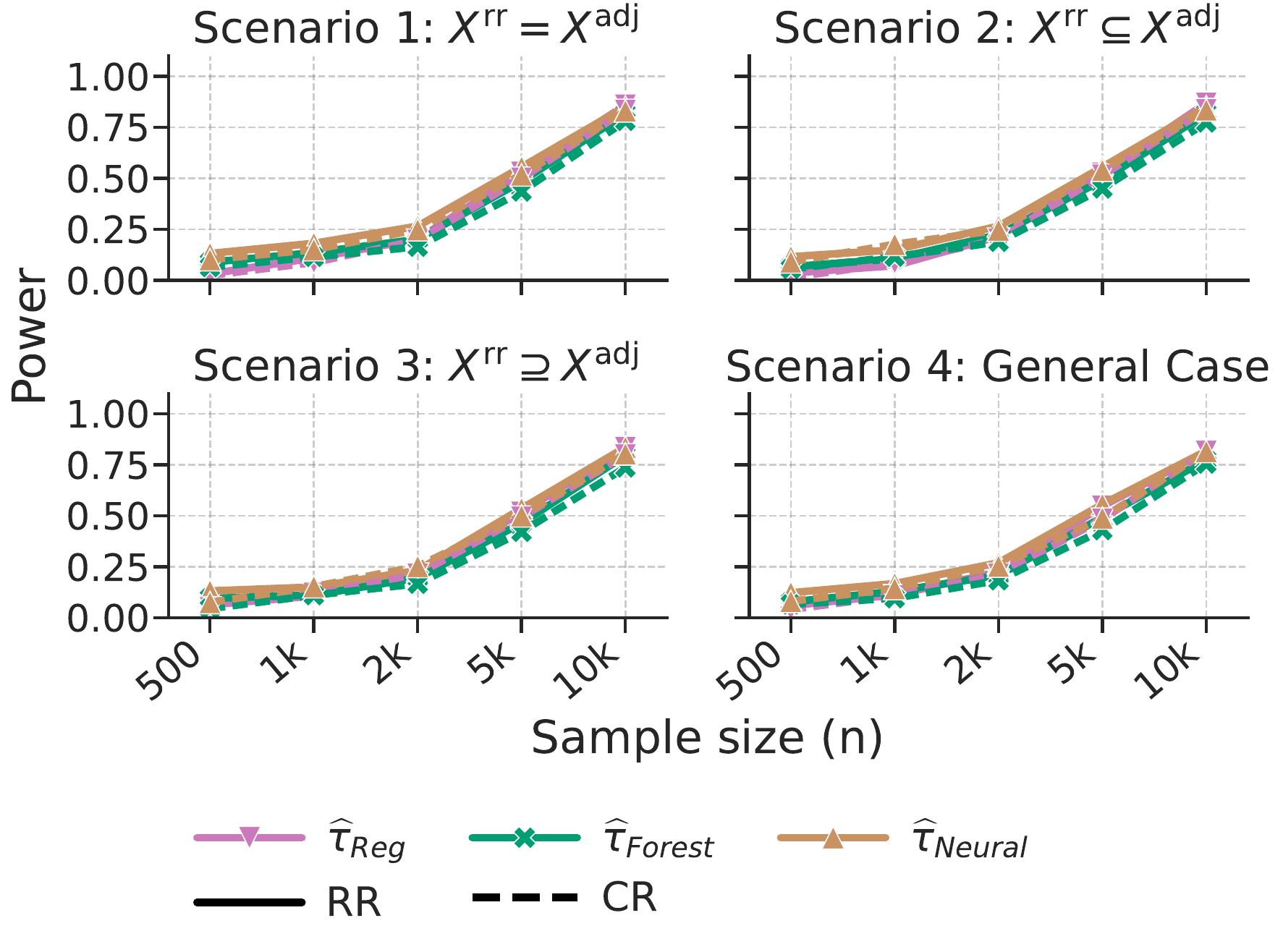}
\caption{Power comparison between the doubly-robust estimator with linear outcome model ($\tauest{Reg}$), the doubly-robust estimator with random forest outcome model ($\tauest{Forest}$), the doubly-robust estimator with neural network outcome model ($\tauest{Neural}$) for different sample sizes and covariates available for design and analysis under complete randomization (CR) and Mahalanobis-based rerandomization (RR) for the linear setting with 100 covariates. The power for all the estimators increase as the sample size increases and they are at least as powerful under RR compared to CR.}
\label{fig:dgp3:DR_models:power}
\end{figure}

Figures~\ref{fig:dgp3:DR_models:precision}, \ref{fig:dgp3:DR_models:coherence}, \ref{fig:dgp3:DR_models:coverage} and \ref{fig:dgp3:DR_models:power} show the results on precision, coherence, coverage and power for the linear setting with 100 covariates when varying the specification of $\mu_z$, for $z \in \{0,1\}$, within the doubly robust estimator. Although there is some instability on the estimators under small samples, there are finite-sample precision benefits for $\tauest{Forest}$. Overall coherence is also significantly improved under rerandomization. All estimators achieve nominal coverage. The power for all the estimators increase as the sample size increases and they are at least as powerful under rerandomization, compared to complete randomization.

\subsection{Covariate selection in the linear setting with 100 covariates}\label{app:subsec:CSL100}

\begin{figure}[ht]
\centering
\includegraphics[width=0.70\textwidth]{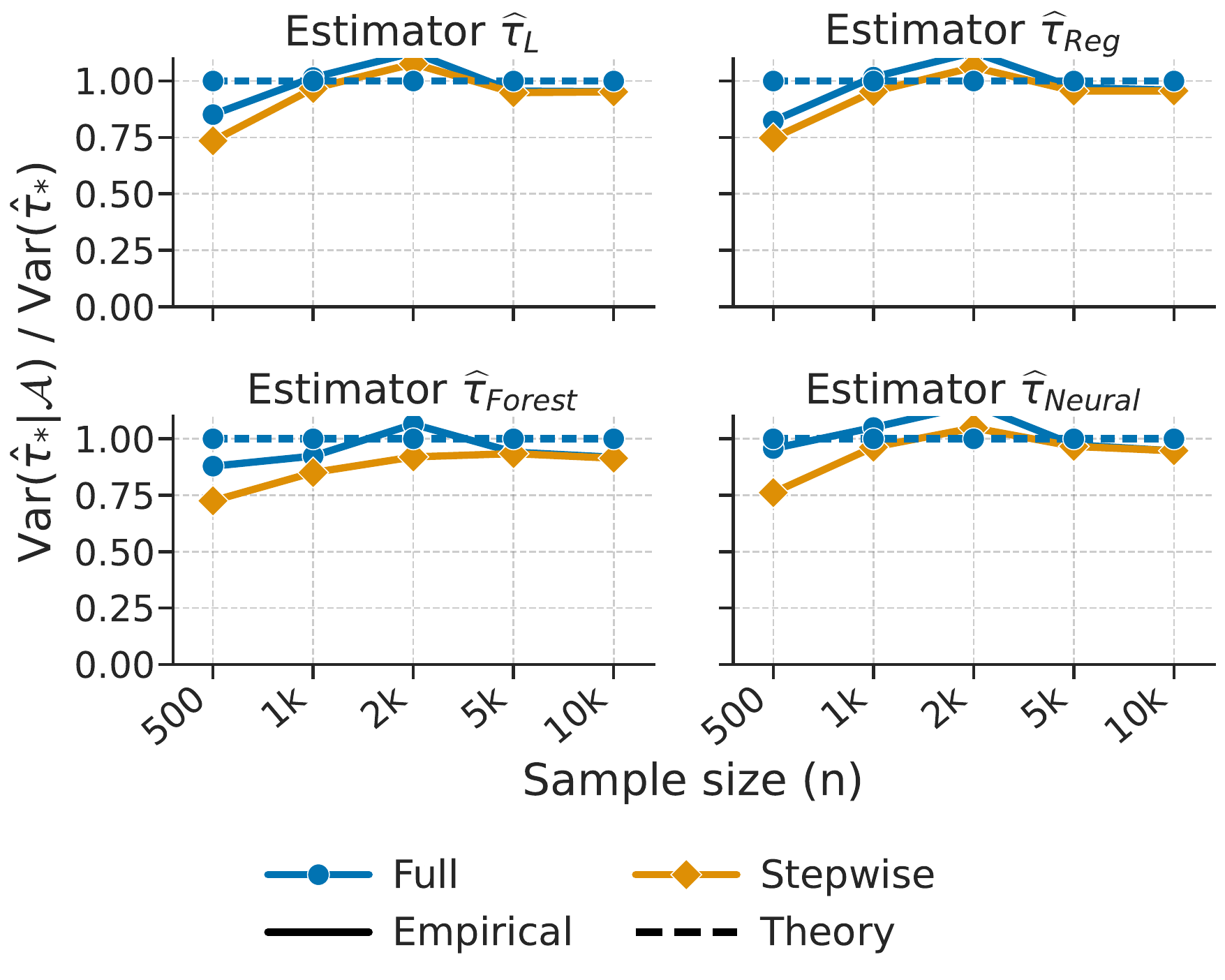}
\caption{Variance ratio comparison for the difference in linear outcome model ($\tauest{L}$), the doubly-robust estimator with linear outcome model ($\tauest{Reg}$), the doubly-robust estimator with random forest outcome model ($\tauest{Forest}$), the doubly-robust estimator with neural network outcome model ($\tauest{Neural}$) for different sample sizes in the linear setting with 100 covariates when different amounts of covariates are considered for adjustment in the analysis stage. \texttt{Stepwise} refers to when the covariate selection is done in the analysis stage prior to estimation, and \texttt{Full} refers to when the analysis is conducted using all covariates. Although there is some instability on the estimators under small samples, all estimators have precision benefits from rerandomization under smaller sample sizes.}
\label{fig:dgp3:covariate_selection:precision}
\end{figure}

\begin{figure}[ht]
\centering
\includegraphics[width=0.70\textwidth]{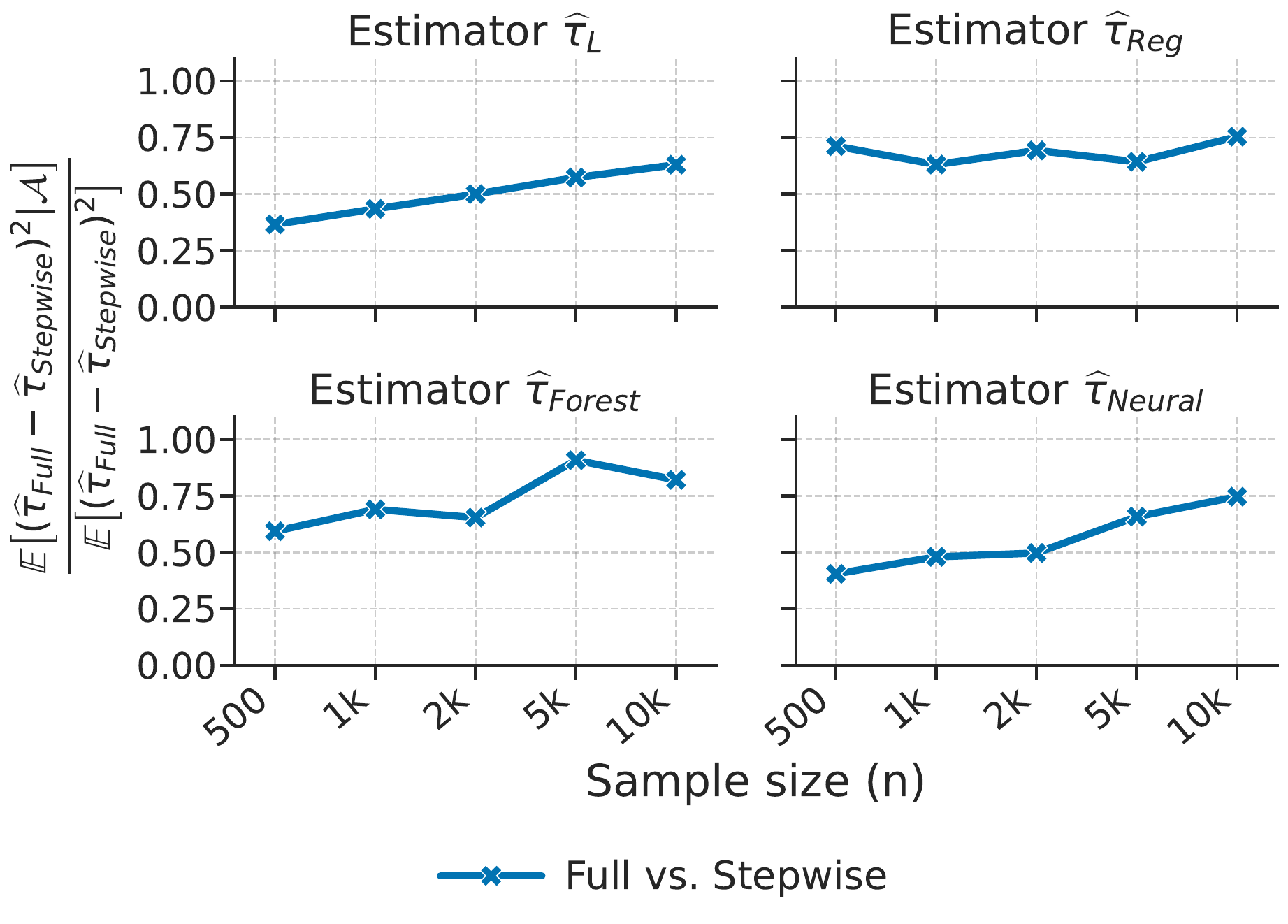}
\caption{Coherence ratio comparison for the difference in linear outcome model ($\tauest{L}$), the doubly-robust estimator with linear outcome model ($\tauest{Reg}$), the doubly-robust estimator with random forest outcome model ($\tauest{Forest}$), the doubly-robust estimator with neural network outcome model ($\tauest{Neural}$) for different sample sizes in the linear setting with 100 covariates when different amounts of covariates are considered for adjustment in the analysis stage. \texttt{Stepwise} refers to when the covariate selection is done in the analysis stage prior to estimation, and \texttt{Full} refers to when the analysis is conducted using all covariates. All estimators coherence are benefited under rerandomization.}
\label{fig:dgp3:covariate_selection:coherence}
\end{figure}

\begin{figure}[ht]
\centering
\includegraphics[width=0.70\textwidth]{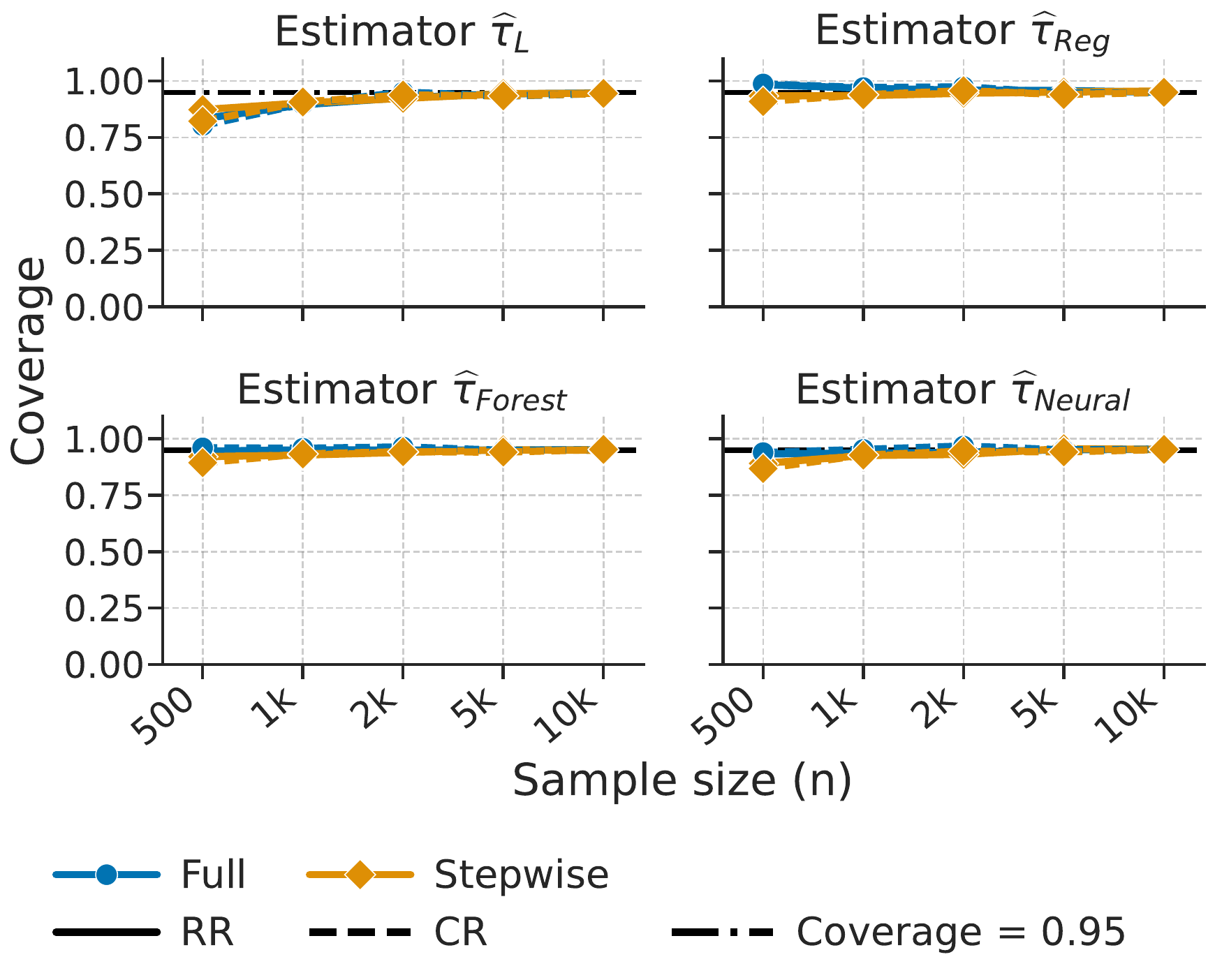}
\caption{Coverage for the difference in linear outcome model ($\tauest{L}$), the doubly-robust estimator with linear outcome model ($\tauest{Reg}$), the doubly-robust estimator with random forest outcome model ($\tauest{Forest}$), the doubly-robust estimator with neural network outcome model ($\tauest{Neural}$) for different sample sizes in the linear setting with 100 covariates when different amounts of covariates are considered for adjustment in the analysis stage. \texttt{Stepwise} refers to when the analysis does covariate selection prior to estimation, and \texttt{Full} refers to when the analysis is conducted using all covariates.  The comparison is between under complete randomization (CR) and Mahalanobis-based rerandomization (RR). Under small sample sizes all estimators are unstable due to the high-dimensionality of the estimation problem. This instability help explain the slight undercoverage for $\tauest{L}$, $\tauest{Forest}$, and $\tauest{Neural}$ and slight overcoverage for $\tauest{Reg}$.}
\label{fig:dgp3:covariate_selection:coverage}
\end{figure}

\begin{figure}[ht]
\centering
\includegraphics[width=0.70\textwidth]{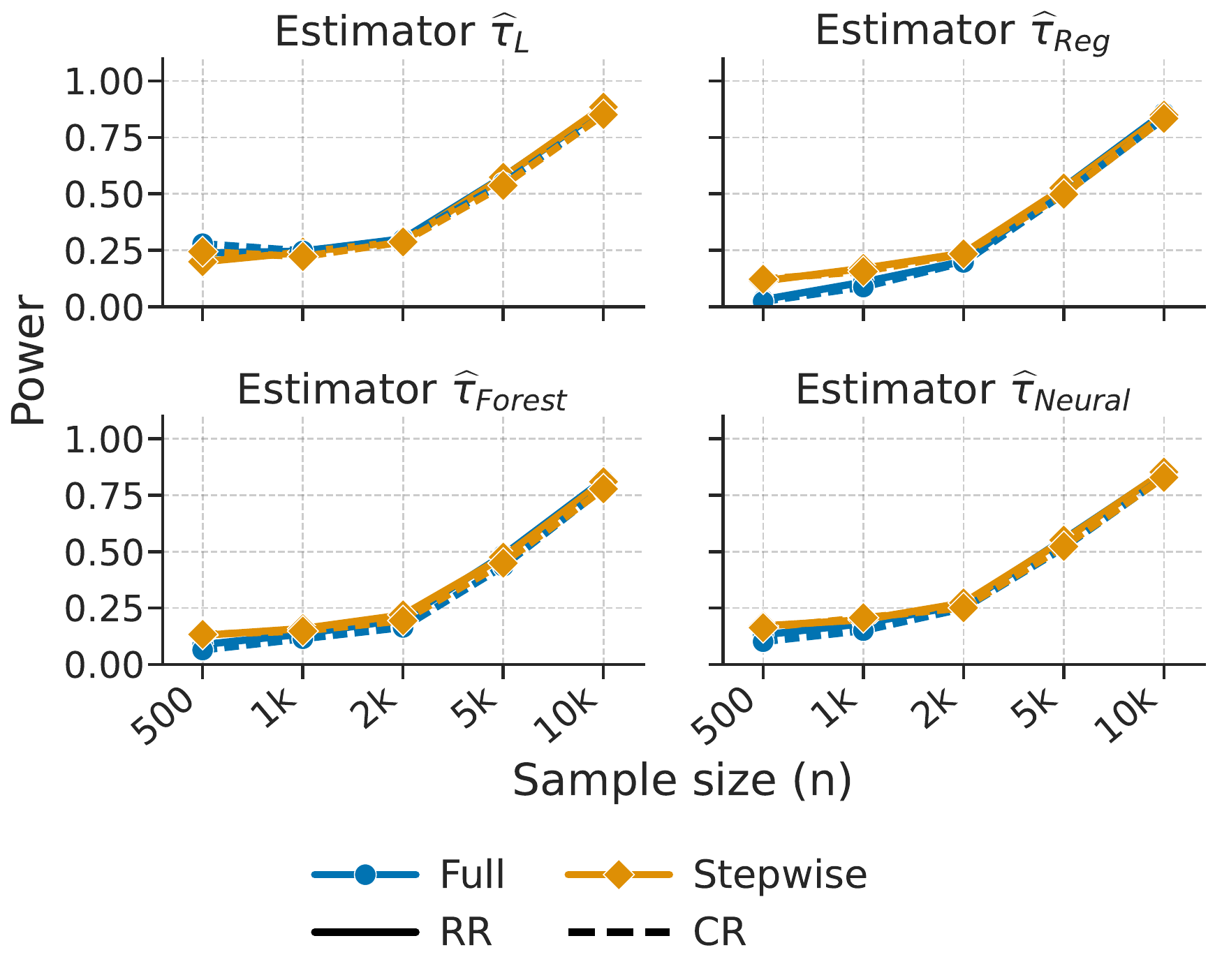}
\caption{Power for the difference in linear outcome model ($\tauest{L}$), the doubly-robust estimator with linear outcome model ($\tauest{Reg}$), the doubly-robust estimator with random forest outcome model ($\tauest{Forest}$), the doubly-robust estimator with neural network outcome model ($\tauest{Neural}$) for different sample sizes in the linear setting with 100 covariates when the analysis uses different amounts of covariates in the estimators. \texttt{Stepwise} refers to when the analysis has covariate selection prior to estimation, and \texttt{Full} refers to when the analysis uses all covariates.  The comparison is between under complete randomization (CR) and Mahalanobis-based rerandomization (RR). Under small sample sizes all estimators are unstable due to the high-dimensionality of the estimation problem, hence the lack of power. However as the sample size increases the power also increases.}
\label{fig:dgp3:covariate_selection:power}
\end{figure}

Figures~\ref{fig:dgp3:covariate_selection:precision}, \ref{fig:dgp3:covariate_selection:coherence}, \ref{fig:dgp3:covariate_selection:coverage} and \ref{fig:dgp3:covariate_selection:power} show the results on precision, coherence, coverage and power for the linear setting with 100 covariates when covariate selection is performed. Under small sample sizes, the adjusted estimators are unstable
due to the high-dimensionality of the estimation problem. Nevertheless, all estimators have precision benefits from rerandomization under smaller sample sizes. All estimators coherence are benefited under rerandomization. 
All estimators have increasing power as the sample size increases. This is due to the high-dimensional estimation problem and the instability in the estimators. This instability also causes some slight undercoverage for $\tauest{L}$, $\tauest{Forest}$, and $\tauest{Neural}$ and slight overcoverage for $\tauest{Reg}$.

\subsection{Linear setting with 10 covariates but rerandomizing based on Euclidean distance}\label{app:subsec:EL10}

\begin{figure}[ht]
\centering
\includegraphics[width=0.70\textwidth]{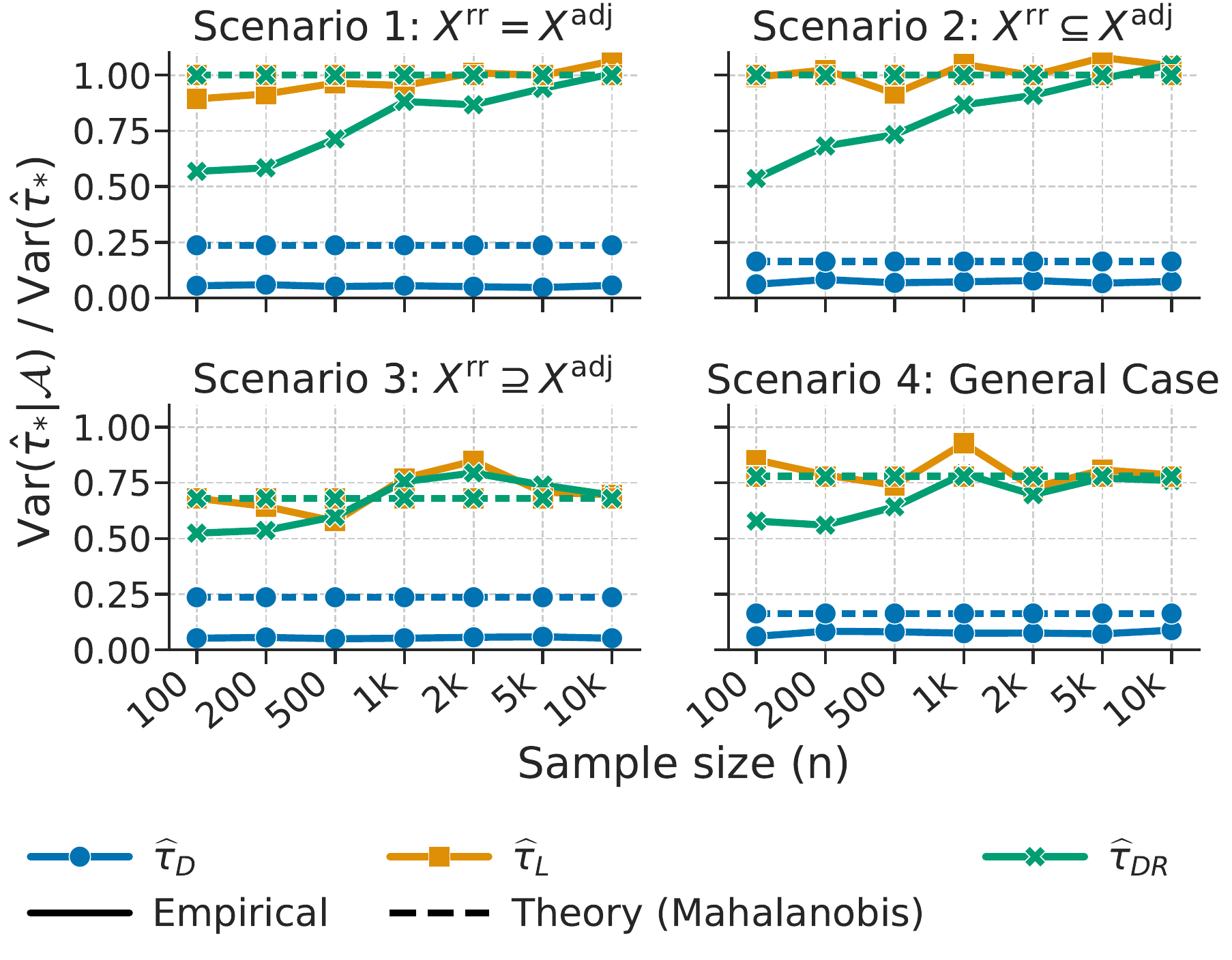}
\caption{Variance ratio comparison between the difference-in-means estimator ($\tauest{D}$), the difference in linear outcome models ($\tauest{L}$), the doubly-robust estimator with random forest outcome model ($\tauest{DR}$) for different sample sizes and covariates available for design and analysis for the linear setting with 10 covariates and rerandomization is done based on the Euclidean distance. The dashed lines show the asymptotic theoretical results based on the Mahalanobis distance, while the solid lines show the empirical results based on Euclidean distance. For $\tauest{D}$ rerandomizing under the Euclidean distance has more precision benefits than balancing over the Mahalanobis distance, this is because the Euclidean distance takes into account the correlation between the covariates \citep{schindl2024}. Meanwhile, for the adjusted estimators, rerandomization improves precision especially under the smaller sample sizes.}
\label{fig:euclidean:precision}
\end{figure}

\begin{figure}[ht]
\centering
\includegraphics[width=0.70\textwidth]{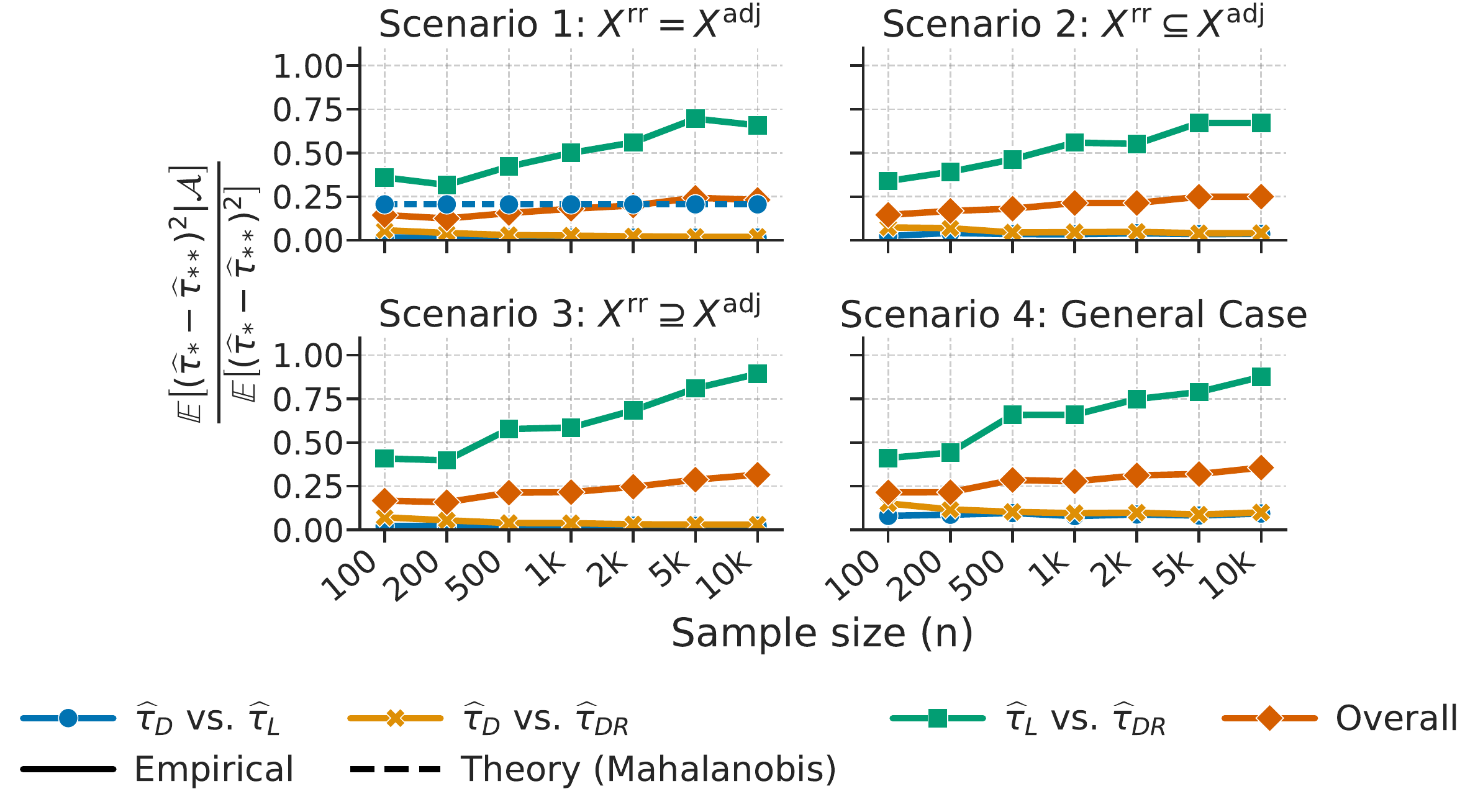}
\caption{Coherence ratio comparison between the difference-in-means estimator ($\tauest{D}$), the difference in linear outcome models ($\tauest{L}$), the doubly-robust estimator with random forest outcome model ($\tauest{DR}$) for different sample sizes and covariates available for design and analysis for the linear setting with 10 covariates and rerandomization is done based on the Euclidean distance. The dashed lines show the asymptotic theoretical results based on the Mahalanobis distance, while the solid lines show the empirical results based on Euclidean distance. Overall, the coherence among the estimators is significantly improved. Furthermore, the coherence gains between $\tauest{D}$ and $\tauest{L}$ are also improved under Euclidean distance compared to Mahalanobis distance rerandomization.}
\label{fig:euclidean:coherence}
\end{figure}

\begin{figure}[ht]
\centering
\includegraphics[width=0.70\textwidth]{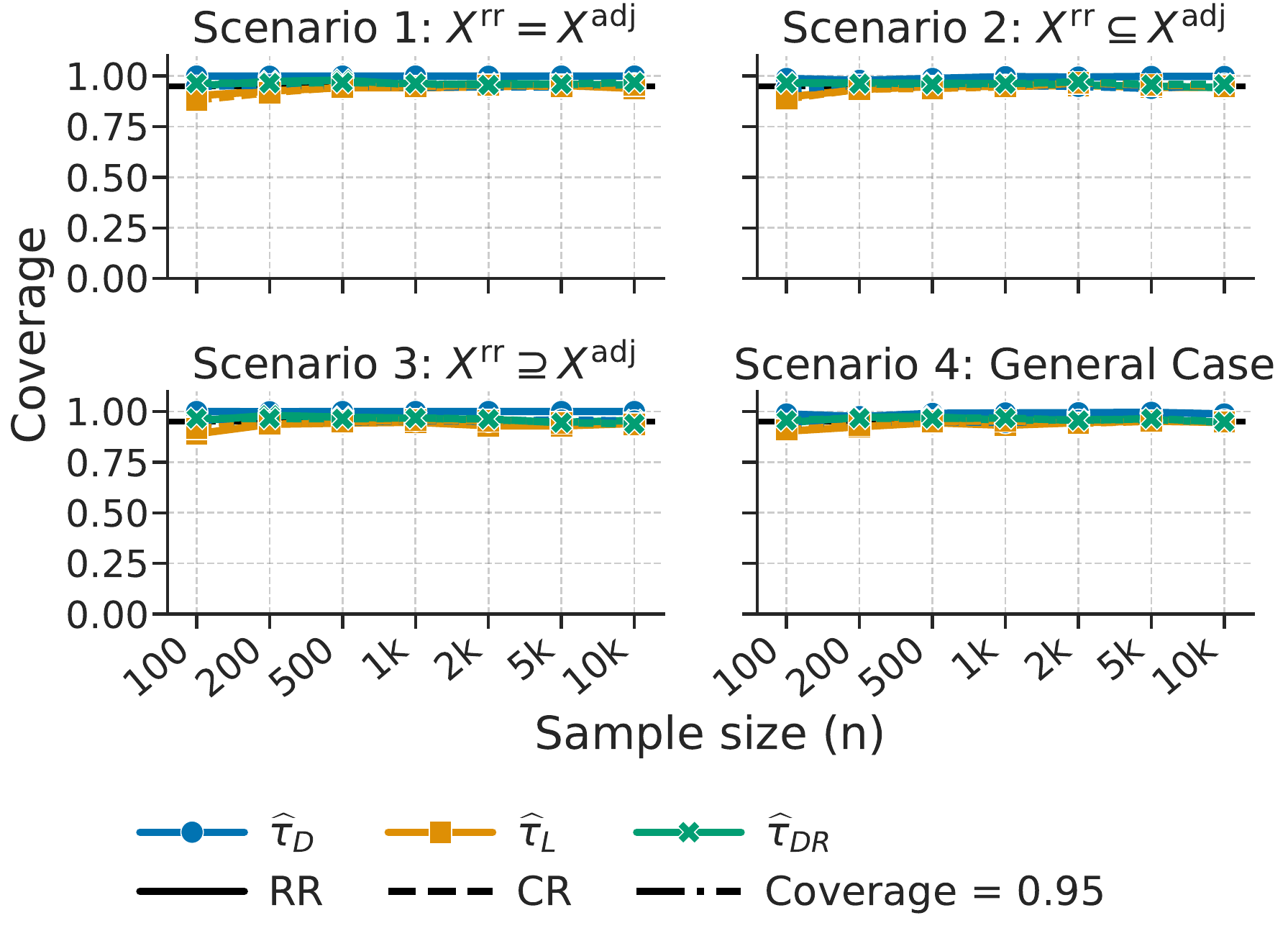}
\caption{Coverage comparison between the difference-in-means ($\tauest{D}$), difference-in-linear outcome models ($\tauest{L}$) and doubly-robust estimator with random forest outcome model ($\tauest{DR}$) for different sample sizes and covariates available for design and analysis under complete randomization (CR) and Euclidean-based rerandomization (RR) for the linear setting with 10 covariates. For all data availability scenarios both adjusted estimators achieve nominal coverage level. However, the difference-in-means estimator has slight overcoverage. Under complete randomization, the unadjusted estimator achieves nominal coverage level.}
\label{fig:euclidean:coverage}
\end{figure}

\begin{figure}[ht]
\centering
\includegraphics[width=0.70\textwidth]{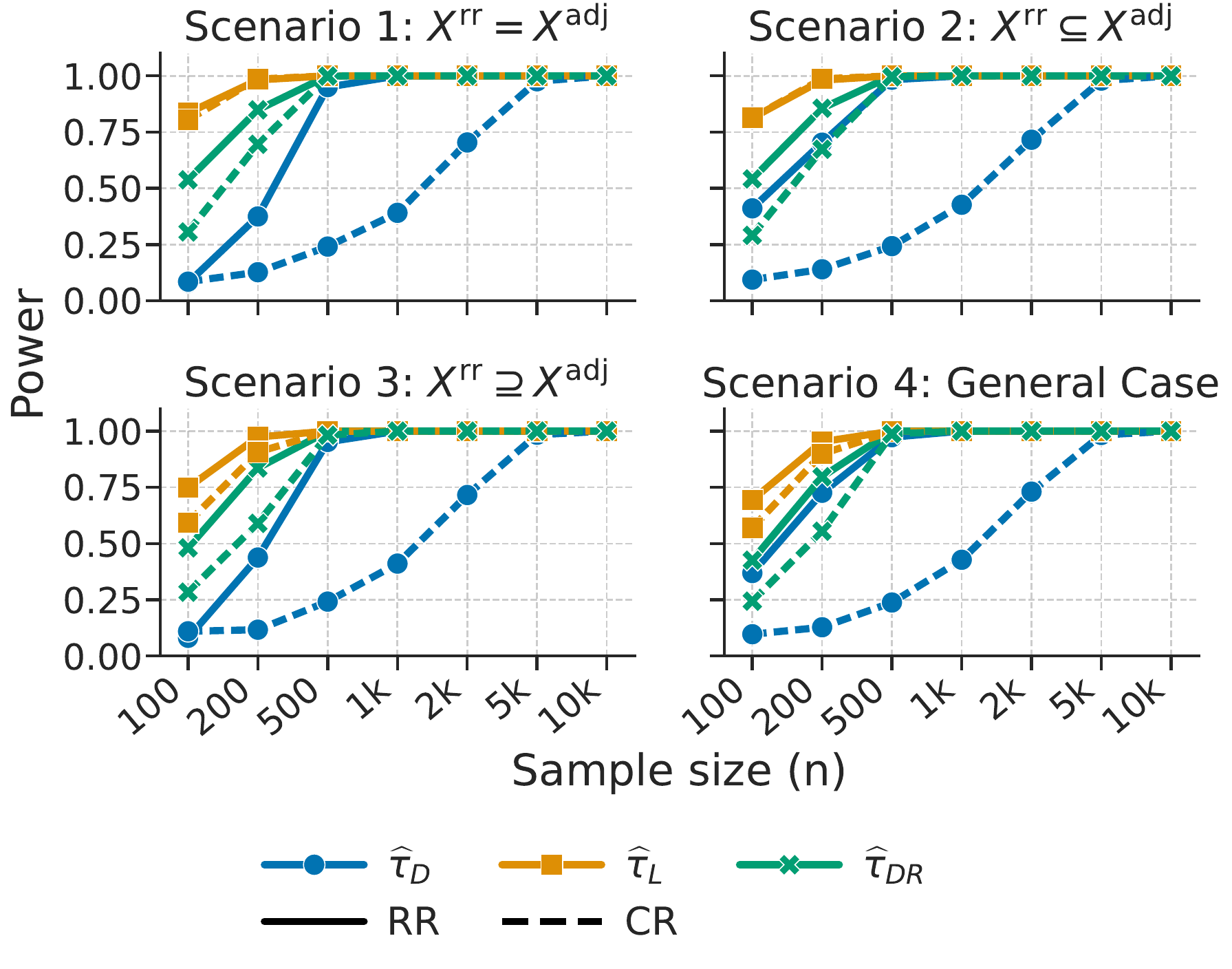}
\caption{Power comparison between the difference-in-means ($\tauest{D}$), difference-in-linear outcome models ($\tauest{L}$) and doubly-robust estimator with random forest outcome model ($\tauest{DR}$) for different sample sizes and covariates available for design and analysis under complete randomization (CR) and Euclidean-based rerandomization (RR) for the linear setting with 10 covariates. All estimators are at least as powerful under rerandomization in comparison to complete randomization for all scenarios of data availability.}
\label{fig:euclidean:power}
\end{figure}

Figures~\ref{fig:euclidean:precision}, \ref{fig:euclidean:coherence}, \ref{fig:euclidean:coverage} and \ref{fig:euclidean:power} show the results on precision, coherence, coverage, and power for the linear setting with 10 covariates, and rerandomization is done based on the Euclidean distance. The dashed lines show the asymptotic theoretical results based on the Mahalanobis distance, while the solid lines show the empirical results based on Euclidean distance. 

For $\tauest{D}$ rerandomizing under the Euclidean distance has more precision benefits than balancing over the Mahalanobis distance, this is because the Euclidean distance prioritizes balancing the top eigenvectors of the covariates \citep{schindl2024}. As before, we notice that for the adjusted estimators, rerandomization improves precision especially under the smaller sample sizes. Coherence is also highly improved under rerandomization under Euclidean distance. Notice that the coherence gains between $\tauest{D}$ and $\tauest{L}$ are also improved under Euclidean distance compared to Mahalanobis distance rerandomization. All estimators are at least as powerful under rerandomization compared to complete randomization. Moreover, both adjusted estimators achieve nominal coverage level under rerandomization. As in the case when rerandomizing based on the Mahalanobis distance, the difference-in-means estimator has slight overcoverage. Again, the overcoverage result is known in the literature \citep{li2018}.

\subsection{Varying the overlap between the design and analysis covariates}\label{app:subsec:overlap}

\begin{figure}[ht]
\centering
\includegraphics[width=0.70\textwidth]{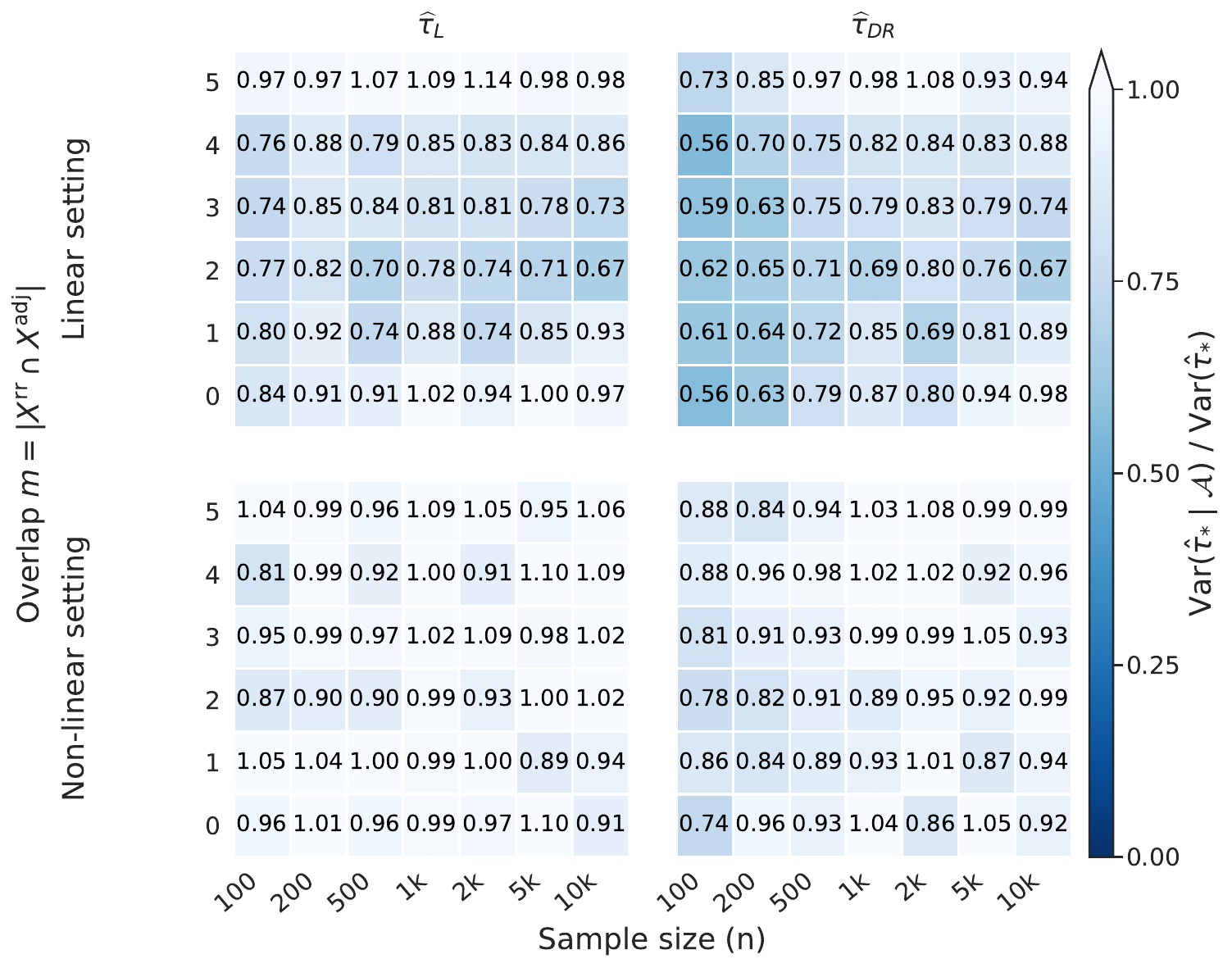}
\caption{Variance ratio of the difference in linear outcome models ($\tauest{L}$) and the doubly-robust estimator with random forest outcome model ($\tauest{DR}$) for different sample sizes and different overlaps $m$ between the 5 covariates used for rerandomization and the 5 covariates used for adjustment, for the linear and non-linear settings. In the linear setting, the precision benefits are largest for a moderate overlap ($m = 2$ or $3$). In the non-linear setting, only $\tauest{DR}$ benefits from rerandomization, under the smaller sample sizes.}
\label{fig:extra:overlap}
\end{figure}

In the four scenarios of Section~\ref{sec:simulations}, the number of covariates used for rerandomization and the overlap between $X^\rr$ and $X^\adj$ change together. Here we fix the number of covariates used in rerandomization or adjustment at $|X^\rr| = |X^\adj| = 5$ and only vary the number of covariates $m$ that the two sets have in common, from $m=5$ (rerandomization and adjustment use the same covariates) to $m=0$ (they share none). Since our simulation has 10 covariates, using 5 covariates in each stage is what allows us to vary the overlap from 0 to 5. Everything else is the same as in Section~\ref{sec:simulations:design}. Figure~\ref{fig:extra:overlap} shows the variance ratios of $\tauest{L}$ and $\tauest{DR}$ in the linear and non-linear settings. In the linear setting, both estimators are more precise under rerandomization for every overlap except $m=5$, where the variance ratio of $\tauest{L}$ is close to 1. The precision benefits are largest for a moderate overlap ($m = 2$ or $3$) and smaller when the two sets of covariates are disjoint ($m = 0$). This matches the asymptotic variance ratio of $\tauest{L}$, $1-(1-v_{5,a})R^2_{L, X^\rr}$, which is $1.00$, $0.86$, $0.78$, $0.76$, $0.81$, and $0.93$ for $m = 5, 4, 3, 2, 1$, and $0$, respectively. Intuitively, after adjusting for $X^\adj$, the outcomes are related to the other covariates only through the part of these covariates that is not explained by $X^\adj$, so rerandomization helps $\tauest{L}$ only by balancing this part. Because our covariates are highly correlated, in this particular case, it is beneficial for $X^\rr$ to contain covariate information not captured by $X^\adj$, as well as some covariate information captured by $X^\adj$. Thus, the optimal choice of overlap will depend on the covariance structure of the covariates (and their relationship to the potential outcomes). For example, in this case, if the covariates were independent, the precision benefits would be largest when the two sets of covariates are disjoint. Choosing the best sets of $X^\rr$ and $X^\adj$ for a given set of covariates is an open problem that we leave for future work. Moreover, $\tauest{DR}$ benefits more from rerandomization under the smaller sample sizes, for every overlap. In the non-linear setting, the variance ratios of $\tauest{L}$ are close to 1 for all overlaps, while $\tauest{DR}$ benefits from rerandomization under the smaller sample sizes.

\subsection{Unequal treatment allocation}\label{app:subsec:allocation}

\begin{figure}[ht]
\centering
\includegraphics[width=1\textwidth]{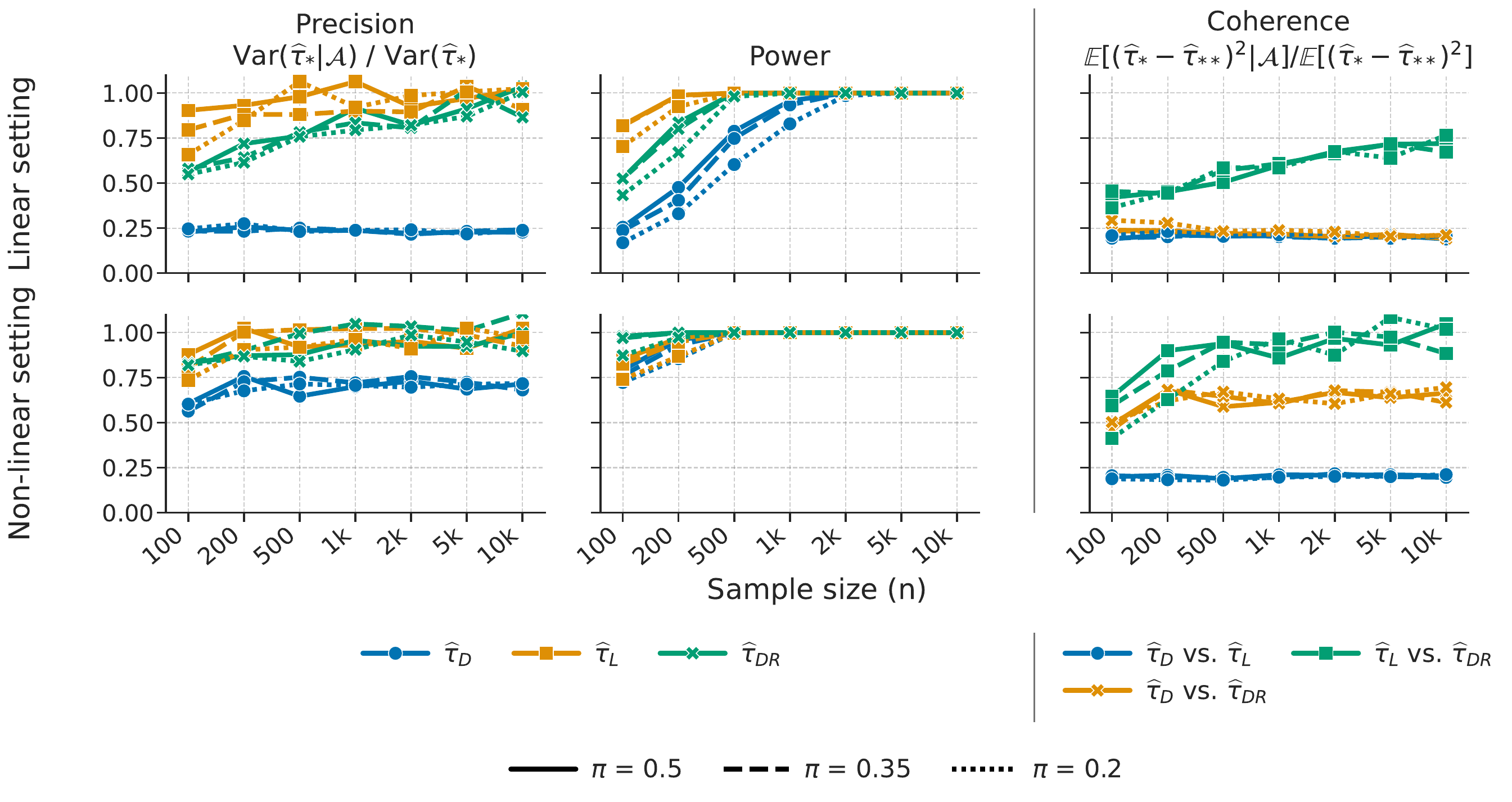}
\caption{Variance ratio, power, and coherence ratio for the difference-in-means estimator ($\tauest{D}$), the difference in linear outcome models ($\tauest{L}$), and the doubly-robust estimator with random forest outcome model ($\tauest{DR}$) under rerandomization with different treatment allocations $\pi = n_1/n$, for the linear and non-linear settings in Scenario 1 ($X^\rr = X^\adj$). The variance ratios and coherence ratios are essentially the same for the three allocations. The power of the estimators decreases for the more unbalanced allocations under the smaller sample sizes, as expected, because the variance of the estimators increases as $\pi$ moves away from 0.5. However for all values of $\pi$ as the sample increases the power also increases.}
\label{fig:extra:allocation}
\end{figure}

In Section~\ref{sec:simulations} we consider $n_1 = n_0 = n/2$. Here we repeat the simulations of Scenario 1 ($X^\rr = X^\adj$) with $\pi = n_1/n \in \{0.2, 0.35, 0.5\}$, for the linear and non-linear settings. Figure~\ref{fig:extra:allocation} shows the results on precision, power, and coherence. The variance ratios and coherence ratios are essentially the same for the three allocations. The power of the estimators decreases for the more unbalanced allocations under the smaller sample sizes, as expected, because the variance of the estimators increases as $\pi$ moves away from 0.5. However for all values of $\pi$ as the sample increases the power also increases.

\subsection{Pair-switching rerandomization and
BRAIN}\label{app:subsec:design}

\begin{figure}[ht]
\centering
\includegraphics[width=1\textwidth]{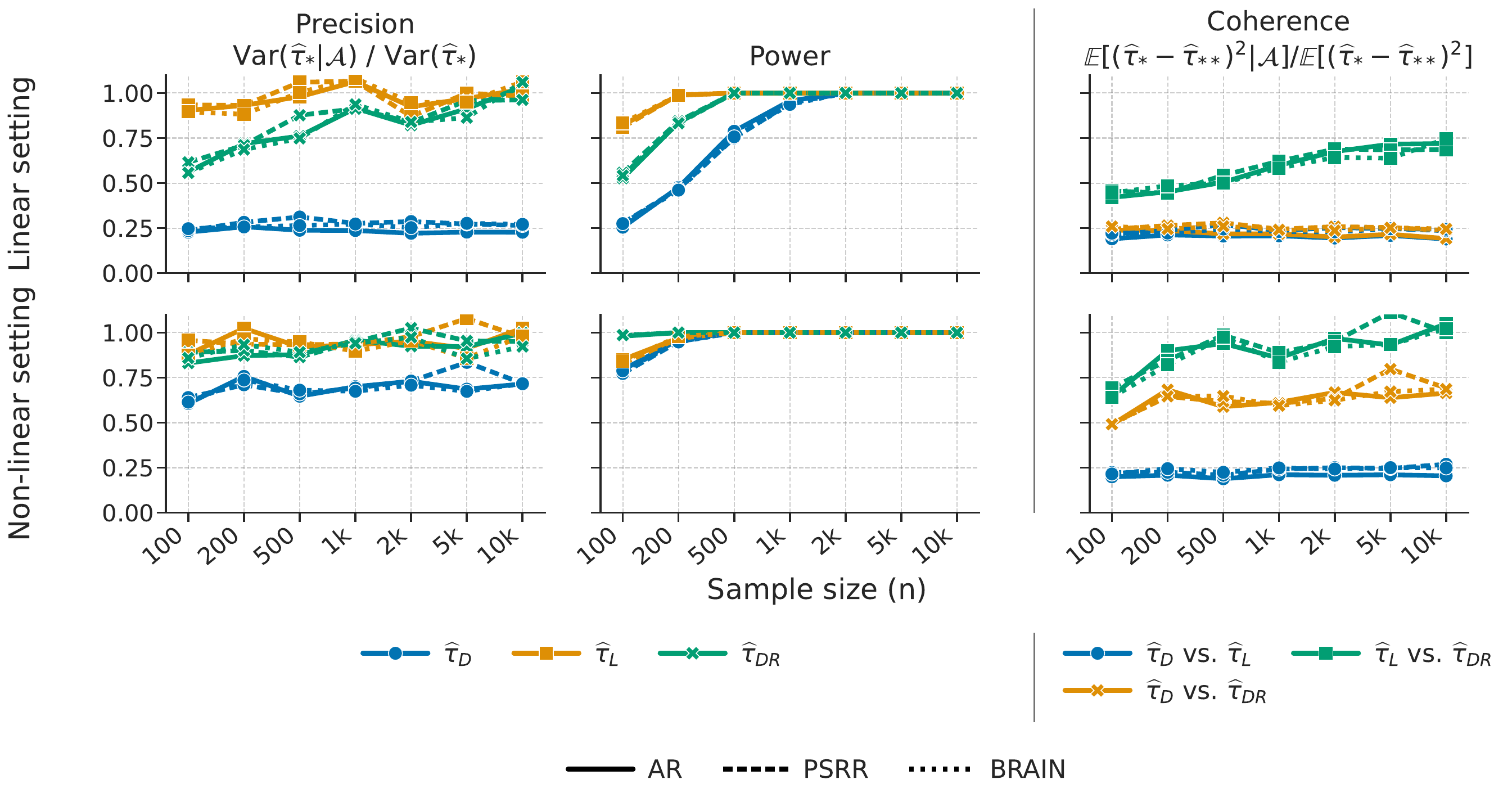}
\caption{Variance ratio, power, and coherence ratio for the difference-in-means estimator ($\tauest{D}$), the difference in linear outcome models ($\tauest{L}$), and the doubly-robust estimator with random forest outcome model ($\tauest{DR}$) under acceptance-rejection rerandomization (AR), pair-switching rerandomization (PSRR), and BRAIN, for the linear and non-linear settings in Scenario 1 ($X^\rr = X^\adj$). The three approaches produce essentially the same results. This suggests that, even though these approaches do not sample uniformly from the set of acceptable randomizations, that does not drastically affect the finite-sample or asymptotic behavior of the estimators, at least for this setting.}
\label{fig:extra:design}
\end{figure}

In Section~\ref{sec:practice:computational} we discuss the pairswitching rerandomization \citep{zhu2023} and BRAIN \citep{lu2025fast} algorithms, which find balanced randomizations more quickly than acceptance-rejection sampling (AR), but do not necessarily sample uniformly from the set of acceptable randomizations. Here we compare the three approaches in Scenario 1 ($X^\rr = X^\adj$), for the linear and non-linear settings. Figure~\ref{fig:extra:design} shows the results on precision, power, and coherence. The three approaches produce essentially the same results. This suggests that, even though these approaches do not sample uniformly from the set of acceptable randomizations, that does not drastically affect the finite-sample or asymptotic behavior of the estimators, at least for this setting.

\clearpage

